\documentclass[letterpaper,english,pra,twocolumn,groupedaddress]{revtex4}
\usepackage[T1]{fontenc}
\usepackage[latin9]{inputenc}
\usepackage{amsmath}
\usepackage{amssymb}
\usepackage{graphicx}
\usepackage{wasysym}
\usepackage{esint}

\makeatletter

\pdfpageheight\paperheight
\pdfpagewidth\paperwidth

\providecommand{\tabularnewline}{\\}

\@ifundefined{textcolor}{}
{%
 \definecolor{BLACK}{gray}{0}
 \definecolor{WHITE}{gray}{1}
 \definecolor{RED}{rgb}{1,0,0}
 \definecolor{GREEN}{rgb}{0,1,0}
 \definecolor{BLUE}{rgb}{0,0,1}
 \definecolor{CYAN}{cmyk}{1,0,0,0}
 \definecolor{MAGENTA}{cmyk}{0,1,0,0}
 \definecolor{YELLOW}{cmyk}{0,0,1,0}
}

\RequirePackage[colorlinks=true,linkcolor=blue]{hyperref}
\usepackage{dsfont}
\usepackage{bbold}
\usepackage{tikz}
\newcommand*\circled[1]{\tikz[baseline=(char.base)]{
            \node[shape=circle,draw,inner sep=2pt] (char) {#1};}}
\usepackage{dsfont}
\RequirePackage[colorlinks=true,linkcolor=blue]{hyperref}

\makeatother

\usepackage{babel}

\begin{document}

\title{High-Fidelity Hot Gates for Generic Spin-Resonator Systems}

\author{M. J. A. Schuetz,$^{1}$ G. Giedke,$^{2,3}$ L. M. K. Vandersypen,$^{4}$
and J. I. Cirac$^{1}$ }

\affiliation{$^{1}$Max-Planck-Institut für Quantenoptik, Hans-Kopfermann-Str.
1, 85748 Garching, Germany}

\affiliation{$^{2}$Donostia International Physics Center, Paseo Manuel de Lardizabal
4, E-20018 San Sebasti\'an, Spain}

\affiliation{$^{3}$Ikerbasque Foundation for Science, Maria Diaz de Haro 3, E-48013
Bilbao, Spain}

\affiliation{$^{4}$Kavli Institute of NanoScience, TU Delft, P.O. Box 5046, 2600
GA Delft, The Netherlands}

\date{\today}
\begin{abstract}
We propose and analyze a high-fidelity hot gate for generic spin-resonator
systems which allows for coherent spin-spin coupling, in the presence
of a thermally populated resonator mode. 
Our scheme is non-perturbative in the spin-resonator coupling strength,
applies to a broad class of physical systems, including for example
spins coupled to circuit-QED and surface acoustic wave resonators
as well as nanomechanical oscillators, and can be implemented readily
with state-of-the-art experimental setups. We provide and numerically
verify simple expressions for the fidelity of creating maximally entangled
states under realistic conditions. 
\end{abstract}

\maketitle

\section{Introduction}

\textit{Motivation}.---The physical realization of a large-scale quantum
information processing (QIP) architecture constitutes a fascinating
problem at the interface between fundamental science and engineering
\cite{hanson08,nielsen10}. With single-qubit control steadily improving
in various physical setups, further advances towards this goal currently
hinge upon realizing \textit{long-range} coupling between the logical
qubits, since coherent interactions at a distance do not only relax
some serious architectural challenges \cite{schreiber14}, but also
allow for applications in quantum communication, distributed quantum
computing and some of the highest tolerances in error-correcting codes
based on long-distance entanglement links \cite{nielsen10,nickerson13,knill05}.
One particularly prominent approach to address this problem is to
interface qubits with a common quantum bus which effectively mediates
long-range interactions between distant qubits, as has been demonstrated
successfully for superconducting qubits \cite{silanp07,majer07} and
trapped ions \cite{schmidt-kaler03}. 

\begin{figure}[b]
\includegraphics[width=0.9\columnwidth]{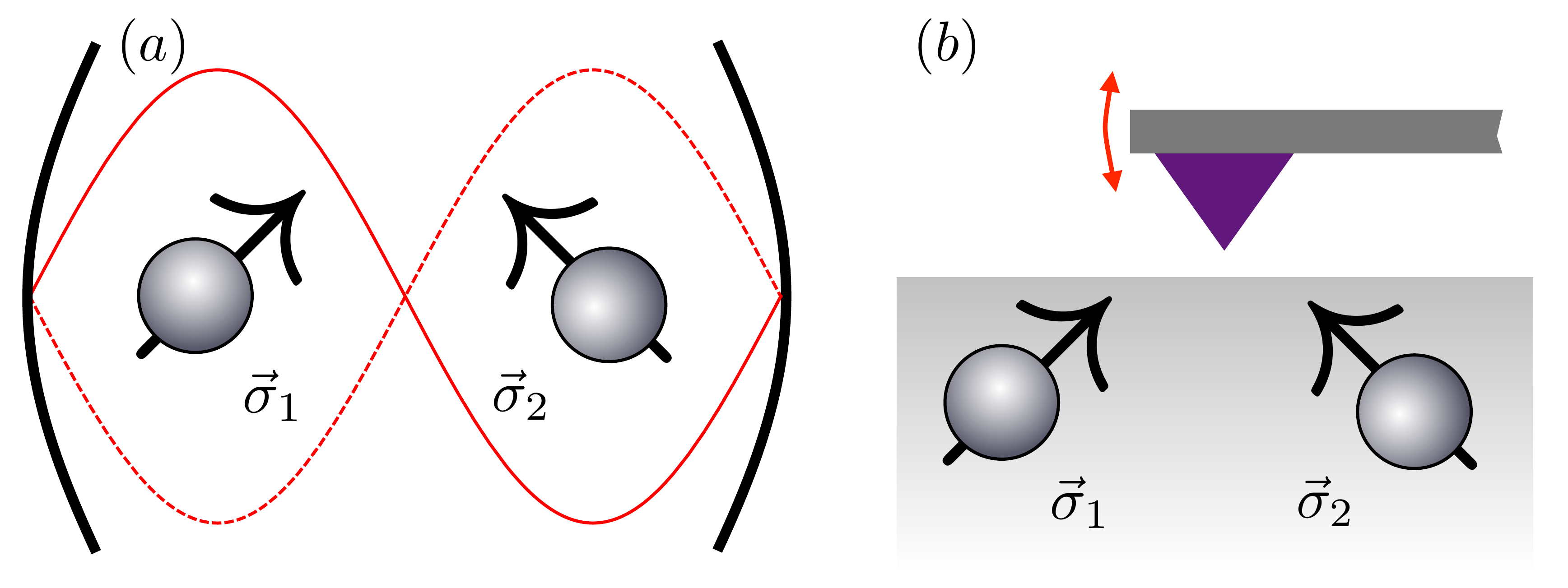}

\caption{\label{fig:implementation}(color online). Schematic illustration
for a generic spin-resonator system, comprising a set of spins $\left\{ \vec{\sigma}_{i}\right\} $
coupled to a common resonator mode 
(as provided by e.g. (a) a transmission line or (b) nanomechanical oscillators),
with a non-vanishing thermal occupation.
}
\end{figure}

\textit{Executive summary}.---In the spirit of the celebrated Sørensen-Mølmer
or similar gates for \textit{hot} trapped ions \cite{soerensen99,kirchmair09,soerensen00,milburn99,moelmer99,poyatos98,porras04,garcia-ripoll03,garcia-ripoll05,cirac00,leibfried03,milburn00},
here we propose and analyze a generic bus-based quantum gate between 
distant (solid-state) qubits coupled to one resonator mode which allows for coherent spin-spin coupling, 
even if the mode is thermally populated.
For certain times the qubits are shown to disentangle entirely from the (thermally
populated) resonator mode, thereby providing a gate that is insensitive
to the state of the resonator, without any need of cooling it to the ground state. 
While a similar gate has been considered for two superconducting qubits and (practically) zero temperature in Refs.\cite{kerman13,royer16}, 
here we show that this gate opens up the prospect of operating and coupling qubits at
elevated temperatures ${\sim}\left(1-4\right)\mathrm{K}$ (as opposed to milli-Kelvin). 
This finding brings about the potential to integrate the qubit plane right next to
the classical cryogenic electronics; therefore, our scheme may
provide a solution to the solid-state QIP interconnect problem between
the quantum (for encoding quantum information) and the classical layer
(for classical control and read-out) \cite{reilly15}.
Our approach should be accessible to a broad class of physical systems \cite{treutlein14}, 
including for example circuit QED setups with
both (i) superconducting qubits \cite{majer07,blais04,royer16,kerman13}, and
(ii) spin qubits \cite{childress04,gullans15,hu12,taylor06,trif08,jin11,kulkarni14,srinivasa16,frey12,petersson12,liu14,toida13,delbecq11,viennot14,viennot15,zou14,beaudoin16,jin12,mi17,stockklauser17},
(iii) spins coupled to surface acoustic wave (SAW) resonators \cite{schuetz15,chen15,golter16},
and (iv) spins coupled to nanomechanical oscillators \cite{rabl10,bennett13,palyi12,kepesidis13,rabl09};
compare Fig.~\ref{fig:implementation}. 
We discuss in detail the dominant sources of errors for our protocol, 
due to rethermalization of the resonator mode and qubit dephasing, 
and numerically verify the expected error scaling. 

\section{The Scheme}

We consider a set of spins
(qubits) $i{=}1,2,\dots$ with transition frequencies $\omega_{q}$
coupled to a common (bosonic) cavity mode of frequency $\omega_{c}$,
as described by the Hamiltonian $\left(\hbar{=}1\right)$
\begin{equation}
H = \omega_{c}a^{\dagger}a+\frac{\omega_{q}}{2}S^{z}+g\mathcal{S}\otimes\left(a+a^{\dagger}\right),\label{eq:generic-spin-resonator-Hamiltonian-multiple-qubits}
\end{equation}
with $\mathcal{S} {=} \sum_{i,\alpha}\eta_{i}^{\alpha}\sigma_{i}^{\alpha},$ $S^{z}{=}\sum_{i}\sigma_{i}^{z}$,
where $\vec{\sigma}_{i}$ refer to the usual Pauli matrices describing
the qubits, and $a$ is the bosonic
annihilation operator for the resonator mode. The operator $\mathcal{S}$
is a generalized (collective) spin operator which accounts for both
transversal $\left(\alpha{=}x,y\right)$ and longitudinal $\left(\alpha{=}z\right)$
spin-resonator coupling; the unit-less parameters $\eta_{i}^{\alpha}$
capture potential anisotropies and inhomogeneities in the single-photon (or single-phonon)
coupling constants $g_{i}^{\alpha}{=}\eta_{i}^{\alpha}g$. 
Similar to existing (low-temperature) schemes \cite{beaudoin16,taylor06}, 
the spin-resonator coupling $g{=}g(t)$ is assumed to be tunable on a timescale ${\ll} \omega^{-1}_{c}$;
for details we refer to Appendix \ref{sec:Time-dependent-Control}.

Typically, for artificial atoms 
such as quantum dots 
the qubit transition frequencies $\omega_{q}$ are highly tunable. 
In what follows, we
consider the regime where $\omega_{q}$ is much smaller than all other
energy scales; therefore, for the purpose of our analytical derivation,
effectively we take $\omega_{q}{=}0$. The robustness of our scheme
against non-zero splittings $\left(\omega_{q}{>}0\right)$ will be discussed
below. In this limit, the Hamiltonian given in Eq.(\ref{eq:generic-spin-resonator-Hamiltonian-multiple-qubits})
can be rewritten as 
\begin{equation}
H=\omega_{c}\left(a+\frac{g}{\omega_{c}}\mathcal{S}\right)^{\dagger}\left(a+\frac{g}{\omega_{c}}\mathcal{S}\right)-\frac{g^{2}}{\omega_{c}}\mathcal{S}^{2}.\label{eq:generic-spin-resonator-Hamiltonian-multiple-qubits-2}
\end{equation}
Using the relation $UaU^{\dagger}{=}a+\left(g/\omega_{c}\right)\mathcal{S},$
with the unitary (polaron) transformation $U{=}\exp\left[g/\omega_{c}\mathcal{S}\left(a-a^{\dagger}\right)\right]$,
Eq.(\ref{eq:generic-spin-resonator-Hamiltonian-multiple-qubits-2})
can be recast into the form 
\begin{eqnarray}
H & = & U\underset{H_{0}}{\underbrace{\left[\omega_{c}a^{\dagger}a-\frac{g^{2}}{\omega_{c}}\mathcal{S}^{2}\right]}}U^{\dagger},\label{eq:H-transformed-U-H0-Ud}
\end{eqnarray}
where we have used that $\mathcal{S}$ commutes with $U$. 
The time-evolution governed by the Hamiltonian $H$ reads
\begin{eqnarray}
e^{-iHt}=e^{-iUH_{0}U^{\dagger}t} & = & Ue^{-i\omega_{c}ta^{\dagger}a}e^{i\frac{g^{2}}{\omega_{c}}t\mathcal{S}^{2}}U^{\dagger},
\end{eqnarray}
where the second equality directly follows from $\exp\left(x\right){=}\sum_{n}x^{n}/n!$
and $U^{\dagger}U{=}\mathds1$. For certain times where $\omega_{c}t_{m}{=}2\pi m$
(with $m$ integer), the first exponential equals the identity, $\exp\left[-i\omega_{c}ta^{\dagger}a\right]{=}\exp\left[-i2\pi ma^{\dagger}a\right]{=}\mathds1$,
since the number operator $\hat{n}{=}a^{\dagger}a$ has an integer spectrum
$0,1,2,\dots$. Thus, for $t_{m}{=}\left(2\pi/\omega_{c}\right)m$,
the full time evolution reduces to 
\begin{equation}
e^{-iHt_{m}}=e^{i\frac{g^{2}}{\omega_{c}}t_{m}\mathcal{S}^{2}}=\exp\left[i2\pi m\left(g/\omega_{c}\right)^{2}\mathcal{S}^{2}\right].\label{eq:effective-stroboscopic-spin-spin-Hamiltonian}
\end{equation}
This relation comes with two major implications:
(i) Our approach is not based on a perturbative argument; therefore,
apart from Eq.(\ref{eq:effective-stroboscopic-spin-spin-Hamiltonian}),
the resonator-mediated qubit-qubit interaction does not lead to any
further undesired, spurious terms. (ii) Since the unitary transformation
given in Eq.(\ref{eq:effective-stroboscopic-spin-spin-Hamiltonian})
does \textit{not} contain any operators acting on the resonator mode,
it is completely insensitive to the state of the resonator \cite{milburn99,soerensen00,soerensen99},
even though the spin-spin interactions present in $\mathcal{S}^{2}$
have been established effectively via the resonator degrees of freedom;
similar considerations have been applied 
for the case of two (superconducting) qubits for a zero temperature mode \cite{royer16} 
and for small finite temperature $T$ in a classically modeled mode \cite{kerman13}.
For specific times, the time-evolution in the polaron and the
lab-frame fully coincide and become truly independent of the resonator
mode,
allowing for the realization of a \textit{thermally} robust gate, 
without any need of cooling the resonator mode to the ground state. 
This statement holds provided that rethermalization of the resonator
mode can be neglected over the relevant gate time. The experimental
implications for this condition will be discussed below. 

\begin{figure}
\includegraphics[width=1.0\columnwidth]{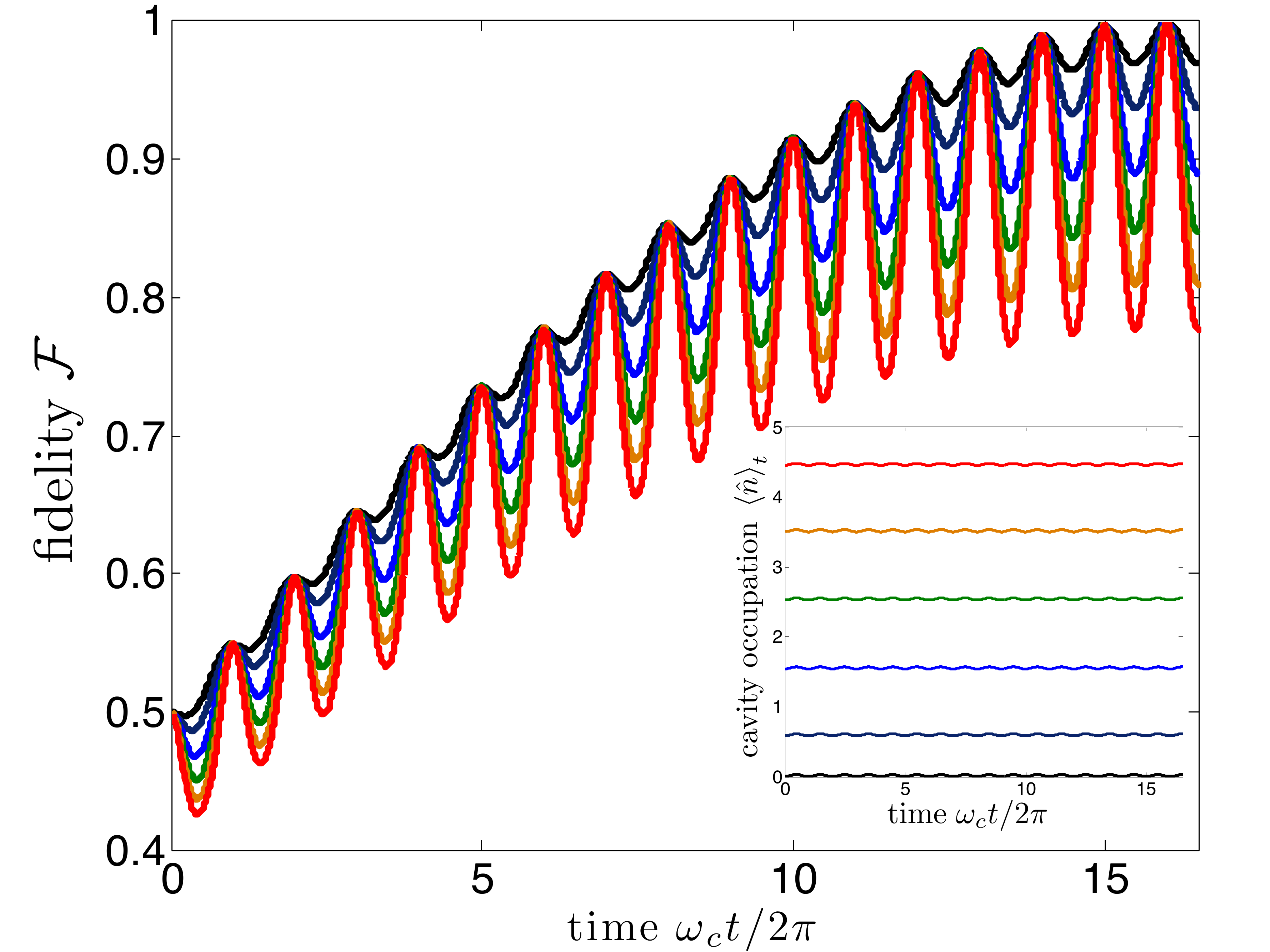}

\caption{\label{fig:fidelity-time-different-temperatures}(color online). Fidelity
$\mathcal{F}$ with the maximally entangled target state 
$\left|\Psi_{\mathrm{tar}}\right\rangle {=}\left(\left|\Uparrow\Downarrow\right\rangle +i\left|\Downarrow\Uparrow\right\rangle \right)/\sqrt{2}$
for transversal coupling $\left(\mathcal{S}{=}\sigma_{1}^{x}+\sigma_{2}^{x}\right)$,
the initial product state $\rho\left(0\right){=}\left|\Uparrow\Downarrow\right\rangle \left\langle \Uparrow\Downarrow\right|\otimes\rho_{\mathrm{th}}\left(T\right)$
and different temperatures $k_{B}T/\omega_{c}{=}0,1,2,3,4,5$. Independently
of the temperature $T$, the spins periodically disentangle from the
(hot) resonator mode and systematically build-up entanglement among
themselves. While the peaks are merely independent of temperature,
the amplitude of the precursory oscillations do increase with temperature.
Inset: Occupation of the resonator $\left\langle \hat{n}\right\rangle _{t}$
showing small oscillations due to weak entanglement between the qubits
and the cavity mode \cite{soerensen00}. Other numerical parameters:
$\omega_{q}/\omega_{c}{=}\Gamma=0$, $g/\omega_{c}{=}1/16$, $\kappa/\omega_{c}{=}Q^{-1}=10^{-5}$. }
\end{figure}

To further illustrate Eq.(\ref{eq:effective-stroboscopic-spin-spin-Hamiltonian}),
let us consider three paradigmatic examples:
(1) For longitudinal coupling ($\eta_{i}^{z}{=}1$, $\eta_{i}^{x}{=}\eta_{i}^{y}{=}0$),
as could be realized (for example) with defect spins coupled to nanomechanical
oscillators \cite{rabl10}, we can identify the effective spin-spin
Hamiltonian $H_{\mathrm{eff}}{=}\Omega_{m}\left(\sigma_{1}^{z}+\sigma_{2}^{z}\right)^{2}$,
which results in a relative phase $\phi{=}4\Omega_{m}$ for the states
$\left|11\right\rangle {=} \left|\Uparrow\Uparrow\right\rangle ,\left|00\right\rangle {=} \left|\Downarrow\Downarrow\right\rangle $
as compared to the states $\left|10\right\rangle $ and $\left|01\right\rangle $,
respectively. By adding a local unitary on both qubits, such that
$\left|0\right\rangle _{i}\rightarrow\exp\left(-i\phi/2\right)\left|0\right\rangle _{i}$
and $\left|1\right\rangle _{i}\rightarrow\exp\left(i\phi/2\right)\left|1\right\rangle _{i}$,
in total for $\phi{=}\pi/2$ we obtain a controlled phase gate $U_{\mathrm{Cphase}}{=}\mathrm{diag}\left(1,1,1,-1\right),$
which gives a phase of $-1$ exclusively to $\left|11\right\rangle $,
while leaving all other states invariant. 
Note that such a controlled phase gate can be implemented even in the presence of non-zero and inhomogeneous qubit 
level splittings $\left(\omega_{q}{>}0\right)$, when applying either fast local single qubit gates 
(to correct the effect of known $\omega_{q}{\neq}0$) or standard spin-echo techniques 
(to compensate unknown detunings), thereby
lifting the requirement of having a small qubit level splitting $\omega_{q}$; 
see Appendix  \ref{sec:Errors-due-to-level-splitting} for details.
(2) Again for longitudinal
coupling ($\eta_{i}^{z}{=}1$, $\eta_{i}^{x}{=}\eta_{i}^{y}=0$) and $N\geq2$
qubits, Eq.(\ref{eq:effective-stroboscopic-spin-spin-Hamiltonian})
results in a unitary transformation $U{=}\exp\left[-i\theta I_{z}^{2}\right]$
generated by a non-linear top Hamiltonian describing precession around
the $I_{z}{=}\sum_{i}\sigma_{i}^{z}$ axis with a rate depending on 
the $z$-component of angular momentum \cite{milburn99}, 
which can be used to simulate nonlinear spin models \cite{milburn99}. 
(3) For transversal coupling with $\mathcal{S}{=}\sigma_{1}^{x}+\sigma_{2}^{x},$
as could be realized (for example) with quantum dot based qubits embedded
in circuit-QED cavities \cite{hu12,beaudoin16} or SAW cavities \cite{chen15,schuetz15},
we have $\mathcal{S}^{2}{=}2\times\mathds1+2\sigma_{1}^{x}\sigma_{2}^{x}$.
Up to an irrelevant global phase $\phi_{\mathrm{gp}}$ due to the
first term $\sim\mathds1$, we get 
\begin{equation}
e^{-iHt_{m}}=e^{-i\phi_{\mathrm{gp}}}\underset{\equiv U_{\mathrm{id}}^{x}\left(m,g/\omega_{c}\right)}{\underbrace{\exp\left[i4\pi m\left(g/\omega_{c}\right)^{2}\sigma_{1}^{x}\sigma_{2}^{x}\right]}},\label{eq:ideal-unitary-time-evolution}
\end{equation}
which for $m\left(g/\omega_{c}\right)^{2}{=}1/16$ yields a maximally entangling gate, 
that is 
$U_{\mathrm{id}}^{x}\left(1,1/4\right)\left|\Uparrow\Downarrow\right\rangle {=}\frac{1}{\sqrt{2}}\left(\left|\Uparrow\Downarrow\right\rangle +i\left|\Downarrow\Uparrow\right\rangle \right)$
etc., i.e., initial qubit product states evolve to maximally entangled
states, irrespectively of the temperature of the resonator mode, on
a timescale $t_{\mathrm{max}}{=}\pi/8g_{\mathrm{eff}}$ (where $g_{\mathrm{eff}}{=}g^{2}/\omega_{c}$);
compare Fig.\ref{fig:fidelity-time-different-temperatures} for an
exemplary time evolution, starting initially from the product state
$\rho\left(0\right){=}\left|\Uparrow\Downarrow\right\rangle \left\langle \Uparrow\Downarrow\right|\otimes\rho_{\mathrm{th}}\left(T\right)$,
with the cavity mode in the thermal state $\rho_{\mathrm{th}}\left(T\right){=}Z^{-1}\exp\left[-\beta\omega_{c}a^{\dagger}a\right]$,
and $\beta{=}1/k_{B}T$. Indeed entanglement peaks are observed at stroboscopic
times $\left(\omega_{c}t_{m}{=}2\pi m\right)$, independent of the temperature $T$, 
culminating in a maximally entangled state at time $t_{\mathrm{max}}$.

\section{Coupling to the Environment}

In the analysis above, we have
ignored the presence of decoherence, which in any realistic setting
will degrade the effects of coherent qubit-resonator interactions.
Therefore, we complement our analytical findings with numerical simulations
of the full master equation for the system's density matrix $\rho$,
\begin{eqnarray}
\dot{\rho} & = & -i\left[H,\rho\right]+\kappa\left(\bar{n}_{\mathrm{th}}+1\right)\mathcal{D}\left[a\right]\rho+\kappa\bar{n}_{\mathrm{th}}\mathcal{D}\left[a^{\dagger}\right]\rho\nonumber \\
 &  & +\frac{\Gamma}{4}\sum_{i=1,2}\mathcal{D}\left[\sigma_{i}^{z}\right]\rho,\label{eq:Master-equation-two-qubits}
\end{eqnarray}
where the generic spin-resonator Hamiltonian $H$ is given in Eq.(\ref{eq:generic-spin-resonator-Hamiltonian-multiple-qubits})
and the last two dissipative terms in the first line of Eq.(\ref{eq:Master-equation-two-qubits}),
with $\mathcal{D}\left[a\right]\rho{=}a\rho a^{\dagger}-\frac{1}{2}\left\{ a^{\dagger}a,\rho\right\} $
and a cavity mode decay rate $\kappa{=}\omega_{c}/Q$, describe rethermalization
of the cavity mode towards the thermal occupation $\bar{n}_{\mathrm{th}}{=}(\mathrm{exp}\left[\hbar\omega_{c}/k_{B}T\right]-1)^{-1}$
at temperature $T$; here, $Q$ is the quality-factor of the cavity.
The last line in Eq.(\ref{eq:Master-equation-two-qubits}) describes
dephasing of the qubits with a dephasing rate $\Gamma{\sim}1/T_{2}^{\star}$,
where $T_{2}^{\star}$ is the time-ensemble-averaged dephasing time.
As discussed in detail in Appendix \ref{sec:Microscopic-Derivation}, 
the noise model underlying Eq.(\ref{eq:Master-equation-two-qubits}) is accurate in 
the experimentally most relevant regime of weak spin-resonator coupling $(g {\ll}\omega_{c})$,
where (within the approximation of independent rates of variation \cite{cohen92}) 
the interactions with the environment can be treated separately for spin and 
resonator degrees of freedom.
In Eq.(\ref{eq:Master-equation-two-qubits}) we have ignored single
spin relaxation processes, since the associated timescale $T_{1}$
is typically much longer than $T_{2}^{\star}$; still, relaxation
processes could be included straightforwardly in our model by adding
the decay terms $\dot{\rho}{=}\dots+T_{1}^{-1}\sum_{i}\mathcal{D}\left[\sigma_{i}^{-}\right]\rho$
and the corresponding error (infidelity) could be analyzed along the
lines of our analysis shown below 
(see Appendix \ref{sec:Relaxation-Induced-Errors} for details). 

\begin{figure}
\includegraphics[width=1\columnwidth]{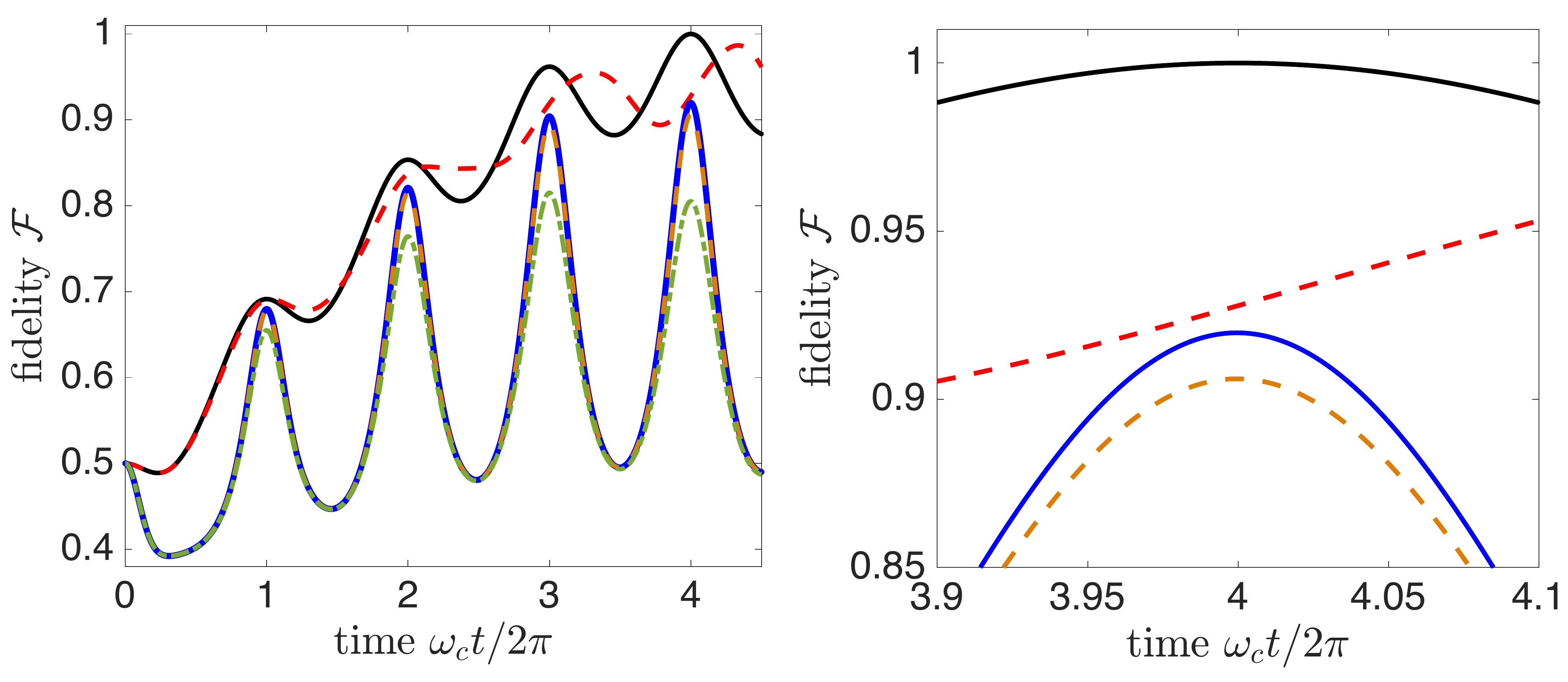}

\caption{\label{fig:log-neg-fid-smaller-coupling}(color online). 
Fidelity $\mathcal{F}$ (left) in the presence of noise,
with a zoom-in around $t_\mathrm{max}$ (right). 
As a benchmark, the solid (topmost) black
line refers to the quasi-ideal limit ($\Gamma=0$, $\kappa/\omega_{c}{=}Q^{-1}=10^{-5}$
and $k_{B}T/\omega_{c}{=}0$), while (only)
the red dashed curve accounts for
a non-zero qubit level splitting $\omega_{q}/\omega_{c}{=}0.1$. 
The solid blue line also accounts for dephasing of the qubits with a (rather large) dephasing rate $\Gamma/\omega_{c}{=}1\%$ and finite thermal
occupation of the resonator mode with $k_{B}T/\omega_{c}{=}5$ ($\bar{n}_{\mathrm{th}}{\approx}4.5$).
The results are relatively insensitive to the quality factor of the
cavity, provided that $\kappa_{\mathrm{eff}}{\ll}\Gamma$; 
the orange dashed line (where $Q{=}10^{3}$) is basically identical to the
$Q{=}10^{5}$ scenario, whereas the green dash-dotted (lowest) one with $Q{=}10^{2}$
(that is, $\kappa/\omega_{c}{=}\Gamma/\omega_{c}{=}1\%$) shows a clear
reduction in $\mathcal{F}$. 
This result can be traced back to the hot-gate requirement given in Eq.(\ref{eq:hot-gate-time-requirement}).
Ideally, maximum entanglement is reached for $f_{c}t{=}4$, with several
precursory oscillation peaks at $f_{c}t{=}1,2,3$. Other numerical parameters:
$g/\omega_{c}{=}1/8$, $\omega_{q}/\omega_{c}{=}0$ (except for the red
dashed curve where $\omega_{q}/\omega_{c}{=}0.1$).}
\end{figure}

\begin{figure*}
\includegraphics[width=2\columnwidth]{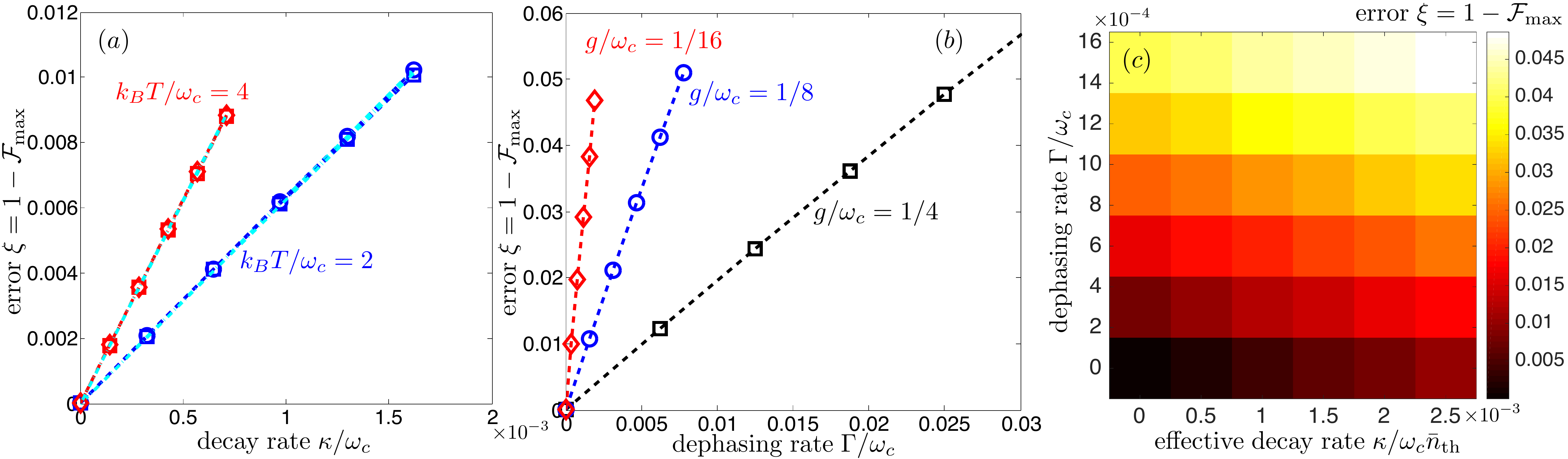}

\caption{\label{fig:error-scaling}(color online). Errors ($\xi{=}1-\mathcal{F}_{\mathrm{max}}$)
due to rethermalization of the cavity mode (a) and qubit dephasing
(b). (a) Rethermalization-induced error for $k_{B}T/\omega_{c}{=}2$
(blue) and $k_{B}T/\omega_{c}{=}4$ (red), and $\Gamma{=}0$. The error
$\xi_{\kappa}$ is found to be independent of $\mu{=}g/\omega_{c}$:
$\mu{=}1/16$ (squares) and $\mu{=}1/8$ (blue circles and red diamonds).
(b) Dephasing induced errors for $\mu{=}1/4$ (squares), $\mu{=}1/8$
(circles) and $\mu{=}1/16$ (diamonds); here, $\kappa/\omega_{c}{=}10^{-6}$
and $k_{B}T/\omega_{c}{=}0.01$. In both cases the linear error scaling
is verified. Other numerical parameters: $\omega_{q}/\omega_{c}{=}0$.
(c) Total error $\xi$ as a function of both the effective rethermalization
rate ${\sim}\kappa/\omega_{c}\bar{n}_{\mathrm{th}}{\sim}\bar{n}_{\mathrm{th}}/Q$
and the spin dephasing rate ${\sim}\Gamma/\omega_{c}$ for $g/\omega_{c}{=}1/16$,
$k_{B}T/\omega_{c}{=}2$ and $\omega_{q}{=}0$.}
\end{figure*}

\textit{Numerical results.}---To quantitatively capture the effects
of decoherence, in the following we provide numerical results of the
Master equation Eq.(\ref{eq:Master-equation-two-qubits}), for the
initial product state $\rho\left(0\right){=}\left|\Uparrow\Downarrow\right\rangle \left\langle \Uparrow\Downarrow\right|\otimes\rho_{\mathrm{th}}\left(T\right)$,
and (transversal) spin-resonator coupling with $\eta_{i}^{x}{=}1$ and
$\eta_{i}^{y}{=}\eta_{i}^{z}=0$. 
As a figure of merit 
for our protocol, 
we quantify 
the state fidelity $\mathcal{F}{=}\left<\Psi_{\mathrm{tar}}|\varrho|\Psi_{\mathrm{tar}}\right>$
with the maximally entangled target state $\left|\Psi_{\mathrm{tar}}\right\rangle {=} \left(\left|\Uparrow\Downarrow\right\rangle +i\left|\Downarrow\Uparrow\right\rangle \right)/\sqrt{2}$;
here, $\varrho{=}\mathrm{tr}_{a}\left[\rho\right]$
refers to the density matrix of the qubits, with $\mathrm{tr}_{a}\left[\dots\right]$ denoting the trace over the resonator degrees of freedom.
As shown in Appendix \ref{sec:Average-Gate-Fidelity},
similar results can be obtained for the average gate fidelity. Typical
results from our numerical simulations in the presence of noise are
displayed in Fig.\ref{fig:log-neg-fid-smaller-coupling}. As expected
from our analytical results, for $\omega_{c}t_{m}{=}2\pi m$ the two
qubits disentangle from the thermally populated resonator mode and
systematically evolve towards the maximally entangled target state
$\left|\Psi_{\mathrm{tar}}\right\rangle $; for example, for $g/\omega_{c}{=}1/8$
(as used in Fig.\ref{fig:log-neg-fid-smaller-coupling}), the spins
evolve towards $U_{\mathrm{id}}^{x}\left(1,1/8\right)\left|\Uparrow\Downarrow\right\rangle {=}\cos\left(\pi/16\right)\left|\Uparrow\Downarrow\right\rangle +i\sin\left(\pi/16\right)\left|\Downarrow\Uparrow\right\rangle $
for $m{=}1$, $U_{\mathrm{id}}^{x}\left(2,1/8\right)\left|\Uparrow\Downarrow\right\rangle {=}\cos\left(\pi/8\right)\left|\Uparrow\Downarrow\right\rangle +i\sin\left(\pi/8\right)\left|\Downarrow\Uparrow\right\rangle $
for $m{=}2$, and $U_{\mathrm{id}}^{x}\left(3,1/8\right)\left|\Uparrow\Downarrow\right\rangle {=}\cos\left(3\pi/16\right)\left|\Uparrow\Downarrow\right\rangle +i\sin\left(3\pi/16\right)\left|\Downarrow\Uparrow\right\rangle $
for $m{=}3$, before the entanglement build-up culminates in the fully-entangling
dynamics $U_{\mathrm{id}}^{x}\left(4,1/8\right)\left|\Uparrow\Downarrow\right\rangle {=}\left(\left|\Uparrow\Downarrow\right\rangle +i\left|\Downarrow\Uparrow\right\rangle \right)/\sqrt{2}$.
For all practical purposes, this statement holds independently of
the temperature $T$ and the associated thermal occupation of the
resonator mode $\bar{n}_{\mathrm{th}} {\approx} k_{B}T/\hbar\omega_{c}$,
provided that the quality factor of the cavity $Q$ is sufficiently
high; a quantitative statement specifying this regime will be given
below. Moreover, while our analytical treatment has assumed $\omega_{q}{=}0$,
we have numerically verified that the proposed protocol is robust
against non-zero level splittings of the qubits $\omega_{q}/\omega_{c}{\lesssim}0.1$;
compare the dashed line in Fig.\ref{fig:log-neg-fid-smaller-coupling}
and further information provided in Appendices \ref{sec:Non-Zero-Qubit-Level-Splitting-1},
\ref{sec:Errors-due-to-level-splitting} and \ref{sec:Additional-Numerical-Results}.

\section{Gate time requirements: Error scaling}

As described by
Eq.(\ref{eq:Master-equation-two-qubits}), coupling to the environment
leads to two dominant error sources: (i) rethermalization of the resonator
mode with an effective rate ${\sim}\kappa\bar{n}_{\mathrm{th}}$, and
(ii) dephasing of the qubits on a timescale ${\sim} T_{2}^{\star}$.
For any \textit{hot} gate, the associated gate time $t_{\mathrm{gate}}{\sim} g_{\mathrm{eff}}^{-1}$,
with $g_{\mathrm{eff}}{=}g^{2}/\omega_{c}{=}\mu^{2}\omega_{c}$, has to
be shorter than the time-scale associated with the effective (thermally-enhanced)
rethermalization rate $\kappa_{\mathrm{eff}}{=}\kappa\bar{n}_{\mathrm{th}}{\approx} k_{B}T/Q$.
For the gate described above, 
this directly leads to the requirement 
\begin{equation}
g^{2}/\omega_{c}\gg k_{B}T/Q\,\,\,\,\,\Leftrightarrow\,\,\,\,\,k_{B}T\ll Q\mu^{2}\omega_{c}.\label{eq:hot-gate-time-requirement}
\end{equation}
Thus, for $T{=}1\mathrm{K}$ $\left(k_{B}T/2\pi{\approx}20\mathrm{GHz}\right)$
and a cavity quality factor $Q{\approx}10^{5}-10^{6}$, we need $g_{\mathrm{eff}}/2\pi{\gg}\left(20-200\right)\mathrm{kHz}$.
Provided that our assumption $\omega_{c}{\gg}\omega_{q}$ is still fulfilled,
for fixed temperature $T$, quality factor $Q$ and coupling $g$,
relation (\ref{eq:hot-gate-time-requirement}) may be conveniently
fulfilled by choosing $\omega_{c}$ sufficiently small, up to the
lower limit $\omega_{c}{\geq}4g$ (which is needed to fulfill $m{\geq}1$; compare Appendix \ref{sec:Gate-Time}) 
and at the cost of a potentially relatively
large device (since the device dimensions scale with ${\sim}\lambda_{c}{\sim}\omega_{c}^{-1}$).
Conversely, for fixed $\mu{=}g/\omega_{c}$ \cite{taylor06,chen15,schoelkopf08},
Eq.(\ref{eq:hot-gate-time-requirement}) can be achieved by choosing
$\omega_{c}$ sufficiently large. In addition, the gate time has to
be short compared to the qubit's dephasing time $T_{2}^{\star}{\sim}\Gamma^{-1}$,
which gives the second requirement 
\begin{equation}
g^{2}/\omega_{c}\gg\Gamma\,\,\,\,\,\Leftrightarrow\,\,\,\,\,\Gamma\ll\mu^{2}\omega_{c}.\label{eq:dephasing-requirement}
\end{equation}
For concreteness, let us consider a specific setup where conditions
(\ref{eq:hot-gate-time-requirement}) and (\ref{eq:dephasing-requirement})
can be met with state-of-the-art technology: Quantum dots (QDs) have
been successfully integrated with superconducting microwave cavities,
with a relatively large charge-cavity coupling of $g_{\mathrm{ch}}/2\pi{\sim}\left(20-100\right)\mathrm{MHz}$
\cite{liu14,frey12,petersson12,viennot14,toida13}. For QD spin qubits
a vacuum Rabi frequency of $g_{\mathrm{sp}}/2\pi{\sim}1\mathrm{MHz}$
has been predicted \cite{hu12,petersson12,jin12}, with the potential
to increase this coupling to $\sim10\mathrm{MHz}$ with new, recently
demonstrated cavity designs \cite{samkharadze15}. Furthermore, for
superconducting transmission line resonators quality factors $Q{\sim}10^{6}$
have been demonstrated \cite{barends08}. Then, taking $g_{\mathrm{sp}}/2\pi{=}10\mathrm{MHz}$,
$\omega_{c}/2\pi{\approx}(0.16-1)\mathrm{GHz}$, i.e., $g_{\mathrm{eff}}/2\pi{\approx}(0.1-0.6)\mathrm{MHz}$,
and $Q{=}10^{6}$, conditions (\ref{eq:hot-gate-time-requirement})
and (\ref{eq:dephasing-requirement}) can be met simultaneously for
temperatures $T{\sim}1\mathrm{K}$ {[}since $T{\ll}5(30)\mathrm{K}$ to
fulfill condition (\ref{eq:hot-gate-time-requirement}) for $g_{\mathrm{eff}}/2\pi{\approx}0.1(0.6)\mathrm{MHz}${]}
and dephasing timescales $T_{2}^{\star}{\sim}100\mu\mathrm{s}$ {[}since
$\Gamma/2\pi{\ll}\left(0.1-0.6\right)\mathrm{MHz}$ to fulfill condition
(\ref{eq:dephasing-requirement}){]}, as has been demonstrated with
isotopically purified Si samples \cite{veldhorst14}. Therefore, a
faithful implementation of our gate will \textit{not} require cooling
to milli-Kelvin temperatures. Similar promising estimates also apply
to spin-qubits coupled to SAW-resonators; compare Appendix \ref{sec:SAW-based-Spin-Resonator-Coupling}. 

In the following, we quantify the infidelities induced by the two
error sources outlined above: Rethermalization of the resonator mode
during the gate leads to errors (infidelities) if the resonator is
entangled with the qubits. Due to leakage of which-way information,
resonator noise leads to qubit dephasing at a rate proportional to
the relevant separation in phase space, that is the square of the
resonator displacement $\mu{=}g/\omega_{c}$ \cite{rabl10}. The effective
rethermalization-induced dephasing rate for the qubits is then $\Gamma_{\mathrm{eff}}{\sim}\kappa\bar{n}_{\mathrm{th}}\left(g/\omega_{c}\right)^{2}$.
To obtain a simple estimate for the rethermalization-induced error,
this effective rate $\Gamma_{\mathrm{eff}}$ is multiplied with the
relevant gate time which scales as $t_{\mathrm{gate}}{\sim}\omega_{c}/g^{2}$,
yielding the error $\xi_{\kappa}{\sim}\left(\kappa/\omega_{c}\right)\bar{n}_{\mathrm{th}}$,
which is \textit{independent} of the spin-resonator coupling strength
$g$ \cite{royer16,rabl10}; 
for a full analytical derivation we refer to Appendix \ref{sec:Analytical-Expression-for-Rethermalization-Induced-Errors}.
However, since the overall
gate time $t_{\mathrm{gate}}{\sim}\omega_{c}/g^{2}$ increases for small
$\mu{=}g/\omega_{c}$, errors will accumulate due to direct qubit decoherence
processes. Accordingly, errors due to qubit dephasing are expected
to scale as $\xi_{\Gamma}{\sim}\Gamma/g_{\mathrm{eff}}{\sim}\mu^{-2}\Gamma/\omega_{c}.$
This simple linear scaling holds for a Markovian noise model where
qubit dephasing is described by a standard pure dephasing term {[}compare
Eq.(\ref{eq:Master-equation-two-qubits}){]} leading to an exponential
loss of coherence $\sim\exp\left[-t/T_{2}^{\star}\right]$; 
for non-Markovian qubit dephasing a better, sub-linear scaling can
be expected \cite{schuetz15,rabl10}. For small infidelities $\left(g_{\mathrm{eff}}{\gg}\kappa_{\mathrm{eff}},\Gamma\right)$,
the individual linear error terms due to cavity rethermalization and
qubit dephasing can be added independently, yielding the total error
\begin{equation}
\xi\approx\alpha_{\kappa}\left(\kappa/\omega_{c}\right)\bar{n}_{\mathrm{th}}+\alpha_{\Gamma}\Gamma/\omega_{c}.\label{eq:total-error-estimate}
\end{equation}
This simple linear error model has been verified numerically; compare
Fig.\ref{fig:error-scaling}. 
Based on these results we extract the
coefficients $\alpha_{\kappa}{\approx}4$ (which is approximately independent
of $g$ \cite{royer16}; compare Appendices \ref{sec:Additional-Numerical-Results} and \ref{sec:Analytical-Expression-for-Rethermalization-Induced-Errors} for details) 
and $\alpha_{\Gamma}{\approx}0.1/\mu^{2}$.
For $g_{\mathrm{sp}}/2\pi{\approx}10\mathrm{MHz}$
\cite{samkharadze15,hu12,jin12}, a relatively low resonator frequency
$\omega_{c}/2\pi{=}16g_{\mathrm{sp}}/2\pi{=}160\mathrm{MHz}$, $T{=}1\mathrm{K}$
(corresponding to $\bar{n}_{\mathrm{th}}{\approx}130$), $Q{=}10^{5}$
\cite{samkharadze15,barends08} and a realistic dephasing rate $\Gamma/2\pi{\approx}0.1\mathrm{MHz}$
\cite{veldhorst14}, that is $\kappa/\omega_{c}\bar{n}_{\mathrm{th}}{\approx}1.3\times10^{-3}$
and $\Gamma/\omega_{c}{\approx}6\times10^{-4}$, our estimates then
predict an overall infidelity of $\xi{\approx}2\%$, with the potential
to reach error rates $\xi{\approx}0.2\%$ below the threshold for quantum
error correction for state-of-the-art experimental parameters ($Q{\approx}10^{6}$,
$\Gamma/2\pi{\approx}10\mathrm{kHz}$) \cite{knill05,veldhorst14,barends08}.
This simple estimate 
compares well with other bus-based, two-qubit (hot) gates reaching fidelities ${\sim}97\%$  \cite{kirchmair09, rabl10,chow12}
and has been corroborated by numerical simulations that
fully account for higher-order errors; compare 
the density plot in
Fig.\ref{fig:error-scaling}(c).
We like to emphasize that, due to the fundamental
temperature-insensitivity of our gate, technological improvements in the
achievable $Q$-factor directly translate to a proportional reduction of
thermalization-induced errors and therefore increase the acceptable
temperature.
Note that the error estimate given
in Eq.(\ref{eq:total-error-estimate}) assumes perfect timing of the
gate, as the maximum fidelity is reached exactly at time $t_{\mathrm{max}}$,
whereas under experimentally realistic conditions there will be a
residual error due to imperfect timing of the gate. However, as shown
in Appendix \ref{sec:Additional-Numerical-Results}, for sufficiently small, but realistic
timing accuracies of $\left(\omega_{c}/2\pi\right)\Delta t{\lesssim}1\%$
and small spin-resonator coupling $g/\omega_{c}{\lesssim}1/16$ (implying
small oscillation amplitudes), the effects of time-jitter become negligible. 

\section{Conclusions \& Outlook}

To conclude, we have proposed and
analyzed a high-fidelity hot gate for generic spin-resonator systems
which allows for coherent spin-spin coupling, even in the presence
of a thermally populated resonator mode. While we have mostly focused
on just two spins, our scheme fully applies to more than two spins,
which should allow for the preparation of maximally entangled multi-partite
states; as shown in Ref.\cite{moelmer99} in the context of trapped
ions, a propagator of the form given in Eq.(\ref{eq:effective-stroboscopic-spin-spin-Hamiltonian})
applied to the initial product state $\left|00\cdots0\right\rangle $
may be used to generate states of the form $1/\sqrt{2}\left(\left|00\cdots0\right\rangle +e^{i\phi}\left|11\cdots1\right\rangle \right)$,
where $\left|00\cdots0\right\rangle $ and $\left|11\cdots1\right\rangle $
are product states with all qubits in the same state $\left|0\right\rangle $
or $\left|1\right\rangle $, respectively. 


\begin{acknowledgments}
M.J.A.S. would like to thank T. Shi for
useful discussions. M.J.A.S., L.M.K.V. and J.I.C. acknowledge support
by the EU project SIQS. M.J.A.S. and J.I.C. also acknowledge support
by the DFG within the Cluster of Excellence NIM. G.G. acknowledges
support by the Spanish Ministerio de Economía y Competitividad through
the Project FIS2014-55987-P. L.M.K.V. acknowledges support by a European
Research Council Synergy grant.
\end{acknowledgments}

\appendix 

\section*{Appendices}

The following Appendices provide additional background material to
specific topics of the main text. They are structured as follows:
In Sec.\ref{sec:Thermal-Occupation} we provide typical thermal occupation
numbers $\bar{n}_{\mathrm{th}}$ for relevant experimental parameter
regimes. In Sec.\ref{sec:Polaron-vs.-Lab} we compare the ideal evolution
in the lab frame to the one in the polaron frame. In Sec.\ref{sec:Gate-Time}
we derive the ideal gate time $t_{\mathrm{max}}$. In Sec.\ref{sec:Time-dependent-Control}
we discuss a prototypical implementation of a spin-resonator system
that allows for time-dependent control of the spin-resonator $g=g\left(t\right)$,
as required for the faithful realization of the proposed hot gate.
In Sec.\ref{sec:Spin-Spin-Coupling-in-Dispersive-Regime} we discuss
the standard approach to coupling spins via a common resonator mode
in the dispersive regime, in which, in contrast to the proposed hot
gate, the spin degrees of freedom do not fully disentangle from the
resonator mode. In Sec.\ref{sec:Schrieffer-Wolff-Transformation}
we compare our general result to a perturbative calculation in the
framework of a Schrieffer-Wolff transformation. In Secs.\ref{sec:Non-Zero-Qubit-Level-Splitting-1}
and \ref{sec:Errors-due-to-level-splitting} we analyze in detail
the effects coming from a non-zero qubit level splitting ($\omega_{q}/\omega_{c}>0$).
In Sec.\ref{sec:SAW-based-Spin-Resonator-Coupling} we provide further
details on how to implement experimental candidate systems governed
by the class of Hamiltonians given in Eq.(1), using quantum dots embedded
in high-quality surface acoustic wave (SAW) resonators. In Sec.\ref{sec:Microscopic-Derivation}
we provide a microscopic derivation of the Master equation given in
Eq.(7) of our manuscript. In Sec.\ref{sec:Additional-Numerical-Results}
we present further results based on the numerical simulation of the
master equation given in Eq.(7) of the main text. In Sec.\ref{sec:Analytical-Expression-for-Rethermalization-Induced-Errors}
we derive an analytical expression for rethermalization-induced errors,
while Sec.\ref{sec:Analytical-Model-for-dephasing-errors} provides
an analytical model for dephasing-induced errors. In Sec.\ref{sec:Relaxation-Induced-Errors}
we address in detail errors induced by relaxation processes. In Sec.\ref{sec:Average-Gate-Fidelity}
we conclude with a discussion on the average gate fidelity.

\section{Thermal Occupation \label{sec:Thermal-Occupation}}

\begin{figure}[b]
\includegraphics[width=1\columnwidth]{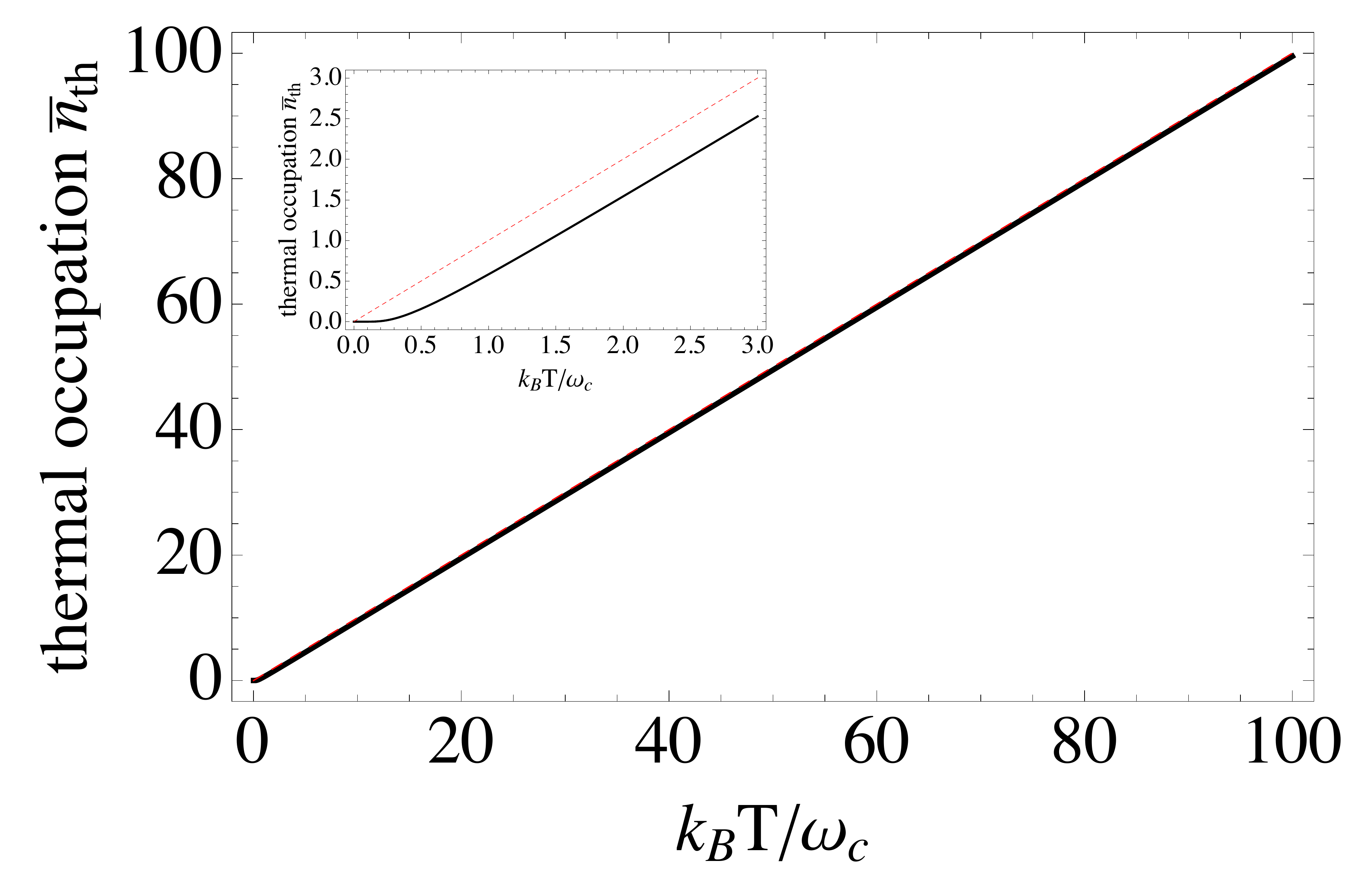}

\caption{\label{fig:thermal-occupation}(color online). Thermal occupation
$\bar{n}_{\mathrm{th}}=(\mathrm{exp}\left[\hbar\omega_{c}/k_{B}T\right]-1)^{-1}$
(black solid line) and high-temperature approximate result $\bar{n}_{\mathrm{th}}\approx k_{B}T/\hbar\omega_{c}$
(red dashed line). For $T=4\mathrm{K}$ and $\omega_{c}/2\pi=1\mathrm{GHz}$
($\omega_{c}/2\pi=10\mathrm{GHz}$), we have $k_{B}T/\hbar\omega_{c}\approx80$
($k_{B}T/\hbar\omega_{c}\approx8$). For $T=1\mathrm{K}$ and $\omega_{c}/2\pi=1\mathrm{GHz}$
($\omega_{c}/2\pi=10\mathrm{GHz}$), we have $\bar{n}_{\mathrm{th}}\approx20$
($\bar{n}_{\mathrm{th}}\approx2$). }
\end{figure}

Here, we first provide typical thermal occupation numbers $\bar{n}_{\mathrm{th}}$
for relevant experimental parameter regimes. At a temperature $T=4\mathrm{K}$,
a (mechanical) oscillator of frequency $\omega_{c}/2\pi\sim\left(1-10\right)\mathrm{GHz}$
has an thermal equilibrium occupation number much larger than one,
$\bar{n}_{\mathrm{th}}\approx8-80$: compare Fig.\ref{fig:thermal-occupation}.

\section{Polaron vs. Lab Frame \label{sec:Polaron-vs.-Lab}}

In this Appendix we show that for stroboscopic times the ideal time
evolution in the lab frame fully coincides with the one in the polaron
frame. 

In the ideal (noise-free) scenario, the evolution of the system in
the lab frame, comprising both spin and resonator degrees of freedom,
is described by Schrödinger's equation 
\begin{equation}
i\frac{d}{dt}\left|\psi\right\rangle _{t}=H\left|\psi\right\rangle _{t}.
\end{equation}
In the polaron frame, the time evolution is governed by
\begin{equation}
i\frac{d}{dt}\tilde{\left|\psi\right\rangle }_{t}=H_{0}\tilde{\left|\psi\right\rangle }_{t},\label{eq:Schroedinger-polaron}
\end{equation}
where $\tilde{\left|\psi\right\rangle }_{t}=U^{\dagger}\left|\psi\right\rangle _{t}$,
$U=\exp\left[\mu S\left(a-a^{\dagger}\right)\right]$, and $H_{0}=U^{\dagger}HU=\omega_{c}a^{\dagger}a-\frac{g^{2}}{\omega_{c}}\mathcal{S}^{2}$;
the polaron transformation $U$ entangles spin with resonator degrees
of freedom. The solution to Eq.(\ref{eq:Schroedinger-polaron}) reads
$\tilde{\left|\psi\right\rangle }_{t}=\exp\left[-iH_{0}t\right]\tilde{\left|\psi\right\rangle }_{0}$.
Using the relation $\exp\left[-i\omega_{c}ta^{\dagger}a\right]=\exp\left[-i2\pi ma^{\dagger}a\right]=\mathds1$
for stroboscopic times ($\omega_{c}t_{m}=2\pi m$, with $m$ integer),
full time evolution in the polaron frame reduces to
\begin{equation}
\tilde{\left|\psi\right\rangle }_{t_{m}}=e^{i2\pi m\mu^{2}\mathcal{S}^{2}}\tilde{\left|\psi\right\rangle }_{0}.
\end{equation}
Transforming back to the lab frame with $\tilde{\left|\psi\right\rangle }_{t}=U^{\dagger}\left|\psi\right\rangle _{t}$,
and using that $U$ commutes with the propagator $\exp\left[i2\pi m\mu^{2}\mathcal{S}^{2}\right]$,
we obtain the (stroboscopic) solution in the lab frame, $\left|\psi\right\rangle _{t_{m}}=e^{i2\pi m\mu^{2}\mathcal{S}^{2}}\left|\psi\right\rangle _{0}$,
which fully coincides with the one in the polaron frame.

\section{Gate Time \label{sec:Gate-Time}}

Ideally, the gate time $t_{\mathrm{gate}}$ has to fulfill two conditions:
(i) it has to be chosen stroboscopically, that is $\omega_{c}t_{\mathrm{gate}}=2\pi m$,
with $m=1,2,\dots$ with (ii) the parameters such that $m\mu^{2}=1/16$
in order to obtain a maximally-entangling gate (in the absence of
noise). Combination of (i) and (ii) then yields the ideal gate time
\begin{equation}
t_{\mathrm{max}}=\frac{\pi}{8g_{\mathrm{eff}}},
\end{equation}
as given in the main text. The gate time $t_{\mathrm{max}}$ should
be short compared to the relevant noise timescales, which yields the
requirement $g_{\mathrm{eff}}\gg\kappa_{\mathrm{eff}},\Gamma$. In
principle, large values of $g_{\mathrm{eff}}=g^{2}/\omega_{c}$ can
be obtained by choosing the resonator frequency $\omega_{c}$ sufficiently
small, provided that $\omega_{c}$ can be tuned independently of $g$.
This can be done up to the lower bound $\omega_{c}\geq4g$ which follows
directly from the requirement $m=1/\left(16\mu^{2}\right)\geq1$.

\section{Time-dependent Control of the Spin-Resonator Coupling \label{sec:Time-dependent-Control}}

In this Appendix we discuss in detail a prototypical implementation
of a spin-resonator system that allows for time-dependent control
of the spin-resonator coupling $g=g\left(t\right)$, as required for
the faithful realization of the proposed hot gate. Here, we first
focus on a \textit{charge} qubit embedded in a lithographically defined
double quantum dot (DQD) containing a single electron, and then extend
our analysis to a singlet-triplet \textit{spin} qubit made out two
electrons in such a DQD. Based on the electric dipole interaction,
this type of device may be coupled either to a microwave transmission
line resonator in a circuit-QED-like setup, as investigated theoretically
and experimentally in (for example) Refs.\cite{frey12,petersson12,viennot14},
or a surface-acoustic-wave resonator, as described in Refs.\cite{schuetz15,chen15}.
Our approach then employs standard all-electrical manipulation strategies,
in which external, tunable gate voltages are used for (basically)
in-situ control of the effective spin-resonator coupling \cite{childress04},
provided that standard adiabaticity conditions are fulfilled \cite{beaudoin16},
with the additional requirement of having a relatively small qubit
transition frequency $\omega_{q}$ when the (hot) gate is turned on;
as shown in Sec.\ref{sec:Errors-due-to-level-splitting}, this condition
can be dropped, however, for longitudinal spin-resonator coupling.

\subsection{Double Quantum Dot Charge Qubit}

\begin{figure}
\includegraphics[width=1\columnwidth]{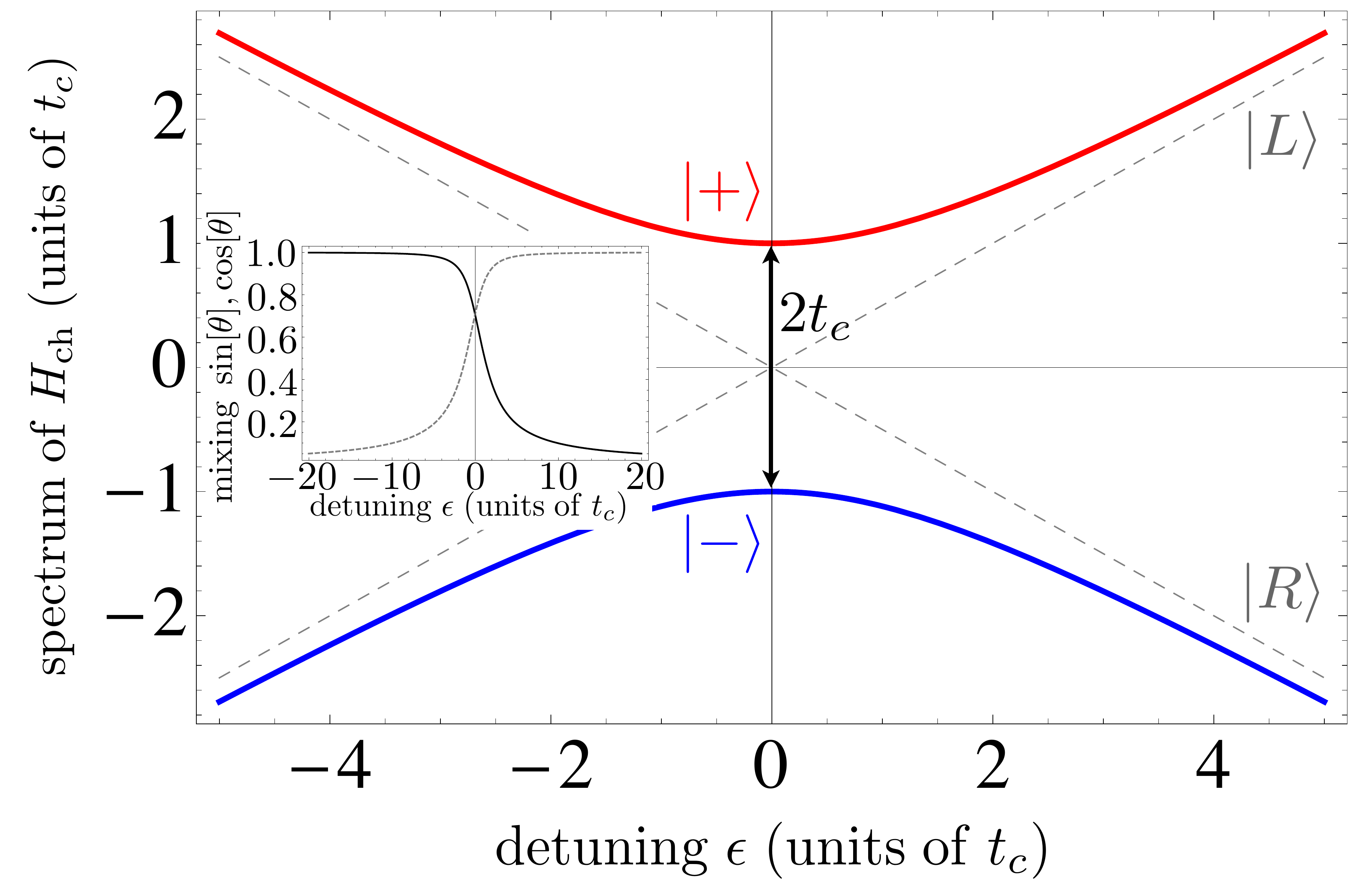}

\caption{\label{fig:charge-qubit-spectrum}(color online). Spectrum of the DQD Hamiltonian
in the single-electron regime, $H_{\mathrm{ch}}=\frac{\epsilon}{2}\tau^{z}+t_{c}\tau^{x}$,
as a function of the interdot detuning parameter $\epsilon$. Inset:
Mixing parameters $\sin\theta$ (black solid) and $\cos\theta$ (gray
dashed) as a function of the interdot detuning parameter $\epsilon$. }
\end{figure}

The Hamiltonian describing a tunnel-coupled DQD in the single-electron
regime coupled to a cavity of frequency $\omega_{c}$ is given by
\cite{jin11,kulkarni14,gullans15} 
\begin{equation}
H=\frac{\epsilon}{2}\tau^{z}+t_{c}\tau^{x}+\omega_{c}a^{\dagger}a+g_{\mathrm{ch}}\tau^{z}\otimes\left(a+a^{\dagger}\right),\label{eq:Hamiltonian-charge-qubit-resonator-bare}
\end{equation}
where $\epsilon$ is the (tunable) level detuning between the dots,
$t_{c}$ gives the (tunable) tunnel coupling, and $g_{\mathrm{ch}}$
refers to the single photon (phonon) coupling strength between the
resonator and the DQD. The electron charge state is described in terms
of orbital Pauli operators defined as $\tau^{z}=\left|L\right\rangle \left\langle L\right|-\left|R\right\rangle \left\langle R\right|$
and $\tau^{x}=\left|L\right\rangle \left\langle R\right|+\left|R\right\rangle \left\langle L\right|$,
respectively, with $\left|L\right\rangle \left(\left|R\right\rangle \right)$
corresponding to the state where the electron is localized in the
left (right) dot, while $a^{\dagger}\left(a\right)$ are the standard
resonator creation (annihilation) operators. 

\begin{figure}
\includegraphics[width=1\columnwidth]{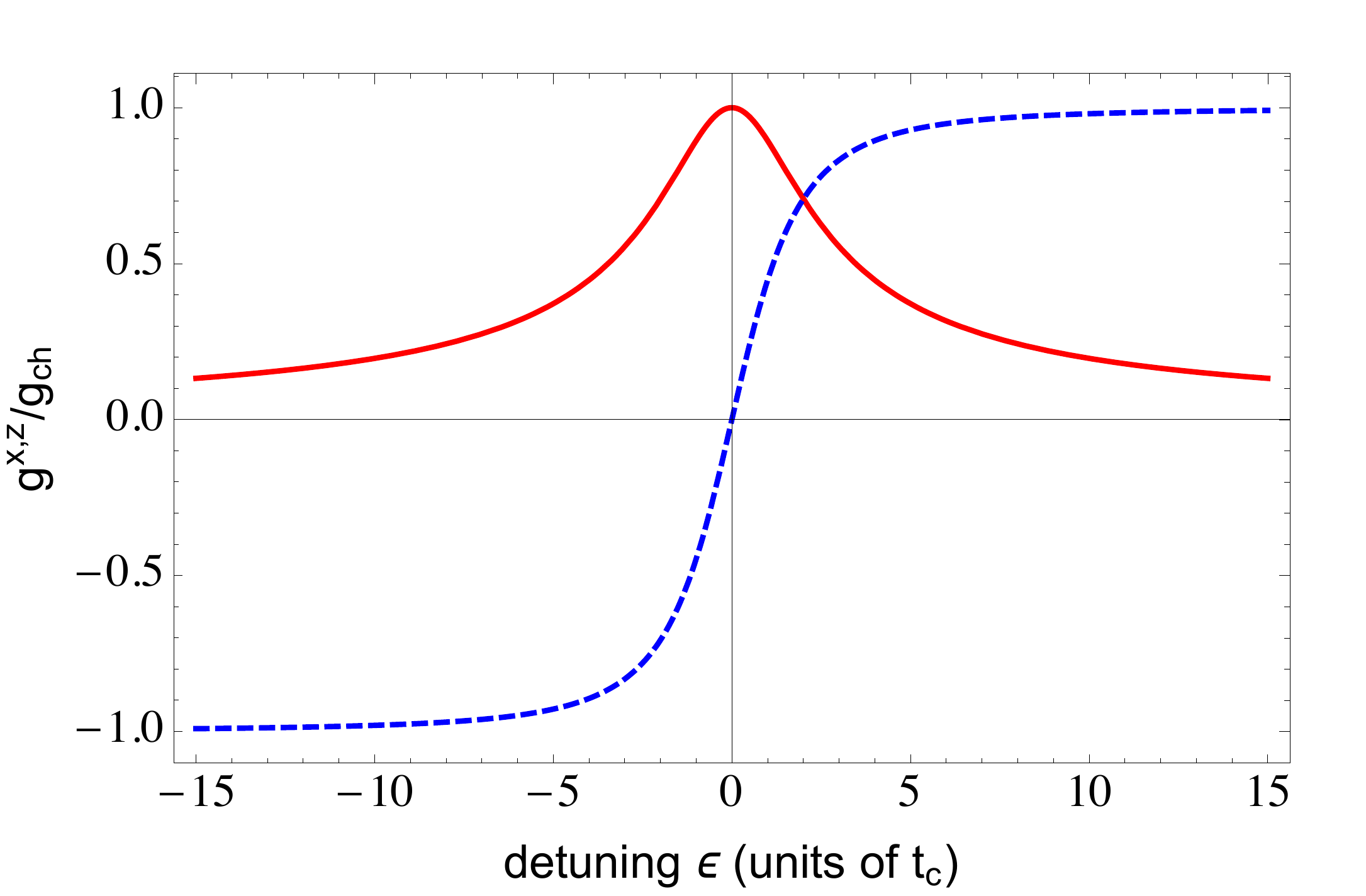}

\caption{\label{fig:charge-qubit-coupling}(color online). Effective spin-resonator coupling
$g^{x}$ (solid) and $g^{z}$ (dashed) as a function of the interdot
detuning parameter $\epsilon$. }
\end{figure}

Diagonalization of the first two terms in the Hamiltonian $H$, that
is $H_{\mathrm{ch}}=\frac{\epsilon}{2}\tau^{z}+t_{c}\tau^{x}$, yields
the electronic charge eigenstates 
\begin{eqnarray}
\left|+\right\rangle  & = & \cos\theta\left|L\right\rangle +\sin\theta\left|R\right\rangle ,\\
\left|-\right\rangle  & = & -\sin\theta\left|L\right\rangle +\cos\theta\left|R\right\rangle ,
\end{eqnarray}
where the mixing angle is given by $\tan\theta=2t_{c}/\left(\epsilon+\omega_{q}\right)$,
and $\omega_{q}=\sqrt{\epsilon^{2}+4t_{c}^{2}}$ refers to the energy
splitting between the eigenstates $\left|\pm\right\rangle $; compare
Fig.\ref{fig:charge-qubit-spectrum}. The logical qubit basis is (by
definition) given by the superposition states $\left|\pm\right\rangle =\left(\left|L\right\rangle \pm\left|R\right\rangle \right)/\sqrt{2}$
at the charge degeneracy point $\left(\epsilon=0\right)$, where to
first order the qubit is insensitive to charge fluctuations $\left(d\omega_{q}/d\epsilon=0\right)$.
In the eigenbasis of $H_{\mathrm{ch}}$, and after a simple gauge
transformation $\left(a\rightarrow-a,a^{\dagger}\rightarrow-a^{\dagger}\right)$,
the spin-resonator Hamiltonian given in Eq.(\ref{eq:Hamiltonian-charge-qubit-resonator-bare})
can be rewritten as 
\begin{eqnarray}
H & = & \frac{\omega_{q}}{2}\sigma^{z}+\omega_{c}a^{\dagger}a+\left(g^{x}\sigma^{x}-g^{z}\sigma^{z}\right)\otimes\left(a+a^{\dagger}\right).
\end{eqnarray}
Here, we have introduced the Pauli operators as $\sigma^{z}=\left(\left|+\right\rangle \left\langle +\right|-\left|-\right\rangle \left\langle -\right|\right)$,
and $\sigma^{x}=\left(\left|+\right\rangle \left\langle -\right|+\left|-\right\rangle \left\langle +\right|\right)$;
the transversal and longitudinal coupling parameters are given by
\begin{eqnarray}
g^{x} & = & g_{\mathrm{ch}}\frac{2t_{c}}{\omega_{q}},\\
g^{z} & = & g_{\mathrm{ch}}\frac{\epsilon}{\omega_{q}}.
\end{eqnarray}
By redefining the interdot detuning parameter as $\epsilon\rightarrow-\epsilon$
(or, equivalently by relabeling $\left|L\right\rangle \leftrightarrow\left|R\right\rangle $),
the spin-resonator Hamiltonian $H$ may be expressed as \cite{childress04,gullans15}
\begin{equation}
H=\frac{\omega_{q}}{2}\sigma^{z}+\omega_{c}a^{\dagger}a+\left(g^{x}\sigma^{x}+g^{z}\sigma^{z}\right)\otimes\left(a+a^{\dagger}\right).
\end{equation}
Both, the effective transversal coupling parameter $g^{x}$ as well
as the longitudinal coupling parameter $g^{z}$ can be controlled
via rapid all-electrical tuning of either the interdot detuning parameter
$\epsilon$ and/or the tunnel splitting $t_{c}$ (recall $\omega_{q}=\sqrt{\epsilon^{2}+4t_{c}^{2}}$)
\cite{beaudoin16,childress04,frey12,gullans15,kulkarni14,trif08}.
As shown in Fig.\ref{fig:charge-qubit-coupling}, the transversal
coupling parameter $g^{x}$ is maximized around $\epsilon=0$ (that
is, when the electron is delocalized in both dots), while it is strongly
suppressed for $\left|\epsilon\right|\gg t_{c}$. Conversely, the
longitudinal coupling parameter $g^{z}$ is maximized for $\left|\epsilon\right|\gg t_{c}$,
while it is strongly suppressed for small detuning $\left|\epsilon\right|\ll t_{c}$.
Note that, outside of our regime of interest, in the limit where $\delta,g_{\mathrm{ch}}\ll\omega_{c}$
(with $\delta=\omega_{q}-\omega_{c}$) one can perform a rotating-wave
approximation yielding the standard Jaynes-Cummings Hamiltonian, as
widely discussed in the literature (see e.g. Refs.\cite{frey12,trif08,schuetz15,childress04,jin11,kulkarni14}). 

Then, since the parameters $\epsilon\left(t\right)$ and $t_{c}\left(t\right)$
can be tuned all-electrically on very fast timescales, the protocol
for the proposed hot gate proceeds as follows: (i) For $\epsilon\sim0$,
the hot gate is turned on, with $g^{x}\approx g_{\mathrm{ch}}$ and
$g^{z}\sim0$ (corresponding to purely transversal spin-resonator
coupling as discussed extensively in the main text). In this regime,
the qubit level splitting is set by the (highly tunable) tunnel-coupling,
according to $\omega_{q}\approx2t_{c}$, which should be chosen to
be much smaller than the cavity frequency $\left(t_{c}\ll\omega_{c}\right)$
in order to satisfy the requirements of the proposed hot gate. (ii)
After some well-controlled (stroboscopic) time $t_{m}=2\pi m/\omega_{c}$,
the hot gate can be turned off by sweeping $\epsilon$ to large detuning
values $\epsilon\gg t_{c}$. 

Both regimes are readily achievable in the quantum dot setting: Due
to the exponential dependence of tunnel coupling strength $t_{c}$
on gate voltage, the interdot barrier characterized by $t_{c}$ can
be varied from about $100\mu\mathrm{eV}$ (verified by the broadening
of the time-averaged charge transition; note that for much larger
tunnel couplings, two neighboring dots become one single dot) all
the way down to less than $10^{-12}\mathrm{eV}\sim10^{-6}\mathrm{GHz}$
(corresponding to a millisecond timescale, as verified by real-time
detection of single charges hopping on or off the dot) \cite{hanson07-sm},
which is five to six orders of magnitude smaller than realistic cavity
frequencies. Similarly, the detuning $\epsilon$ between the dots
can be varied anywhere between zero and a positive or negative detuning
equal to the addition energy, at which point additional electrons
are pulled into the dot. The typical energy scale for the addition
energy is very large $\left(\sim1-3\mathrm{meV}\right)$ \cite{hanson07-sm}. 

Note that in the proposed off-setting {[}step (ii){]} the qubits and
the cavity are not strictly decoupled due to the non-vanishing longitudinal
term (compare Fig.\ref{fig:charge-qubit-coupling}). For $g_{\mathrm{ch}}\ll\omega_{c}$,
this coupling is usually neglected within a rotating-wave approximation
\cite{childress04,frey12,jin11}. However, here we provide
an exact treatment, that takes into account the energy shifts and
couplings arising from the (fast rotating) qubit-cavity coupling term.
For $g^{x}=0$, the Hamiltonian $H$ can be diagonalized exactly,
yielding the eigenstates $\left|\sigma\right\rangle \otimes D^{\dagger}(\sigma\frac{g^{z}}{\omega_{c}})\left|n\right\rangle $
with the corresponding eigenenergies $\epsilon\left(\sigma,n\right)=\sigma\omega_{q}/2-g_{z}^{2}/\omega_{c}+n\omega_{c}$,
with $\sigma=\pm$ for spin-up and spin-down, respectively, the displacement
operator $D\left(\alpha\right)=\exp\left[\alpha a^{\dagger}-\alpha^{*}a\right]$
and $\left|n\right\rangle $ denoting the usual Fock states. This
treatment can be extended straightforwardly to more than one qubit. 

While the analysis above has focused on a single charge qubit, in
the following we consider \textit{two} qubits of this type, coupled
to a common resonator mode. Then, for two qubits and purely longitudinal
spin-resonator coupling, in the presence of a non-zero (and potentially
large, $\omega_{q}\sim\left|\epsilon\right|$) level splitting $\omega_{q}$
the time-evolution generated by the Hamiltonian $H$ reads 
\begin{equation}
U\left(t_{m}\right)=e^{-iHt_{m}}=e^{-i\frac{\omega_{q}}{2}S^{z}t_{m}}U_{\mathrm{id}}^{z}\left(t_{m}\right),
\end{equation}
with the ideal evolution $U_{\mathrm{id}}^{z}\left(t_{m}\right)=\exp\left[i4\pi m\mu^{2}\sigma_{1}^{z}\sigma_{2}^{z}\right]$,
up to an irrelevant global phase. Therefore, in the regime $\left|\epsilon\right|\gg t_{c}$,
a general two-qubit state $\left|\Psi_{2q}\right\rangle =c_{00}\left|\Downarrow\Downarrow\right\rangle +c_{01}\left|\Downarrow\Uparrow\right\rangle +c_{10}\left|\Uparrow\Downarrow\right\rangle +c_{11}\left|\Uparrow\Uparrow\right\rangle $
evolves as 
\begin{eqnarray}
U\left(t_{m}\right)\left|\Psi_{2q}\right\rangle  & = & e^{+2im\pi\frac{\omega_{q}}{\omega_{c}}}c_{00}\left|\Downarrow\Downarrow\right\rangle +e^{-2im\pi\frac{\omega_{q}}{\omega_{c}}}c_{11}\left|\Uparrow\Uparrow\right\rangle \nonumber \\
 &  & +e^{-8im\pi\mu^{2}}\left(c_{01}\left|\Downarrow\Uparrow\right\rangle +c_{10}\left|\Uparrow\Downarrow\right\rangle \right)
\end{eqnarray}
When tuning the qubit level splitting on resonance $\left(\omega_{q}\approx\left|\epsilon\right|=\omega_{c}\right)$,
such that $\exp\left[\pm2im\pi\omega_{q}/\omega_{c}\right]=1$ for
all $m=1,2,3\dots$, for certain times $t^{\star}=2\pi m^{\star}/\omega_{c}=\pi/2g_{\mathrm{eff}}$,
this unitary returns the original state, since $U_{\mathrm{id}}^{z}\left(t^{\star}\right)=\mathds1$,
and therefore, absent any other noise sources, leaves the (typically
entangled) state prepared by the first step (i) with $g^{x}=g_{\mathrm{ch}}$,
$g^{z}=0$ unaffected; recall that $\mu=g_{\mathrm{ch}}/\omega_{c}=1/4,1/8,\dots$
is chosen commensurately. While this statement holds for any two qubit
state $\left|\Psi_{2q}\right\rangle $, this effect becomes even simpler
to see when the qubits are initialized in any of the four computational
basis states $\left\{ \left|\sigma,\sigma'\right\rangle \right\} $.
Here, the ideal transversal gate (i) first prepares maximally entangled
states, according to
\begin{eqnarray}
\left|\Downarrow,\Downarrow\right\rangle  & \rightarrow & \frac{1}{\sqrt{2}}\left(\left|\Downarrow,\Downarrow\right\rangle +i\left|\Uparrow\Uparrow\right\rangle \right),\\
\left|\Uparrow,\Uparrow\right\rangle  & \rightarrow & \frac{1}{\sqrt{2}}\left(\left|\Uparrow\Uparrow\right\rangle +i\left|\Downarrow,\Downarrow\right\rangle \right),\\
\left|\Uparrow,\Downarrow\right\rangle  & \rightarrow & \frac{1}{\sqrt{2}}\left(\left|\Uparrow,\Downarrow\right\rangle +i\left|\Downarrow\Uparrow\right\rangle \right),\label{eq:trafo-ideal-transversal-up-down}\\
\left|\Downarrow,\Uparrow\right\rangle  & \rightarrow & \frac{1}{\sqrt{2}}\left(\left|\Downarrow\Uparrow\right\rangle +i\left|\Uparrow,\Downarrow\right\rangle \right),\label{eq:trafo-ideal-transversal-down-up}
\end{eqnarray}
which subsequently in stage (ii) where $\left(g^{x}=0,g^{z}=g_{\mathrm{ch}}\right)$
are left invariant $\forall m=1,2,\dots$; Eqs.(\ref{eq:trafo-ideal-transversal-up-down})
and (\ref{eq:trafo-ideal-transversal-down-up}) even hold independently
of $\omega_{q}$.

The charge-qubit-based scheme discussed above can be extended to (switchable)
coupling between the resonator mode and the electron\textquoteright s
spin, by making use of various mechanisms which hybridize spin and
charge degrees of freedom, as provided by spin-orbit interaction or
inhomogeneous magnetic fields \cite{hu12,viennot15,beaudoin16,trif08}.
Such an implementation that easily generalizes to $N$ qubits and
would allow to fully turn off any coupling to the cavity mode (and
to do so selectively for any chosen subset of qubits) is discussed
in the next section.

\subsection{Double Quantum Dot Spin Qubit \label{sub:Double-Quantum-Dot}}

Let us now extend our treatment to singlet-triplet \textit{spin} qubits
in quantum dots, where logical qubits are encoded in a two-dimensional
subspace of a higher-dimensional two-electron spin system, as investigated
theoretically and experimentally (for example) in Refs.\cite{hanson07-sm,levy02}.
This approach successfully combines spin and charge manipulation,
making use of the very long coherence times associated with spin states
and, at the same time, enabling efficient readout and coherent manipulation
of coupled spin states based on intrinsic interactions \cite{taylor06}. 

In contrast to the charge qubit setting discussed above (where the
electron's charge will always couple to the resonator mode with the
type of coupling depending on the particular parameter regime), in
this setting the coupling to the cavity mode can be turned off completely,
since the dipole-moment associated with the singlet-triplet qubit
(which in this case determines the spin-resonator coupling) vanishes
in the so-called $\left(1,1\right)$ regime; here, $\left(m,n\right)$
refers to a configuration with $m(n)$ electrons in the left (right)
dot, respectively. 

\begin{figure}
\includegraphics[width=1\columnwidth]{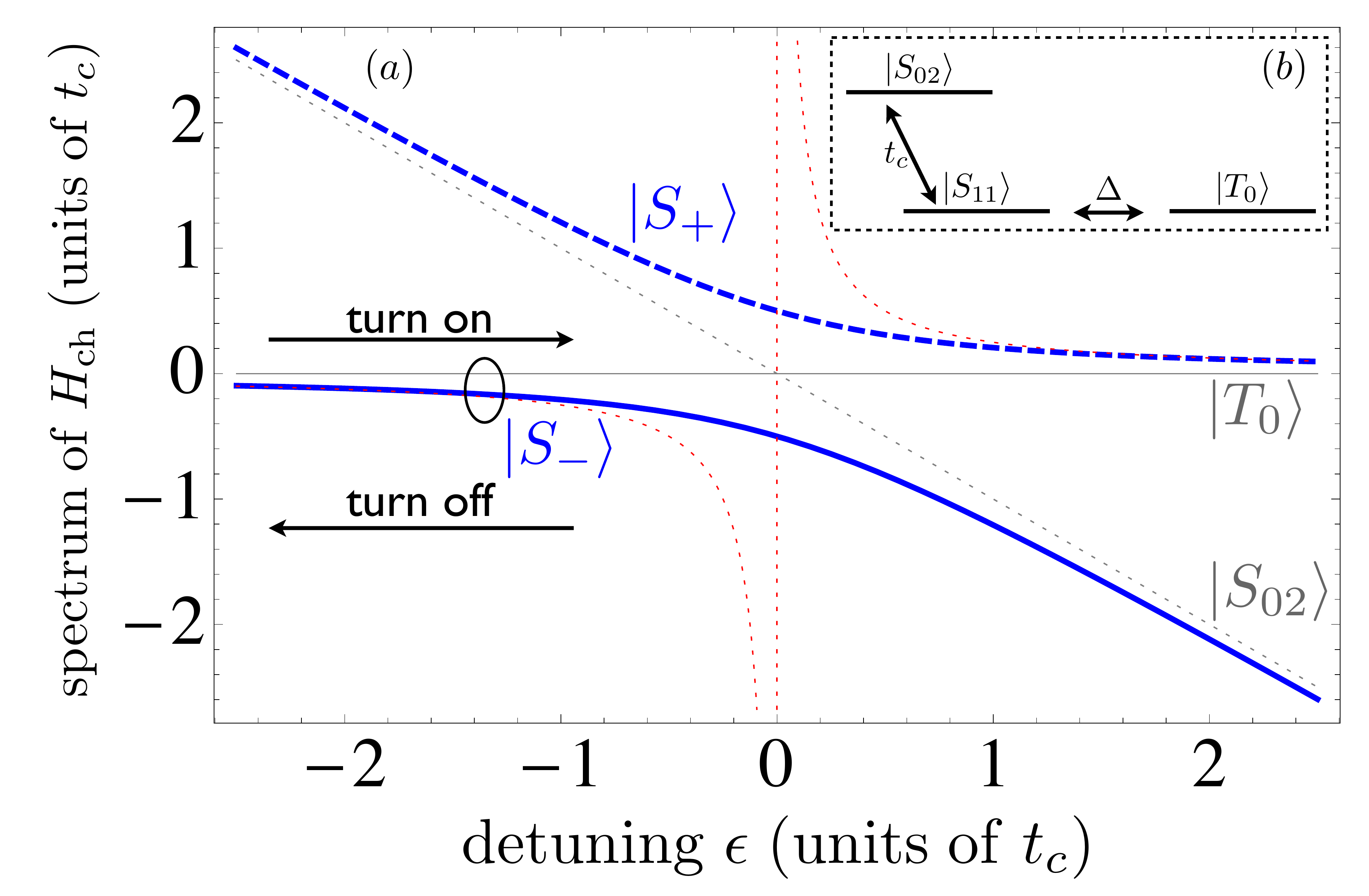}

\caption{\label{fig:spin-qubit-spectrum}(color online). (a) Spectrum of the DQD Hamiltonian
in the two-electron regime, as given in Eq.(\ref{eq:Hamiltonian-DQD-spin-qubit}),
as a function of the interdot detuning parameter $\epsilon$ for $\Delta=0$.
Tunnel coupling between the singlet states $\left|S_{11}\right\rangle $
with (1,1) charge occupation and $\left|S_{02}\right\rangle $ with
(0,2) charge occupation yields the hybridized singlet states $\left|S_{\pm}\right\rangle $.
The ellipse refers to the qubit subspace, spanned by $\left|T_{0}\right\rangle $
and $\left|S_{-}\right\rangle $, while the dotted line (red) refers
to the effective exchange coupling $J\left(\epsilon\right)=t_{c}^{2}/4\epsilon$.
The arrows indicate schematically how to turn on and off the effective
spin resonator coupling, by changing the effective dipole moment associated
with the qubit. Inset (b): Relevant level diagram in the subspace
$\left\{ \left|T_{0}\right\rangle ,\left|S_{11}\right\rangle ,\left|S_{02}\right\rangle \right\} $. }
\end{figure}

We focus on the typical regime of interest, where (following the standard
notation) the relevant electronic levels are given by the triplet
states $\left|T_{+}\right\rangle =\left|\Uparrow\Uparrow\right\rangle $,
$\left|T_{-}\right\rangle =\left|\Downarrow\Downarrow\right\rangle $,
and $\left|T_{0}\right\rangle =\left(\left|\Uparrow\Downarrow\right\rangle +\left|\Downarrow\Uparrow\right\rangle \right)/\sqrt{2}$,
as well as the singlet states $\left|S_{11}\right\rangle =\left(\left|\Uparrow\Downarrow\right\rangle -\left|\Downarrow\Uparrow\right\rangle \right)/\sqrt{2}$
and $\left|S_{02}\right\rangle =d_{R\uparrow}^{\dagger}d_{R\downarrow}^{\dagger}\left|0\right\rangle $
with $\left|\sigma\sigma'\right\rangle =d_{L\sigma}^{\dagger}d_{R\sigma'}^{\dagger}\left|0\right\rangle $;
the fermionic creation (annihilation) operators $d_{i\sigma}^{\dagger}\left(d_{i\sigma}\right)$
create (annihilate) an electron with spin $\sigma=\uparrow,\downarrow$
in the orbital $i=L,R$. For sufficiently large magnetic field $B$,
the levels $\left|T_{+}\right\rangle $ and $\left|T_{-}\right\rangle $
are far detuned and can be neglected for the remainder of the discussion.
Therefore, in the following, we restrict ourselves to the subspace
$\left\{ \left|T_{0}\right\rangle ,\left|S_{11}\right\rangle ,\left|S_{02}\right\rangle \right\} $,
as schematically depicted in the inset of Fig.\ref{fig:spin-qubit-spectrum}.
In the relevant regime of interest, the electronic DQD system is described
by the Hamiltonian \cite{taylor06}
\begin{eqnarray}
H_{\mathrm{DQD}} & = & \frac{t_{c}}{2}\left(\left|S_{02}\right\rangle \left\langle S_{11}\right|+\mathrm{h.c.}\right)+\Delta\left(\left|T_{0}\right\rangle \left\langle S_{11}\right|+\mathrm{h.c.}\right)\nonumber \\
 &  & -\epsilon\left|S_{02}\right\rangle \left\langle S_{02}\right|,\label{eq:Hamiltonian-DQD-spin-qubit}
\end{eqnarray}
where (as before) $t_{c}$ refers to the interdot tunneling amplitude,
$\epsilon$ is the interdot detuning parameter, and $\Delta$ is a
static magnetic field gradient between the two dots which couples
singlet and triplet states. State preparation, measurement, single-qubit
gates and local two-qubit gates can be achieved by tuning the bias
$\epsilon$ \cite{hanson07-sm}. Tunnel coupling between the singlet
states $\left|S_{11}\right\rangle $ with (1,1) charge occupation
and $\left|S_{02}\right\rangle $ with (0,2) charge occupation (here,
$\left(m,n\right)$ refers to a configuration with $m(n)$ electrons
in the left (right) dot, respectively) yields the hybridized singlet
states $\left|S_{\pm}\right\rangle $, given by 
\begin{eqnarray}
\left|S_{+}\right\rangle  & = & \cos\theta\left|S_{11}\right\rangle +\sin\theta\left|S_{02}\right\rangle ,\\
\left|S_{-}\right\rangle  & = & -\sin\theta\left|S_{11}\right\rangle +\cos\theta\left|S_{02}\right\rangle ,
\end{eqnarray}
with $\tan\theta=t_{c}/\left(\epsilon+\Omega\right)$, $\Omega=\sqrt{\epsilon^{2}+t_{c}^{2}}$
and the associated eigenenergies $\epsilon_{\pm}=1/2\left(-\epsilon\pm\sqrt{\epsilon^{2}+t_{c}^{2}}\right)$.
For large, negative detuning values ($\left|\epsilon\right|\gg t_{c}$),
the splitting between the triplet $\left|T_{0}\right>$ and the hybridized
singlet $\left|S_{-}\right\rangle $ can be approximated very well
by the effective (tunable) exchange splitting $J\left(t_{c},\epsilon\right)=t_{c}^{2}/4\epsilon$;
compare Fig.\ref{fig:spin-qubit-spectrum}. As schematically denoted
by the ellipse in Fig.\ref{fig:spin-qubit-spectrum}, we focus on
the regime where the singlet $\left|S_{+}\right\rangle $ is far off-resonance,
yielding the effective qubit subspace $\left\{ \left|T_{0}\right\rangle ,\left|S_{-}\right\rangle \right\} $
with a qubit level splitting $\omega_{q}\approx J\left(t_{c},\epsilon\right)$. 

\begin{figure}
\includegraphics[width=1\columnwidth]{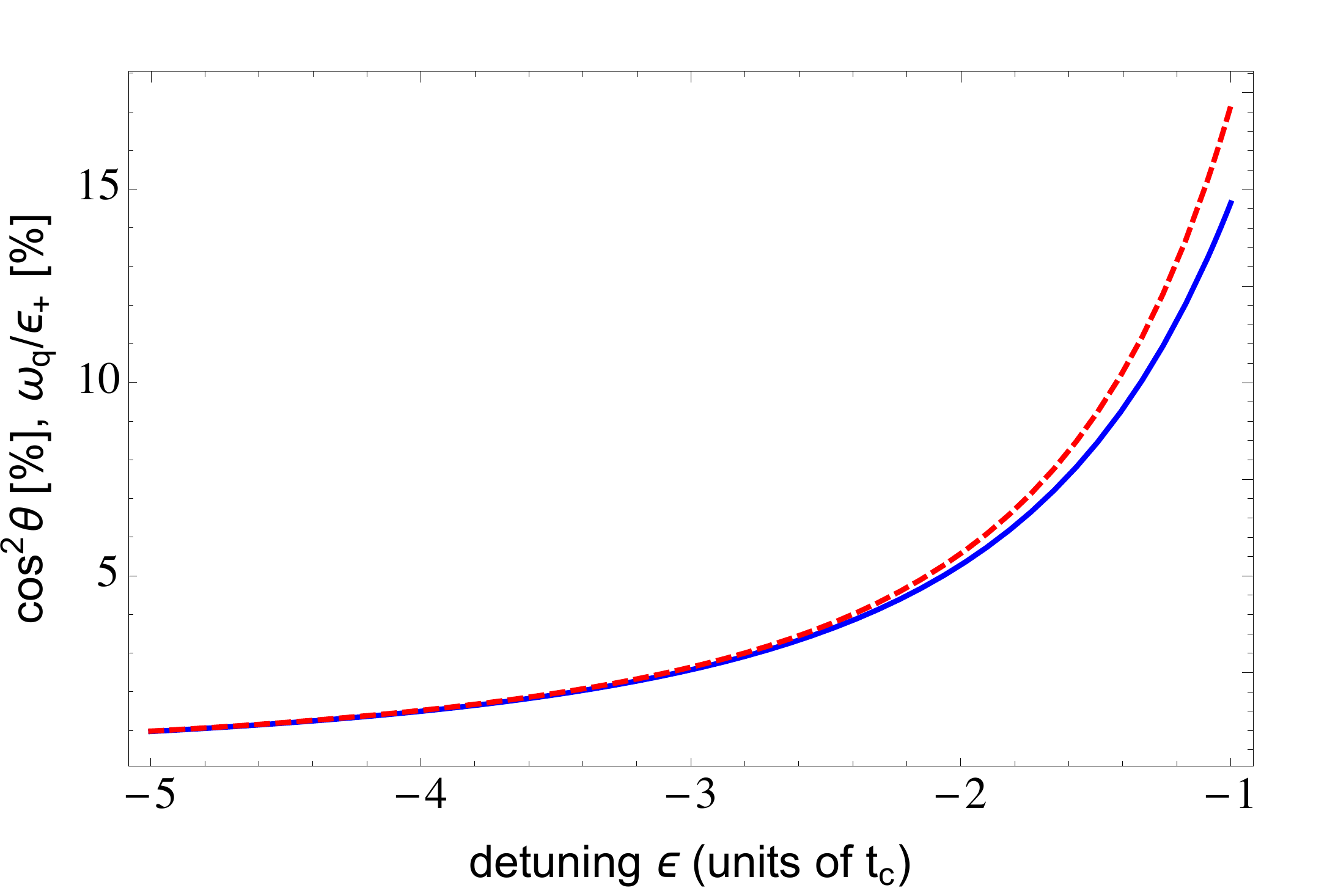}

\caption{\label{fig:spin-qubit-coupling}(color online). Effective spin resonator coupling
$g_{\mathrm{sp}}/g_{0}=\cos^{2}\theta$ (solid blue line) and qubit
level splitting $\omega_{q}\approx\left|J\right|$ relative to $\epsilon_{+}$
(dashed red line) as a function of the interdot detuning parameter
$\epsilon$. The spin resonator coupling may reach a few percent of
the bare charge resonator coupling $g_{0}$, with a qubit frequency
$\omega_{q}$ that is much smaller than the energy of the level $\left|S_{+}\right\rangle $. }
\end{figure}

Again we consider a resonator with a single relevant mode of frequency
$\omega_{c}$, as modeled by the Hamiltonian
\begin{equation}
H_{\mathrm{cav}}=\omega_{c}a^{\dagger}a.
\end{equation}
In order to couple the electric field associated with the resonator
mode to the electron spin states, the essential idea is to make use
of an effective electric dipole moment associated with the exchange-coupled
spin states of the DQD \cite{taylor06}. The resonator mode interacts
capacitively with the double quantum dot \cite{taylor06}, as described
by the interaction Hamiltonian $H_{I}=g_{0}\left|S_{02}\right\rangle \left\langle S_{02}\right|\otimes\left(a+a^{\dagger}\right)$.
Projection onto the electronic low-energy subspace $\left\{ \left|T_{0}\right\rangle ,\left|S_{-}\right\rangle \right\} $
(i.e., projecting out the high-energy level $\left|S_{+}\right\rangle $)
then leads (to lowest order in $\sim g_{0}/\epsilon_{+}$) to the
effective spin resonator system 
\begin{eqnarray}
H & = & J\left|S_{-}\right\rangle \left\langle S_{-}\right|-\Delta\sin\theta\left(\left|T_{0}\right\rangle \left\langle S_{-}\right|+\mathrm{h.c.}\right)+\omega_{c}a^{\dagger}a\nonumber \\
 &  & +g_{0}\cos^{2}\theta\left|S_{-}\right\rangle \left\langle S_{-}\right|\otimes\left(a+a^{\dagger}\right),
\end{eqnarray}
which includes a tunable spin resonator coupling, explicitly given
by
\begin{equation}
g_{\mathrm{sp}}/g_{0}=\cos^{2}\theta=\frac{1}{2}\left(1+\frac{\epsilon}{\sqrt{\epsilon^{2}+t_{c}^{2}}}\right).
\end{equation}
As demonstrated in Fig.\ref{fig:spin-qubit-coupling}, the effective
coupling $g_{\mathrm{sp}}$ may be turned on and off by sweeping the
detuning parameter $\epsilon$ (closely following the functional dependence
of $\omega_{q}/\epsilon_{+}$), i.e. by controlling the admixture
of $\left|S_{02}\right\rangle $ to the hybridized singlet level $\left|S_{-}\right\rangle $.
For large, negative values of $\epsilon$ this admixture vanishes
$\left(\cos^{2}\theta\rightarrow0\right)$, such that the effective
dipole moment associated with the qubit vanishes and therefore the
spin-resonator coupling is switched off. The type of spin-resonator
coupling (transversal versus longitudinal) may be controlled by the
magnetic gradient $\Delta$, as can be done using e.g. a nanomagnet
or nuclear Overhauser fields \cite{hanson07-sm,beaudoin16}. While
for longitudinal spin-resonator coupling the resonator frequency $\omega_{c}$
may be comparable or even smaller than the effective qubit level splitting
$J$ (see Sec.\ref{sec:Errors-due-to-level-splitting} for details),
in the case of transversal coupling the effective qubit level splitting
needs to be much smaller than the cavity frequency, that is $|J\left(t_{c},\epsilon\right)|\approx|t_{c}^{2}/4\epsilon|\ll\omega_{c}$,
but, at the same time, $\epsilon_{+}\approx\left|\epsilon\right|+t_{c}^{2}/4\left|\epsilon\right|\gg\omega_{c}$
should be fulfilled in order to neglect the high-energy level $\left|S_{+}\right\rangle $.
Still, both requirements can be satisfied by choosing the parameters
as $t_{c},\omega_{c}\ll\left|\epsilon\right|$.

\section{Spin-Spin Coupling in Dispersive Regime \label{sec:Spin-Spin-Coupling-in-Dispersive-Regime}}

We consider two identical spins homogeneously coupled to a common
resonator mode. The dynamics are assumed to be governed by the Jaynes-Cummings
Hamiltonian 
\begin{equation}
H=\Delta\left(S_{1}^{z}+S_{2}^{z}\right)+g\left[a\left(S_{1}^{+}+S_{2}^{+}\right)+a^{\dagger}\left(S_{1}^{-}+S_{2}^{-}\right)\right],\label{eq:JC-Hamiltonian-two-spins}
\end{equation}
which is valid within the rotating-wave approximation for $\sqrt{\bar{n}_{\mathrm{th}}}g,\Delta\ll\omega_{c}$,
with the detuning $\Delta=\omega_{q}-\omega_{c}$. In the following
we consider the \textit{dispersive regime}, where the spin-resonator
coupling is strongly detuned $\left(\sqrt{\bar{n}_{\mathrm{th}}}g\ll\Delta\right)$.
In this regime, the spin-resonator coupling can be treated perturbatively.
To stress the perturbative treatment we write 
\begin{eqnarray}
H & = & H_{0}+H_{1},\label{eq:SW-JC-Hamiltonian}\\
H_{0} & = & \Delta S^{z},\\
H_{1} & = & g\left(aS^{+}+a^{\dagger}S^{-}\right),
\end{eqnarray}
where $S^{\alpha}=S_{1}^{\alpha}+S_{2}^{\alpha}$ (for $\alpha=\pm,z$)
are collective spin operators. We perform a standard Schrieffer-Wolff
transformation 
\begin{eqnarray}
\tilde{H} & = & e^{A}He^{-A}\\
 & \approx & H_{0}+H_{1}+\left[A,H_{0}+H_{1}\right]+\frac{1}{2}\left[A,\left[A,H_{0}\right]\right],
\end{eqnarray}
where the operator $A$ (with $A^{\dagger}=-A$) is assumed to have
a perturbative expansion in $g$, i.e., $A=0+\mathcal{O}\left(g\right)+\dots$
By choosing
\begin{equation}
\left[A,H_{0}\right]=-H_{1},\label{eq:SW-requirement}
\end{equation}
one obtains a Hamiltonian $\tilde{H}$ without linear coupling in
$g$, 
\begin{equation}
\tilde{H}\approx H_{0}+\frac{1}{2}\left[A,H_{1}\right].
\end{equation}
For the Hamiltonian given in Eq.(\ref{eq:SW-JC-Hamiltonian}), the
condition in Eq.(\ref{eq:SW-requirement}) is fulfilled by the choice
\begin{equation}
A=\frac{g}{\Delta}\left(aS^{+}-a^{\dagger}S^{-}\right),
\end{equation}
which yields the Hamiltonian 
\begin{equation}
\tilde{H}\approx\left(\Delta+\frac{g^{2}}{\Delta}+2\frac{g^{2}}{\Delta}a^{\dagger}a\right)S^{z}+\frac{g^{2}}{\Delta}\left(S_{1}^{+}S_{2}^{-}+S_{1}^{-}S_{2}^{+}\right).\label{eq:JC-Hamiltonian-in-dispersive-regime}
\end{equation}
Here, the last two terms describe a cavity-state dependent dispersive
shift of the qubit transition frequencies and spin-spin coupling via
virtual occupation of the cavity mode, respectively. The strength
of the effective spin-spin coupling is given by 
\begin{equation}
g_{\mathrm{eff}}=\frac{g^{2}}{\Delta}=\frac{\epsilon}{\sqrt{\bar{n}_{\mathrm{th}}}}g,
\end{equation}
where we have set $\sqrt{\bar{n}_{\mathrm{th}}}g/\Delta=\epsilon\ll1$
in order to reach the regime of validity for Eq.(\ref{eq:JC-Hamiltonian-in-dispersive-regime}),
given by 
\begin{equation}
\sqrt{\bar{n}_{\mathrm{th}}}g\ll\Delta\ll\omega_{c}.
\end{equation}
By transforming the Hamiltonian given in Eq.(\ref{eq:JC-Hamiltonian-in-dispersive-regime})
back into the lab-frame, we recover the result presented in Ref.\cite{blais04},
namely 
\begin{eqnarray}
H & \approx & \left[\omega_{c}+2\frac{g^{2}}{\Delta}\left(S_{1}^{z}+S_{2}^{z}\right)\right]a^{\dagger}a+\left(\omega_{q}+\frac{g^{2}}{\Delta}\right)\left(S_{1}^{z}+S_{2}^{z}\right)\nonumber \\
 &  & +\frac{g^{2}}{\Delta}\left(S_{1}^{+}S_{2}^{-}+S_{1}^{-}S_{2}^{+}\right).
\end{eqnarray}
Here, spins and cavity mode are still coupled by the ac Stark shift
term $\sim a^{\dagger}a$. Accordingly, one obtains an effective pure
spin Hamiltonian with flip-flop interactions provided that one can
neglect any fluctuations of the photon number $a^{\dagger}a\rightarrow\bar{n}=\left\langle a^{\dagger}a\right\rangle $,
where $\bar{n}$ is the average number of photons in the cavity mode
\cite{trif08}. 

Since the operator $S^{z}a^{\dagger}a$ in Eq.(\ref{eq:JC-Hamiltonian-in-dispersive-regime})
has an integer spectrum, one may wonder whether for stroboscopic times
the spins disentangle from the resonator mode here as well. Thus,
let us consider the full time evolution generated by Eq.(\ref{eq:JC-Hamiltonian-two-spins})
\begin{eqnarray}
e^{-iHt} & = & e^{-iU^{\dagger}\tilde{H}Ut}=U^{\dagger}e^{-i\tilde{H}t}U\\
 & \approx & U^{\dagger}\left[\exp\left[-it\left(\delta+\tilde{\delta}a^{\dagger}a\right)S^{z}\right.\right.\label{eq:time-evolution-dispersive-regime-approx1}\\
 &  & \left.\left.-i\tilde{g}t\left(S_{1}^{+}S_{2}^{-}+S_{1}^{-}S_{2}^{+}\right)\right]\right]U,\nonumber 
\end{eqnarray}
with $U=\exp\left(A\right)$, $\delta=\Delta+g^{2}/\Delta$, $\tilde{\delta}=2g^{2}/\Delta$
and $\tilde{g}=g^{2}/\Delta$. Note that Eq.(\ref{eq:time-evolution-dispersive-regime-approx1})
is an approximate statement, relying on a perturbative expansion in
the coupling $g$. Since the flip-flop interaction conserves $S^{z}$,
we find 
\begin{equation}
e^{-iHt}\approx U^{\dagger}e^{-i\delta tS^{z}}e^{-i\tilde{\delta}tS^{z}a^{\dagger}a}e^{-i\tilde{g}t\left(S_{1}^{+}S_{2}^{-}+S_{1}^{-}S_{2}^{+}\right)}U.
\end{equation}
For stroboscopic times $\tilde{\delta}t=2\pi m$, $e^{-i\tilde{\delta}tS^{z}a^{\dagger}a}=\mathds1$,
yielding 
\begin{equation}
e^{-iHt}\approx U^{\dagger}e^{-iH_{\mathrm{spin}}t}U,
\end{equation}
where $H_{\mathrm{spin}}=\delta S^{z}+\tilde{g}\left(S_{1}^{+}S_{2}^{-}+S_{1}^{-}S_{2}^{+}\right)$
is a pure spin Hamiltonian, without any coupling to the resonator
mode. However, in contrast to our scheme presented in the main text,
the full time evolution does not reduce to a pure spin problem, since
the Schrieffer-Wolff transformation $U=\exp\left[\frac{g}{\Delta}\left(aS^{-}-a^{\dagger}S^{+}\right)\right]$
does not commute with $e^{-iH_{\mathrm{spin}}t}$, but rather entangles
the qubits with the resonator mode.

\section{Schrieffer-Wolff Transformation \label{sec:Schrieffer-Wolff-Transformation}}

If one restricts oneself to the regime $g\ll\omega_{c}$, the result
stated in Eqn.(6) may also be derived in the perturbative framework
of a Schrieffer-Wolff transformation. For concreteness, assuming $\omega_{q}=0$,
we consider the Hamiltonian 
\begin{equation}
H=\underset{H_{0}}{\underbrace{\omega_{c}a^{\dagger}a}}+\underset{V}{\underbrace{gS^{x}\otimes\left(a+a^{\dagger}\right)}},
\end{equation}
where $S^{x}=\sum_{i}\eta_{i}^{x}\sigma_{i}^{x}$ is a collective
operator. In the following, and contrary to our general analysis in
the main text, we restrict ourselves to the regime where the spin-resonator
coupling $V$ can be treated perturbatively with respect to $H_{0}$,
that is $g\ll\omega_{c}$. Performing a Schrieffer-Wolff transformation
$\tilde{H}=e^{A}He^{-A}$ as presented in Sec. \ref{sec:Spin-Spin-Coupling-in-Dispersive-Regime},
with $A=-\frac{g}{\omega_{c}}S^{x}\left(a-a^{\dagger}\right)$, we
obtain an effective Hamiltonian $\tilde{H}$ where the slow subspace
is decoupled from the fast subspace up to second order in $g$. Explicitly
it reads {[}compare Eq.(5){]} 
\begin{equation}
\tilde{H}\approx\omega_{c}a^{\dagger}a-\frac{g^{2}}{\omega_{c}}S_{x}^{2}.
\end{equation}

\section{Non-Zero Qubit Level Splitting \label{sec:Non-Zero-Qubit-Level-Splitting-1}}

In our derivation of Eq.(5), starting from the generic spin-resonator
Hamiltonian given in Eq.(1), we have assumed $\omega_{q}=0$. As demonstrated
also numerically in Section \ref{sec:Additional-Numerical-Results}
below, small level splittings with $\omega_{q}\approx0.1\omega_{c}$
may still be tolerated without a significant loss in the amount of
generated entanglement and the fidelity with the maximally entangled
target state. 

In this Appendix we investigate analytically the effects associated
with a finite splitting $\omega_{q}>0$. In this case, Eq.(3) can
be generalized straightforwardly to 
\begin{equation}
H=U[\underset{H_{0}}{\underbrace{\omega_{c}a^{\dagger}a-\frac{g^{2}}{\omega_{c}}\mathcal{S}^{2}}}+\frac{\omega_{q}}{2}\tilde{S}^{z}]U^{\dagger},\label{eq:transformed-Hamiltonian-with-finite-splitting-perturbation}
\end{equation}
where $\tilde{S}^{z}=U^{\dagger}S^{z}U$, with $U=\exp\left[\frac{g}{\omega_{c}}\mathcal{S}\left(a-a^{\dagger}\right)\right]$.
In what follows, we restrict ourselves to the (experimentally) most
relevant regime where $\mu=g/\omega_{c}\ll1$, which allows for a
simple perturbative treatment. Expansion in the small parameter $\mu$
yields 
\begin{equation}
\tilde{S}^{z}\approx S^{z}-\mu\left(a-a^{\dagger}\right)\left[\mathcal{S},S^{z}\right]+\frac{\mu^{2}}{2}\left(a-a^{\dagger}\right)^{2}\left[\mathcal{S},\left[\mathcal{S},S^{z}\right]\right].
\end{equation}
Specifically, for $\mathcal{S}=\sum_{i}\sigma_{i}^{x}$ (as considered
in the main text) we then obtain 
\begin{equation}
\tilde{S}^{z}\approx S^{z}+2i\frac{g}{\omega_{c}}S^{y}\left(a-a^{\dagger}\right)+2\left(\frac{g}{\omega_{c}}\right)^{2}S^{z}\left(a-a^{\dagger}\right)^{2},
\end{equation}
which leads to an additional (undesired) contribution in Eq.(\ref{eq:transformed-Hamiltonian-with-finite-splitting-perturbation})
of the form 
\begin{equation}
\frac{\omega_{q}}{2}\tilde{S}^{z}\approx\frac{\omega_{q}}{2}S^{z}+\epsilon\left[igS^{y}\left(a-a^{\dagger}\right)+\frac{g^{2}}{\omega_{c}}S^{z}\left(a-a^{\dagger}\right)^{2}\right].
\end{equation}
Here, in contrast to the ideal Hamiltonian $H_{0}$ in Eq.(\ref{eq:transformed-Hamiltonian-with-finite-splitting-perturbation})
the spins are not decoupled from the (hot) resonator mode. However,
apart from being detuned by at least $\omega_{c}-\omega_{q}$, the
undesired terms---that lead to entanglement of the spins with the
(hot) resonator mode---are suppressed by the small parameter $\epsilon=\omega_{q}/\omega_{c}\ll1$.
In the limit $\omega_{q}\rightarrow0$ $\left(\epsilon\rightarrow0\right)$
we recover the ideal dynamics.

\section{Errors due to Non-Zero Qubit-Level Splitting \label{sec:Errors-due-to-level-splitting}}

In this Appendix we analyze errors induced by a non-zero qubit level
splitting $\left(\omega_{q}/\omega_{c}>0\right)$. In the case of
longitudinal spin-resonator coupling, we show that controlled phase
gates can be implemented (as described in the main text for $\omega_{q}=0$),
even in the presence of non-zero and inhomogeneous qubit level splittings
$\left(\omega_{q}>0\right)$, when applying either fast local single
qubit gates (to correct the effect of known $\omega_{q}\neq0$) or
standard spin-echo techniques (to compensate unknown detunings); see
section \ref{sub:Longitudinal-Spin-Resonator-Coupling}. Therefore,
for longitudinal spin-resonator coupling, our approach yields a high-fidelity
hot gate, that is \textit{independent} of the qubit level splitting
$\omega_{q}/\omega_{c}\geq0$. As detailed in section \ref{sub:Transversal-Spin-Resonator-Coupling},
this is not the case for transversal coupling, where $\omega_{q}\neq0$
causes second order errors, which, however, are suppressed in certain
decoherence-free subspaces. Thus, as opposed to the limiting regime
where $\omega_{q}=0$, the distinction between longitudinal and transversal
spin-resonator coupling indeed becomes meaningful. 

\textit{The model.}---In the absence of other error sources $\left(\kappa=\Gamma=0\right)$,
the system's dynamics are governed by the Hamiltonian 
\begin{eqnarray}
H & = & H_{0}+V,\\
H_{0} & = & \omega_{c}a^{\dagger}a+g\mathcal{S}\otimes\left(a+a^{\dagger}\right),\\
V & = & \frac{\omega_{q}}{2}S^{z},
\end{eqnarray}
with $S^{z}=\sum_{i}\sigma_{i}^{z}$ and $\mathcal{S}=\sum_{i,\alpha}\eta_{i}^{\alpha}\sigma_{i}^{\alpha}$.
Below, we will set $S^{\alpha}=S_{\alpha}$ $\left(\alpha=x,z\right)$
interchangeably. Also, note that $S^{x},S^{z}$ as defined here refer
to the usual spin operators muliplied by 2.

\subsection{Longitudinal Spin-Resonator Coupling \label{sub:Longitudinal-Spin-Resonator-Coupling}}

\textit{Controlled phase gate}.---Let us first focus on the case of
longitudinal spin-resonator coupling, where $\mathcal{S}=\sum_{i}\sigma_{i}^{z}=S^{z}$
and accordingly $\left[H_{0},V\right]=0$. In this scenario, controlled
phase gates can be implemented (as described in the main text for
$\omega_{q}=0$), even in the presence of non-zero qubit level splittings
$\left(\omega_{q}>0\right)$, when applying either fast local single
qubit phase-gates (to correct the effect of known $\omega_{q}\neq0$)
or standard spin-echo techniques (to compensate unknown detunings).
By flipping the qubits (for example) halfway the evolution and at
the end of the gate, the effect of $V$ is canceled exactly. Denoting
such a global flip of all qubits around the axis $\alpha=x,y,z$ as
$U_{\alpha}\left(\varphi\right)=\exp\left[-i\varphi/2\sigma_{1}^{\alpha}\right]\dots\exp\left[-i\varphi/2\sigma_{N}^{\alpha}\right]=\exp\left[-i\varphi/2\sum\sigma_{i}^{\alpha}\right]$,
for two qubits the full evolution (in the computational basis \{$\left|00\right\rangle ,\left|10\right\rangle ,\left|01\right\rangle ,\left|11\right\rangle $\}),
intertwined by spin echo pulses, reads 
\begin{eqnarray}
U\left(2t_{m}\right) & = & U_{x}\left(\pi\right)e^{-iHt_{m}}U_{x}\left(\pi\right)e^{-iHt_{m}},\\
 & = & \mathrm{diag}\left(e^{i\phi},1,1,e^{i\phi}\right),
\end{eqnarray}
with $\phi=16m\pi\mu^{2}$. The gate $U\left(2t_{m}\right)$ is independent
of the resonator mode and, as a consequence of the spin-echo $\pi$-pulses
$U_{x}\left(\pi\right)$, independent of $\omega_{q}$; accordingly,
the qubit level splittings do not have to be necessarily small. When
complementing the propagator $U\left(2t_{m}\right)$ with local unitaries,
such that $\left|0\right\rangle _{i}\rightarrow e^{-i\phi/2}\left|0\right\rangle _{i}$
and $\left|1\right\rangle _{i}\rightarrow e^{i\phi/2}\left|1\right\rangle _{i}$,
we obtain 
\begin{eqnarray}
U_{\mathrm{Cphase}} & = & U_{z}\left(-\phi\right)U_{x}\left(\pi\right)e^{-iHt_{m}}U_{x}\left(\pi\right)e^{-iHt_{m}}\\
 & = & \mathrm{diag}\left(1,1,1,e^{2i\phi}\right),
\end{eqnarray}
which yields a controlled phase gate for $\phi=\pi/2$ (corresponding
to a gate time $t_{\max}=\pi/16g_{\mathrm{eff}}$), that is insensitive
to the qubit level splittings $\omega_{q}>0$. 

For longitudinal spin-resonator coupling, Eq.(5) of the main text
simply reads 
\begin{equation}
e^{-iHt_{m}}=\exp\left[i2\pi m\mu^{2}\tilde{\mathcal{S}}^{2}\right],
\end{equation}
with (the generalized expression) $\tilde{\mathcal{S}}^{2}=\mathcal{S}^{2}-\left(\omega_{q}/2g_{\mathrm{eff}}\right)S^{z}$,
where $\mathcal{S}=\sum_{i}\eta_{i}\sigma_{i}^{z}$, while the operator
$S^{z}$ can also be generalized to account for possible inhomogeneities
in the qubit level splittings (with $\omega_{q,i}=\delta_{i}\omega_{q}$),
i.e. $S^{z}\rightarrow\sum\delta_{i}\sigma_{i}^{z}$. This gate differs
from the ideal one ($\exp\left[i2\pi m\mu^{2}\mathcal{S}^{2}\right]$)
only by the local phases $\exp\left[-it_{m}(\omega_{q}/2)S^{z}\right]$
and thus has the same computational power.

\subsection{Transversal Spin-Resonator Coupling \label{sub:Transversal-Spin-Resonator-Coupling}}

\textit{Transversal spin-resonator coupling}.---In the following we
turn to systems with transversal spin resonator coupling, where $\mathcal{S}=S^{x}=\sum_{i}\sigma_{i}^{x}$.
In this case, the theoretical treatment is more involved as compared
to our previous discussion on longitudinal spin resonator coupling,
because the ideal free evolution does not commute with the perturbation
($\left[H_{0},V\right]\neq0$). We use perturbative techniques to
derive an analytic expression for the error $\xi_{q}$ induced by
non-zero qubit splittings $\omega_{q}>0$. For the sake of readability,
here we restrict ourselves to two qubits, while our analysis can be
generalized readily to more than two qubits. 

\textit{Perturbative series}.---Up to second order in the perturbation
$V$, the unitary evolution operator associated with $H$ is approximately
given by 
\begin{eqnarray}
U\left(t\right) & \approx & e^{-iH_{0}t}\left[\mathbb{1}-i\int_{0}^{t}d\tau\tilde{V}\left(\tau\right)\right.\nonumber \\
 &  & \left.-\int_{0}^{t}d\tau_{2}\int_{0}^{\tau_{2}}d\tau_{1}\tilde{V}\left(\tau_{2}\right)\tilde{V}\left(\tau_{1}\right)\right],\label{eq:perturbative-expansion-evolution-operator}
\end{eqnarray}
with 
\begin{equation}
\tilde{V}\left(\tau\right)=e^{iH_{0}\tau}Ve^{-iH_{0}\tau}.\label{eq:perturbation-interaction-picture}
\end{equation}
Initially, the resonator mode is assumed to be in a thermal state
$\rho_{\mathrm{th}}=\rho_{\mathrm{th}}\left(T\right)=Z^{-1}\exp\left[-\beta\omega_{c}a^{\dagger}a\right]$.
Then, starting from the initial state $\rho\left(0\right)=\varrho\left(0\right)\otimes\rho_{\mathrm{th}}$,
the system (comprising both spin and resonator degrees of freedom)
evolves as 
\begin{equation}
\rho\left(t\right)=U\left(t\right)\varrho\left(0\right)\rho_{\mathrm{th}}U^{\dagger}\left(t\right).
\end{equation}
Inserting the perturbative expansion given in Eq.(\ref{eq:perturbative-expansion-evolution-operator}),
up to second order in $V$ we obtain 
\begin{eqnarray}
\rho\left(t\right) & \approx & e^{-iH_{0}t}\left\{ \rho\left(0\right)-i\int_{0}^{t}d\tau\left[\tilde{V}\left(\tau\right),\rho\left(0\right)\right]\right.\nonumber \\
 &  & +\int_{0}^{t}d\tau\int_{0}^{t}d\tau'\tilde{V}\left(\tau\right)\rho\left(0\right)\tilde{V}\left(\tau'\right)\nonumber \\
 &  & -\int_{0}^{t}d\tau_{2}\int_{0}^{\tau_{2}}d\tau_{1}\tilde{V}\left(\tau_{2}\right)\tilde{V}\left(\tau_{1}\right)\rho\left(0\right)\nonumber \\
 &  & \left.-\int_{0}^{t}d\tau_{2}\int_{0}^{\tau_{2}}d\tau_{1}\rho\left(0\right)\tilde{V}\left(\tau_{1}\right)\tilde{V}\left(\tau_{2}\right)\right\} e^{iH_{0}t}.\label{eq:perturbative-series-rho}
\end{eqnarray}

\textit{Eigensystem of unperturbed Hamiltonian}.---In the first step,
it it instructive to find the eigensystem of $H_{0}$. Following the
same strategy as outlined in the main text, $H_{0}$ can be written
as 
\begin{equation}
H_{0}=D^{\dagger}(\mu S^{x})\left[\omega_{c}a^{\dagger}a-g_{\mathrm{eff}}S_{x}^{2}\right]D(\mu S^{x}),
\end{equation}
where $\mu=g/\omega_{c}$, $g_{\mathrm{eff}}=g^{2}/\omega_{c}=\mu^{2}\omega_{c}$
and $D\left(\alpha\right)=\exp\left[\alpha a^{\dagger}-\alpha^{*}a\right]$
is a displacement operator. Accordingly, the eigensystem of $H_{0}$
is found to be 
\begin{equation}
H_{0}\widetilde{\left|n,\vec{\sigma}_{x}\right\rangle }=E_{n,s}\widetilde{\left|n,\vec{\sigma}_{x}\right\rangle },
\end{equation}
where the eigenvectors are given by product states of spins aligned
along the transversal direction $x$ and displaced resonator states
with a displacement proportional to the total spin projection $s$
along $x$, 
\begin{equation}
\widetilde{\left|n,\vec{\sigma}_{x}\right\rangle }=D^{\dagger}\left(\mu s\right)\left|n\right\rangle \otimes\left|\vec{\sigma}_{x}\right\rangle ,
\end{equation}
with $s=s_{1}^{x}+s_{2}^{x}$, $S^{x}\left|\vec{\sigma}_{x}\right\rangle =\left(s_{1}^{x}+s_{2}^{x}\right)\left|\vec{\sigma}_{x}\right\rangle $
and $\left|n\right\rangle $ denoting the usual Fock states. The corresponding
eigenenergies 
\begin{equation}
E_{n,s}=n\omega_{c}-s^{2}g_{\mathrm{eff}},
\end{equation}
refer to manifolds with fixed resonator excitation number $n=0,1,2,\dots$
and two-qubit spin states with a resonator-induced splitting of $4g_{\mathrm{eff}}$
between the states $\left\{ \left|\uparrow_{x},\downarrow_{x}\right\rangle ,\left|\downarrow_{x},\uparrow_{x}\right\rangle \right\} $
with $s^{2}=0$ and $\left\{ \left|\uparrow_{x},\uparrow_{x}\right\rangle ,\left|\downarrow_{x},\downarrow_{x}\right\rangle \right\} $
with $s^{2}=4$, respectively. 

\textit{Perturbation in the interaction picture}.---In the following
we focus on the perturbative regime where the perturbation $\sim\omega_{q}$
is small compared to the resonator-induced splitting of $S_{x}^{2}$-eigenstates,
that is $\omega_{q}\ll8g_{\mathrm{eff}}=8\mu^{2}\omega_{c}$. Rewriting
the perturbation in the unperturbed eigenbasis yields 
\begin{equation}
V=\sum_{n,n'}\sum_{\vec{\sigma},\vec{\sigma}'}\left<n'|D\left[\mu\left(s'-s\right)\right]|n\right>\left<\vec{\sigma}_{x}'|V|\vec{\sigma}_{x}\right>\widetilde{\left|n',\vec{\sigma}'_{x}\right\rangle }\widetilde{\left\langle n,\vec{\text{\ensuremath{\sigma}}}_{x}\right|}.
\end{equation}
Using the relation \cite{cahill69} 
\begin{equation}
\left<m|D\left[\alpha\right]|n\right>=\sqrt{\frac{n!}{m!}}\alpha^{m-n}e^{-\left|\alpha\right|^{2}/2}L_{n}^{(m-n)}\left(\left|\alpha\right|^{2}\right),\label{eq:Glauber-relation}
\end{equation}
with $L_{n}^{(m-n)}$ denoting the associated Laguerre polynominals,
in the experimentally most relevant regime of weak spin-resonator
coupling (that is, $\mu\ll1$) we can neglect the off-diagonal contributions
where $n\neq m$, since eigenstates with different boson number are
very weakly coupled $\left(\sim\omega_{q}\mu^{|n-m|}\right)$ and
far off-resonance $\left(\omega_{q}\ll8g_{\mathrm{eff}}\ll\omega_{c}\right)$,
with rapidly decaying contributions as the number difference increases.
In this limit, the perturbation in the interaction picture {[}compare
Eq.(\ref{eq:perturbation-interaction-picture}){]} reads
\begin{eqnarray}
\tilde{V}\left(\tau\right) & \approx & \tilde{V}_{q}\left(\tau\right)\otimes\sum_{n}\chi_{n}\left(\mu\right)\left|n\right\rangle \left\langle n\right|,\label{eq:perturbation-interaction-picture-qubit-resonator}\\
\tilde{V}_{q}\left(\tau\right) & = & \frac{\omega_{q}}{2}\left[e^{i4g_{\mathrm{eff}}\tau}Q+e^{-i4g_{\mathrm{eff}}\tau}Q^{\dagger}\right],\label{eq:perturbation-interaction-picture-explicit}
\end{eqnarray}
where 
\begin{equation}
\chi_{n}\left(\mu\right)=\left<n|D\left[\pm2\mu\right]|n\right>=e^{-2\mu^{2}}L_{n}^{(0)}\left(4\mu^{2}\right),
\end{equation}
Since the perturbation $\sim S^{z}$ is purely off-diagonal in the
$S^{x}$ eigenbasis, the operator 
\begin{eqnarray}
Q & = & \left|\uparrow_{x}\downarrow_{x}\right\rangle \left\langle \downarrow_{x}\downarrow_{x}\right|+\left|\downarrow_{x}\uparrow_{x}\right\rangle \left\langle \downarrow_{x}\downarrow_{x}\right|\nonumber \\
 &  & +\left|\uparrow_{x}\downarrow_{x}\right\rangle \left\langle \uparrow_{x}\uparrow_{x}\right|+\left|\downarrow_{x}\uparrow_{x}\right\rangle \left\langle \uparrow_{x}\uparrow_{x}\right|,
\end{eqnarray}
describes \textit{only} transitions from the $s=\pm2$ subspace to
the $s=0$ subspace (and vice versa for the Hermitian conjugate operator
$Q^{\dagger}$), which in the interaction picture underlying Eq.(\ref{eq:perturbation-interaction-picture-explicit})
rotate with the corresponding transition frequency $\pm4g_{\mathrm{eff}}$.
While Eq.(\ref{eq:perturbation-interaction-picture-qubit-resonator})
is purely off-diagonal in spin-space, in the limit $\mu\ll1$ it is
(approximately) diagonal in the excitation number $\left|n\right\rangle $,
as the coupling $V$ between different $n$-subspaces is strongly
detuned by the corresponding large energy splitting $\sim\omega_{c}$. 

\textit{Quasi-decoherence-free subspace}.---In our numerical simulations,
the initial qubit states have been chosen to be aligned along the
$z$-direction, defining the computational basis states and corresponding
to eigenstates of the perturbation $V\sim S^{z}$. Therefore, it is
didactic to rewrite $\tilde{V}\left(\tau\right)$ in the eigenbasis
of $S^{z}$. With $\left|\uparrow_{x}\right\rangle =\left(\left|\uparrow_{z}\right\rangle +\left|\downarrow_{z}\right\rangle \right)/\sqrt{2}$,
and $\left|\downarrow_{x}\right\rangle =\left(\left|\uparrow_{z}\right\rangle -\left|\downarrow_{z}\right\rangle \right)/\sqrt{2}$,
we obtain 
\begin{eqnarray}
Q & = & \left|\uparrow_{z}\uparrow_{z}\right\rangle \left\langle \uparrow_{z}\uparrow_{z}\right|-\left|\downarrow_{z}\downarrow_{z}\right\rangle \left\langle \downarrow_{z}\downarrow_{z}\right|\nonumber \\
 &  & +\left|\uparrow_{z}\uparrow_{z}\right\rangle \left\langle \downarrow_{z}\downarrow_{z}\right|-\left|\downarrow_{z}\downarrow_{z}\right\rangle \left\langle \uparrow_{z}\uparrow_{z}\right|.\label{eq:Q-operator-z-basis}
\end{eqnarray}
As can be seen readily from this expression, the subspace $\left\{ \left|\uparrow_{z}\downarrow_{z}\right\rangle ,\left|\downarrow_{z}\uparrow_{z}\right\rangle \right\} $
with $S^{z}=0$ defines a decoherence-free subspace, since $Q$ and
$Q^{\dagger}$ {[}and therefore $\tilde{V}\left(\tau\right)${]} vanish
on this subspace, with $Q\left|\uparrow_{z}\downarrow_{z}\right\rangle =Q\left|\downarrow_{z}\uparrow_{z}\right\rangle =0$.
In the following this finding is elaborated in more detail: To do
so, we first rewrite $\tilde{V}\left(\tau\right)$ as 
\begin{eqnarray}
\tilde{V}\left(\tau\right) & = & \frac{\omega_{q}}{2}D^{\dagger}(\mu S^{x})e^{i\omega_{c}a^{\dagger}a\tau}e^{-ig_{\mathrm{eff}}\tau S_{x}^{2}}D(\mu S^{x})S^{z}\nonumber \\
 &  & \times D^{\dagger}(\mu S^{x})e^{-i\omega_{c}a^{\dagger}a\tau}e^{ig_{\mathrm{eff}}\tau S_{x}^{2}}D(\mu S^{x}).\label{eq:perturbation-interaction-picture-exact}
\end{eqnarray}
This expression is exact. Defining triplet and singlet states in the
spin-eigenbasis of $H_{0}$ as 
\begin{eqnarray}
\left|T_{+}^{x}\right\rangle  & = & \left|\uparrow_{x}\uparrow_{x}\right\rangle ,\\
\left|T_{0}^{x}\right\rangle  & = & \left(\left|\uparrow_{x}\downarrow_{x}\right\rangle +\left|\downarrow_{x}\uparrow_{x}\right\rangle \right)/\sqrt{2},\\
\left|T_{-}^{x}\right\rangle  & = & \left|\downarrow_{x}\downarrow_{x}\right\rangle ,\\
\left|S^{x}\right\rangle  & = & \left(\left|\uparrow_{x}\downarrow_{x}\right\rangle -\left|\downarrow_{x}\uparrow_{x}\right\rangle \right)/\sqrt{2},
\end{eqnarray}
the (by definition) computational basis states (taken as initial states
in our numerical simulations) are given by 
\begin{eqnarray}
\left|\uparrow_{z}\uparrow_{z}\right\rangle  & = & \frac{1}{2}\left[\left|T_{+}^{x}\right\rangle +\sqrt{2}\left|T_{0}^{x}\right\rangle +\left|T_{-}^{x}\right\rangle \right],\label{eq:up-up-z}\\
\left|\uparrow_{z}\downarrow_{z}\right\rangle  & = & \frac{1}{2}\left[\left|T_{+}^{x}\right\rangle -\sqrt{2}\left|S^{x}\right\rangle -\left|T_{-}^{x}\right\rangle \right],\label{eq:up-dn-z}\\
\left|\downarrow_{z}\uparrow_{z}\right\rangle  & = & \frac{1}{2}\left[\left|T_{+}^{x}\right\rangle +\sqrt{2}\left|S^{x}\right\rangle -\left|T_{-}^{x}\right\rangle \right],\label{eq:dn-up-z}\\
\left|\downarrow_{z}\downarrow_{z}\right\rangle  & = & \frac{1}{2}\left[\left|T_{+}^{x}\right\rangle -\sqrt{2}\left|T_{0}^{x}\right\rangle +\left|T_{-}^{x}\right\rangle \right].\label{eq:dn-dn-z}
\end{eqnarray}
For a general resonator state $\left|\mathrm{cav}\right\rangle $,
the first-order error term will be proportional to\begin{widetext}
\begin{eqnarray}
\tilde{V}\left(\tau\right)\left|T_{+}^{x}\right\rangle \left|\mathrm{cav}\right\rangle  & = & \frac{\omega_{q}}{\sqrt{2}}e^{4ig_{\mathrm{eff}}\tau}\left|T_{0}^{x}\right\rangle \otimes e^{i\omega_{c}a^{\dagger}a\tau}D^{\dagger}(2\mu)e^{-i\omega_{c}a^{\dagger}a\tau}D(2\mu)\left|\mathrm{cav}\right\rangle ,\\
\tilde{V}\left(\tau\right)\left|T_{0}^{x}\right\rangle \left|\mathrm{cav}\right\rangle  & = & \frac{\omega_{q}}{\sqrt{2}}e^{-4ig_{\mathrm{eff}}\tau}\left[\left|T_{+}^{x}\right\rangle \otimes D^{\dagger}(2\mu)e^{i\omega_{c}a^{\dagger}a\tau}D(2\mu)e^{-i\omega_{c}a^{\dagger}a\tau}\left|\mathrm{cav}\right\rangle \right. \nonumber \\
 &  & \left.+\left|T_{-}^{x}\right\rangle \otimes D^{\dagger}(-2\mu)e^{i\omega_{c}a^{\dagger}a\tau}D(-2\mu)e^{-i\omega_{c}a^{\dagger}a\tau}\left|\mathrm{cav}\right\rangle \right],\\
\tilde{V}\left(\tau\right)\left|T_{-}^{x}\right\rangle \left|\mathrm{cav}\right\rangle  & = & \frac{\omega_{q}}{\sqrt{2}}e^{4ig_{\mathrm{eff}}\tau}\left|T_{0}^{x}\right\rangle \otimes e^{i\omega_{c}a^{\dagger}a\tau}D^{\dagger}(-2\mu)e^{-i\omega_{c}a^{\dagger}a\tau}D(-2\mu)\left|\mathrm{cav}\right\rangle \\
\tilde{V}\left(\tau\right)\left|S^{x}\right\rangle \left|\mathrm{cav}\right\rangle  & = & 0.
\end{eqnarray}
In the spirit of our previous discussion {[}recall Eq.(\ref{eq:Glauber-relation})
with $D^{\dagger}\left(\alpha\right)=D\left(-\alpha\right)${]}, these
exact statements can be simplified in the limit $\mu\ll1$ as 
\begin{eqnarray}
e^{i\omega_{c}a^{\dagger}a\tau}D^{\dagger}(\pm2\mu)e^{-i\omega_{c}a^{\dagger}a\tau} & = & \sum_{n,n'}e^{i\omega_{c}\tau\left(n'-n\right)}\left<n'|D^{\dagger}\left(\pm2\mu\right)|n\right>\left|n'\right\rangle \left\langle n\right|,\\
 & \approx & \sum_{n}\chi_{n}\left(\mu\right)\left|n\right\rangle \left\langle n\right|,
\end{eqnarray}
yielding the approximate results {[}for a Fock state $\left|\mathrm{cav}\right\rangle =\left|n\right\rangle ${]}
\begin{eqnarray}
\tilde{V}\left(\tau\right)\left|T_{+}^{x}\right\rangle \left|n\right\rangle  & \approx & \frac{\omega_{q}}{\sqrt{2}}e^{4ig_{\mathrm{eff}}\tau}\chi_{n}^{2}\left(\mu\right)\left|T_{0}^{x}\right\rangle \left|n\right\rangle ,\\
\tilde{V}\left(\tau\right)\left|T_{0}^{x}\right\rangle \left|n\right\rangle  & \approx & \frac{\omega_{q}}{\sqrt{2}}e^{-4ig_{\mathrm{eff}}\tau}\chi_{n}^{2}\left(\mu\right)\left[\left|T_{+}^{x}\right\rangle +\left|T_{-}^{x}\right\rangle \right]\left|n\right\rangle ,\\
\tilde{V}\left(\tau\right)\left|T_{-}^{x}\right\rangle \left|n\right\rangle  & \approx & \frac{\omega_{q}}{\sqrt{2}}e^{4ig_{\mathrm{eff}}\tau}\chi_{n}^{2}\left(\mu\right)\left|T_{0}^{x}\right\rangle \left|n\right\rangle .
\end{eqnarray}
\end{widetext}With these (approximate) relations, one can readily
verify $\tilde{V}\left(\tau\right)\left|\uparrow_{z}\downarrow_{z}\right\rangle \left|n\right\rangle \approx0$
and $\tilde{V}\left(\tau\right)\left|\downarrow_{z}\uparrow_{z}\right\rangle \left|n\right\rangle \approx0$,
in agreement with our result based on Eq.(\ref{eq:Q-operator-z-basis}),
while the subspace $\left\{ \left|\uparrow_{z}\uparrow_{z}\right\rangle ,\left|\downarrow_{z}\downarrow_{z}\right\rangle \right\} $
is directly affected by the perturbation $\tilde{V}\left(\tau\right)$.
As long as transitions between different $n$-subspaces can be neglected,
the bosonic part of the Hamiltonian can be ignored and the free part
of the Hamiltonian reduces to $H_{0}\approx-g_{\mathrm{eff}}S_{x}^{2}$.
Then, since the perturbation $V=(\omega_{q}/2)S^{z}$ leaves the subspace
$\left\{ \left|\uparrow_{z}\downarrow_{z}\right\rangle ,\left|\downarrow_{z}\uparrow_{z}\right\rangle \right\} $
invariant, $V$ cannot induce errors, since it vanishes on this subspace.
As a perspective, this finding opens up the possibility to define
a logical qubit in the quasi-decoherence-free subspace $\left\{ \left|\uparrow_{z}\downarrow_{z}\right\rangle ,\left|\downarrow_{z}\uparrow_{z}\right\rangle \right\} $
as $\left|\mathrm{qubit}\right\rangle =\alpha\left|\uparrow_{z}\downarrow_{z}\right\rangle +\beta\left|\downarrow_{z}\uparrow_{z}\right\rangle $,
which is largely protected from splitting-induced errors in the limit
$\mu\ll1$ (provided that the perturbative condition $\omega_{q}\ll8g_{\mathrm{eff}}$
is still satisfied). 

\begin{figure}
\includegraphics[width=1\columnwidth]{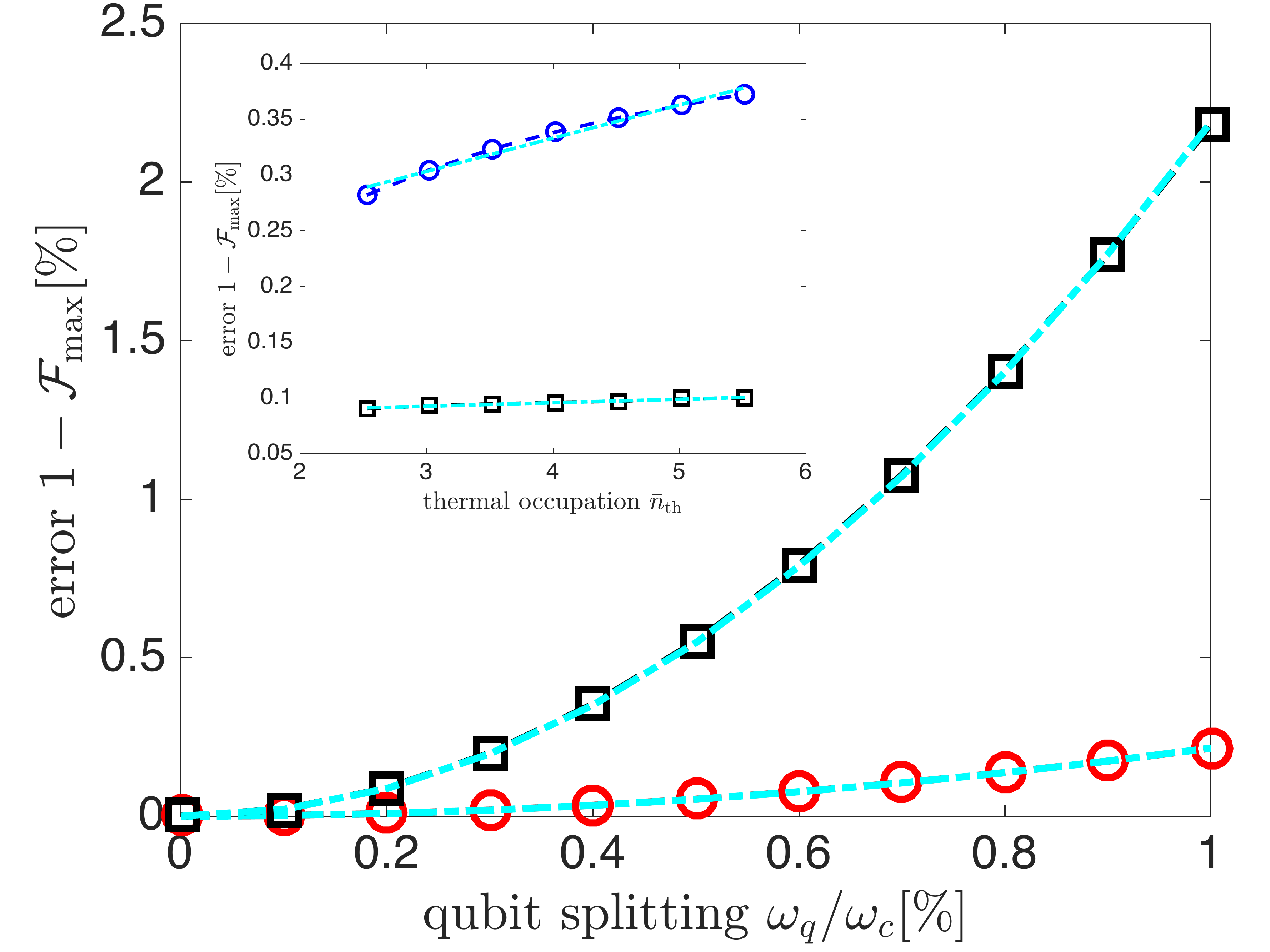}

\caption{\label{fig:error-qubit-splitting}(color online). The error $\xi_{q}$
induced by a non-zero qubit splitting $\omega_{q}>0$, for $\mathcal{S}=S^{x}=\sum_{i}\sigma_{i}^{x}$
(transversal coupling), and for different initial qubit states $\left|\Psi(0)\right\rangle =\left|\uparrow_{z}\downarrow_{z}\right\rangle $
(red circles) and $\left|\Psi(0)\right\rangle =\left|\downarrow_{z}\downarrow_{z}\right\rangle $
(black squares); here, $g/\omega_{c}=1/8$ and $k_{B}T/\omega_{c}=1$.
Quadratic fits (cyan, dash-dotted lines) verify a quadratic error
scaling $\sim\omega_{q}^{2}$, with the numerical pre-factor $\alpha_{q}$
depending on both the spin-resonator coupling $g$ and temperature
$T$. Inset: The error $\xi_{q}$ as a function of the thermal occupation
number $\bar{n}_{\mathrm{th}}$ for $g/\omega_{c}=1/8$ (black squares)
and $g/\omega_{c}=1/\left(8\sqrt{2}\right)$ (blue circles) for $\left|\Psi(0)\right\rangle =\left|\uparrow_{z}\downarrow_{z}\right\rangle $
and $\omega_{q}/\omega_{c}=0.5\%$. Other numerical parameters: $\Gamma=\kappa=0$.}
\end{figure}

\textit{Splitting-induced error}.---Based on Eqs.(\ref{eq:perturbative-series-rho})
and (\ref{eq:perturbation-interaction-picture-qubit-resonator}),
in the following we derive an approximate analytic expression for
the splitting-induced error $\xi_{q}$. Taking the trace over the
resonator mode, for stroboscopic times $t_{m}=2\pi m/\omega_{c}$
(where the ideal evolution reduces to a pure spin gate, leaving the
resonator mode unaffected) the fidelity $\mathcal{F}$ with the target
qubit state $\left|\Psi_{\mathrm{tar}}\right\rangle =\exp\left[-iH_{0}t_{m}\right]\left|\Psi(0)\right\rangle $
is found to be 
\begin{equation}
\mathcal{F}\left(t_{m}\right)=1-\left<\Psi_{\mathrm{tar}}\right|\varrho^{(2)}\left(t_{m}\right)\left|\Psi_{\mathrm{tar}}\right>,
\end{equation}
where we have used that first-order terms vanish; moreover, we have
introduced the second-order contribution 
\begin{eqnarray*}
\varrho^{(2)} & = & -\varUpsilon{}_{q}e^{-iH_{0}t_{m}}\left\{ \int_{0}^{t_{m}}d\tau\int_{0}^{t_{m}}d\tau'\tilde{V}_{q}\left(\tau\right)\varrho\left(0\right)\tilde{V}_{q}\left(\tau'\right)\right.\\
 &  & -\int_{0}^{t_{m}}d\tau_{2}\int_{0}^{\tau_{2}}d\tau_{1}\tilde{V}_{q}\left(\tau_{2}\right)\tilde{V}_{q}\left(\tau_{1}\right)\varrho\left(0\right)\\
 &  & \left.-\int_{0}^{t_{m}}d\tau_{2}\int_{0}^{\tau_{2}}d\tau_{1}\varrho\left(0\right)\tilde{V}_{q}\left(\tau_{1}\right)\tilde{V}_{q}\left(\tau_{2}\right)\right\} e^{iH_{0}t_{m}},
\end{eqnarray*}
with $\varrho\left(0\right)=\left|\Psi(0)\right\rangle \left\langle \Psi(0)\right|$
and the pre-factor 
\begin{equation}
\varUpsilon{}_{q}=\varUpsilon{}_{q}\left(\mu,k_{B}T\right)=\frac{1}{Z}\sum_{n}e^{-\beta\omega_{c}n}\chi_{n}^{2}\left(\mu\right).\label{eq:prefactor-1}
\end{equation}
The latter depends on both the spin-resonator coupling $\mu=g/\omega_{c}$
and temperature $T$ (with $\beta=1/k_{B}T$) and can be readily evaluated
numerically. After some manipulations, we then arrive at an analytic
expression for the error $\xi_{q}=1-\mathcal{F}\left(t_{\mathrm{max}}\right)$
at the (nominally) optimal time $t_{\mathrm{max}}=\pi/8g_{\mathrm{eff}}$.
For $\left|\Psi(0)\right\rangle \in\left\{ \left|\uparrow_{z}\uparrow_{z}\right\rangle ,\left|\downarrow_{z}\downarrow_{z}\right\rangle \right\} $,
it reads explicitly 
\begin{eqnarray}
\xi_{q} & = & \varUpsilon{}_{q}\left(\mu,k_{B}T\right)\frac{\omega_{q}^{2}}{16g_{\mathrm{eff}}^{2}},\\
 & = & \alpha{}_{q}\times\left(\omega_{q}/\omega_{c}\right)^{2},
\end{eqnarray}
showing a quadratic scaling with the splitting $\sim\omega_{q}^{2}$.
In the last step, we have introduced the pre-factor $\alpha{}_{q}=\varUpsilon{}_{q}\left(\mu,k_{B}T\right)/\left(16\mu^{4}\right)$.

\textit{Numerical results}.---As shown in Fig.\ref{fig:error-qubit-splitting},
we have numerically verified our analytical results (as discussed
above): (i) The error $\xi_{q}$ scales quadratically with the qubit
splitting, i.e., $\xi_{q}\sim\left(\omega_{q}/\omega_{c}\right)^{2}$,
with (ii) a numerical pre-factor $\alpha_{q}$ depending on both the
spin-resonator coupling $g$ and temperature $T$, and (iii) (all
other parameters equal) the error $\xi_{q}$ is found to be significantly
smaller for initial states in the quasi-decoherence-free subspace
$\left\{ \left|\uparrow_{z}\downarrow_{z}\right\rangle ,\left|\downarrow_{z}\uparrow_{z}\right\rangle \right\} $
than for initial qubit states in the orthogonal subspace $\left\{ \left|\uparrow_{z}\uparrow_{z}\right\rangle ,\left|\downarrow_{z}\downarrow_{z}\right\rangle \right\} $.

\section{SAW-based Spin-Resonator System \label{sec:SAW-based-Spin-Resonator-Coupling} }

Here, we provide further details on how to implement experimental
candidate systems governed by the class of Hamiltonians given in Eq.(1),
using quantum dots embedded in high-quality surface acoustic wave
(SAW) resonators \cite{chen15,schuetz15}. For similar considerations
based on (for example) transmission-line resonators or nanomechanical
oscillators, we refer to Refs.\cite{hu12} and \cite{rabl10},
respectively. 

\textit{Charge qubit}.---A single electron in a double quantum dot
(DQD) coupled to a SAW resonator can be described by 
\begin{equation}
H_{\mathrm{charge}}=\frac{\epsilon}{2}\sigma^{z}+t_{c}\sigma^{x}+\omega_{c}a^{\dagger}a+g_{\mathrm{ch}}\sigma^{z}\otimes\left(a+a^{\dagger}\right),\label{eq:spin-resonator-Hamiltonian-charge-qubit}
\end{equation}
where $\epsilon$ is the interdot detuning parameter, $t_{c}$ the
tunnel coupling between the dots, $g_{\mathrm{ch}}=e\phi_{0}\mathcal{F}\left(kd\right)\sin\left(kl/2\right)$
the bare single-phonon coupling strength (assuming a sine-like mode
function of the piezoelectric potential, with a node tuned between
the two dots separated by a distance $l$), and the (orbital) Pauli
operators are defined as $\sigma^{z}=\left|L\right\rangle \left\langle L\right|-\left|R\right\rangle \left\langle R\right|$
and $\sigma^{x}=\left|L\right\rangle \left\langle R\right|+\left|R\right\rangle \left\langle L\right|$,
respectively \cite{schuetz15}. In our expression for $g_{\mathrm{ch}}$,
$e$ refers to the electron's charge, and $\phi_{0}$ to the piezoelectric
potential associated with a single SAW phonon; the decay of the SAW
resonator mode into the bulk is captured by the factor $\mathcal{F}\left(kd\right)$,
where $d$ is the distance between the DQD and the surface and $k=2\pi/\lambda_{c}$
the wavenumber of the resonator mode \cite{schuetz15}. In the
computational basis, where the dot Hamiltonian $H_{\mathrm{dot}}=\frac{\epsilon}{2}\sigma^{z}+t_{c}\sigma^{x}$
is diagonal, with the electronic eigenstates 
\begin{eqnarray}
\left|+\right\rangle  & = & \cos\theta\left|L\right\rangle +\sin\theta\left|R\right\rangle ,\\
\left|-\right\rangle  & = & -\sin\theta\left|L\right\rangle +\cos\theta\left|R\right\rangle ,
\end{eqnarray}
where the mixing angle is given by $\tan\theta=2t_{c}/\left(\epsilon+\Omega\right)$,
$\Omega=\sqrt{\epsilon^{2}+4t_{c}^{2}}$, the spin-resonator Hamiltonian
given in Eq.(\ref{eq:spin-resonator-Hamiltonian-charge-qubit}) can
be rewritten as 
\begin{eqnarray}
H_{\mathrm{charge}} & = & \frac{\Omega}{2}S^{z}+\omega_{c}a^{\dagger}a+g^{x}S^{x}\otimes\left(a+a^{\dagger}\right)\nonumber \\
 &  & +g^{z}S^{z}\otimes\left(a+a^{\dagger}\right),\label{eq:spin-resonator-Hamiltonian-charge-qubit-computational-basis}
\end{eqnarray}
where the Pauli operators in the logical qubit basis are $S^{z}=\left(\left|+\right\rangle \left\langle +\right|-\left|-\right\rangle \left\langle -\right|\right)$,
$S^{x}=\left(\left|+\right\rangle \left\langle -\right|+\left|-\right\rangle \left\langle +\right|\right)$
and 
\begin{eqnarray}
g^{x} & = & g_{\mathrm{ch}}\frac{2t_{c}}{\Omega},\\
g^{z} & = & -g_{\mathrm{ch}}\frac{\epsilon}{\Omega}.
\end{eqnarray}
In the last step, we have made use of the relations $2\sin\theta\cos\theta=\sin\left(2\theta\right)=2t_{c}/\Omega$
and $\cos^{2}\theta-\sin^{2}\theta=\cos\left(2\theta\right)=\epsilon/\Omega$.
In the limit where $\delta,g_{\mathrm{ch}}\ll\omega_{c}$, with $\delta=\Omega-\omega_{c}$,
one can perform a rotating-wave approximation yielding the standard
Jaynes-Cummings Hamiltonian \cite{frey12}. Finally, the spin-resonator
Hamiltonian given in Eq.(\ref{eq:spin-resonator-Hamiltonian-charge-qubit-computational-basis})
belongs to the general class of Hamiltonians defined in Eq.(1). In
particular, at the charge degeneracy point $\epsilon=0$, where $\sin\theta=\cos\theta=1/\sqrt{2}$,
the Hamiltonian given in Eq.(\ref{eq:spin-resonator-Hamiltonian-charge-qubit-computational-basis})
reduces to 
\begin{equation}
H_{\mathrm{charge}}=t_{c}S^{z}+\omega_{c}a^{\dagger}a+g_{\mathrm{ch}}S^{x}\otimes\left(a+a^{\dagger}\right).
\end{equation}
Accordingly, the (pseudo-) spin-resonator coupling is maximized at
this charge-degeneracy point, i.e., when there is no bias between
the two dots, and decreases as one moves away from this point \cite{frey12,jin11,hu12}. 

\textit{Coupling strength}.---Following Ref.\cite{chen15}, the
single phonon coupling strength $g_{\mathrm{ch}}$ may be expressed
as 
\begin{equation}
\text{\ensuremath{\frac{g_{\mathrm{ch}}}{\omega_{c}}}=\ensuremath{\zeta_{\mathrm{ch}}}=\ensuremath{\sqrt{\alpha_{\mathrm{eff}}}}}\sqrt{\frac{l^{2}\lambda}{V}},\label{eq:charge-coupling-SAW-relative}
\end{equation}
where $V$ is the mode volume associated with the resonator mode and
$\alpha_{\mathrm{eff}}=\alpha K^{2}c/v_{s}\epsilon_{r}$ is an effective
fine-structure constant, defined in terms of the fine structure constant
$\alpha\sim1/137$, the (material-specific) electromechanical coupling
coefficient $K^{2}$ (as a widely used measure to quantify the piezoelectric
coupling strength), the speed of light $c$, the SAW speed of sound
$v_{s}$ and the relative dielectric constant $\epsilon_{r}$. The
coupling parameter $K^{2}$ describes piezoelectric stiffening and
may be expressed as $K^{2}=e_{14}^{2}/\underline{c}\epsilon$, where
$e_{14}$, $\underline{c}$, and $\epsilon$ refer to representative
values of the piezoelectric, the elasticity and the dielectric tensor,
respectively. Typical values for $\alpha_{\mathrm{eff}}/\alpha$ range
from $\alpha_{\mathrm{eff}}/\alpha\sim10$ for GaAs up to $\alpha_{\mathrm{eff}}/ \alpha \gtrsim 100$
for strongly piezoelectric materials such as $\mathrm{LiNbO}_{3}$
or ZnO, underlining the potential of SAW based systems to reach the
ultra-strong coupling regime \cite{chen15}. For a typical SAW
penetration length $\sim0.3\lambda$ close to the surface, Eq.(\ref{eq:charge-coupling-SAW-relative})
further simplifies to $g_{\mathrm{ch}}/\omega_{c}\approx\left(0.5-1.5\right)\sqrt{l^{2}/A}$,
where $A$ refers to the surface mode area. When expressing $\alpha_{\mathrm{eff}}$
in terms of the fundamental material parameters, Eq.(\ref{eq:charge-coupling-SAW-relative})
can be rewritten as 
\begin{equation}
\frac{g_{\mathrm{ch}}}{\omega_{c}}\approx\frac{ee_{14}}{\epsilon v_{s}}\sqrt{\frac{1}{\rho v_{s}}}\sqrt{\frac{l^{2}\lambda}{V}}.
\end{equation}
This estimate also follows from the expression given above, $g_{\mathrm{ch}}=e\phi_{0}\mathcal{F}\left(kd\right)\sin\left(kl/2\right)$,
with $\phi_{0}\approx\left(e_{14}/\epsilon\right)\sqrt{\hbar/2\rho V\omega_{c}}$
\cite{schuetz15}, close to the surface $\mathcal{F}\left(kd\right)\sim1$,
and with $\sin\left(kl/2\right)\approx kl/2$ for $kl/2\ll1$ (in
the spirit of circuit QED setups). 

\textit{Spin qubit}.---In the two-electron regime of a DQD, one can
couple the effective dipole-moment of singlet-triplet subspace to
the resonator mode \cite{schuetz15,taylor06}. Within the two-level
subspace (all other levels are far detuned), the dynamics are described
by 
\begin{equation}
H_{\mathrm{spin}}=\frac{\text{\ensuremath{\Omega}}}{2}\sigma^{z}+\omega_{c}a^{\dagger}a+g^{x}\sigma^{x}\otimes\left(a+a^{\dagger}\right)+g^{z}\sigma^{z}\otimes\left(a+a^{\dagger}\right),\label{eq:spin-resonator-Hamiltonian-spin-qubit-general-xz-coupling}
\end{equation}
where $\sigma^{z}=\left|1\right\rangle \left\langle 1\right|-\left|0\right\rangle \left\langle 0\right|$,
$\sigma^{x}=\left|1\right\rangle \left\langle 0\right|+\left|0\right\rangle \left\langle 1\right|$
and 
\begin{eqnarray}
g^{x} & = & e\phi_{0}\mathcal{F}\left(kd\right)\eta_{\mathrm{geo}}\kappa_{0}\kappa_{1},\\
g^{z} & = & e\phi_{0}\mathcal{F}\left(kd\right)\eta_{\mathrm{geo}}\left[\kappa_{1}^{2}-\kappa_{0}^{2}\right]/2.
\end{eqnarray}
Here, $\eta_{\mathrm{geo}}=\sin\left(kx_{R}\right)-\sin\left(kx_{L}\right)$
accounts for the positioning of the DQD with respect to the piezoelectric
mode function. The coupling is reduced by the admixtures of the qubit's
states $\left\{ \left|0\right\rangle ,\left|1\right\rangle \right\} $
with the localized singlet $\kappa_{n}=\left<n|S_{02}\right>$. Again,
for $\Omega\approx\omega_{c}$ and $g^{\alpha}\ll\omega_{c}$, we
recover the prototypical Jaynes-Cummings dynamics. Moreover, the spin-resonator
Hamiltonian given in Eq.(\ref{eq:spin-resonator-Hamiltonian-spin-qubit-general-xz-coupling})
belongs to the general class of Hamiltonians defined in Eq.(1).

\textit{Hot gat}e.---For such a spin qubit a spin-resonator coupling
strength of $g_{\mathrm{sp}}/2\pi\equiv g^{x}/2\pi=\left(g_{0}/2\pi\right)\kappa_{0}\kappa_{1}\approx3.2\mathrm{MHz}$
$\left(g^{z}/2\pi\approx0.64\mathrm{MHz}\right)$ has been predicted
for typical parameters in GaAs \cite{schuetz15}. For a typical
resonator frequency $\omega_{c}/2\pi\approx1.5\mathrm{GHz}$, this
amounts to a relative coupling strength $\mu_{\mathrm{sp}}=g_{\mathrm{sp}}/\omega_{c}\approx0.2\%$
and an effective coupling $g_{\mathrm{eff}}/2\pi=\mu_{\mathrm{sp}}g_{\mathrm{sp}}/2\pi\approx65\mathrm{kHz}$,
which could be increased substantially by additionally depositing
a strongly piezoelectric material such as $\mathrm{LiNbO}_{3}$ or
$\mathrm{ZnO}$ on the GaAs substrate \cite{schuetz15,chen15,gustafsson12-sm}.
The condition $\omega_{c}\gg\Omega$ can be satisfied by choosing
the magnetic gradient $\Delta$ between the dots appropriately, $\Delta\lesssim0.1\mu\mathrm{eV}$.
Recently, SAW resonators with quality-factors approaching $\sim10^{6}$
have been realized experimentally \cite{manenti16-sm}. Then, taking
an optimistic quality-factor of $Q=10^{6}$, according to the hot-gate
requirement $k_{B}T\ll Q\times g_{\mathrm{eff}}$, we find $T\ll3.1\mathrm{K}$;
therefore, for spin qubits coupled to high-quality SAW-resonators,
our scheme can tolerate temperatures approaching the Kelvin regime,
where the thermal occupation number is much larger than one. For example,
for $\omega_{c}/2\pi\approx\left(1.0-1.5\right)\mathrm{GHz}$ and
$T\approx0.5\mathrm{K}$, we have $\bar{n}_{\mathrm{th}}\approx6.5-10$.
The second requirement for small errors, $\Gamma\ll g_{\mathrm{eff}}$,
yields $\Gamma/2\pi\ll65\mathrm{kHz}$, which may be satisfied in
GaAs with recently demonstrated echo techniques, where decoherence
timescales $T_{2}\approx1\mathrm{ms}$ have been demonstrated \cite{malinowski16-sm}.
Finally, with $\bar{n}_{\mathrm{th}}/Q\approx10/10^{6}$ and $\Gamma/\omega_{c}\approx1\mathrm{kHz}/1.5\mathrm{GHz}$,
and using the relation $\xi\approx\alpha_{\kappa}\left(\kappa/\omega_{c}\right)\bar{n}_{\mathrm{th}}+\alpha_{\Gamma}\Gamma/\omega_{c},$
we can estimate the overall gate error as $\xi\approx4\times10^{-5}+2.5\times10^{-2}\approx2.5\%$,
which is largely limited by dephasing-induced errors (for the parameters
chosen here). Again, to counteract this source of error, a strongly
piezoelectric material such as $\mathrm{LiNbO}_{3}$ may be used on
the GaAs substrate. Alternatively, one could also investigate silicon
quantum dots: while this setup also requires a more sophisticated
heterostructure including some piezoelectric layer, it should benefit
from prolonged dephasing times $T_{2}^{\star}>100\mu\mathrm{s}$ \cite{veldhorst14},
which is not longer than the dephasing time $T_{2}$ quoted above
for GaAs, but relaxes the need for dynamical decoupling.

\section{Microscopic Derivation of the Noise Model\label{sec:Microscopic-Derivation}}

In this Appendix we provide a microscopic derivation of the Master
equation given in Eq.(7) of our manuscript. Here, we focus on the
relevant decoherence processes induced by coupling between the resonator
mode and its environment and restrict ourselves to the regime of interest
where $\omega_{q}\rightarrow0$. Our analysis is built upon the master
equation formalism, a tool widely used in quantum optics for studying
the irreversible dynamics of a quantum system coupled to a macroscopic
environment. We detail the assumptions of our approach and discuss
in detail the relevant approximations.

\subsection{The Model}

\begin{figure}
\includegraphics[width=1\columnwidth]{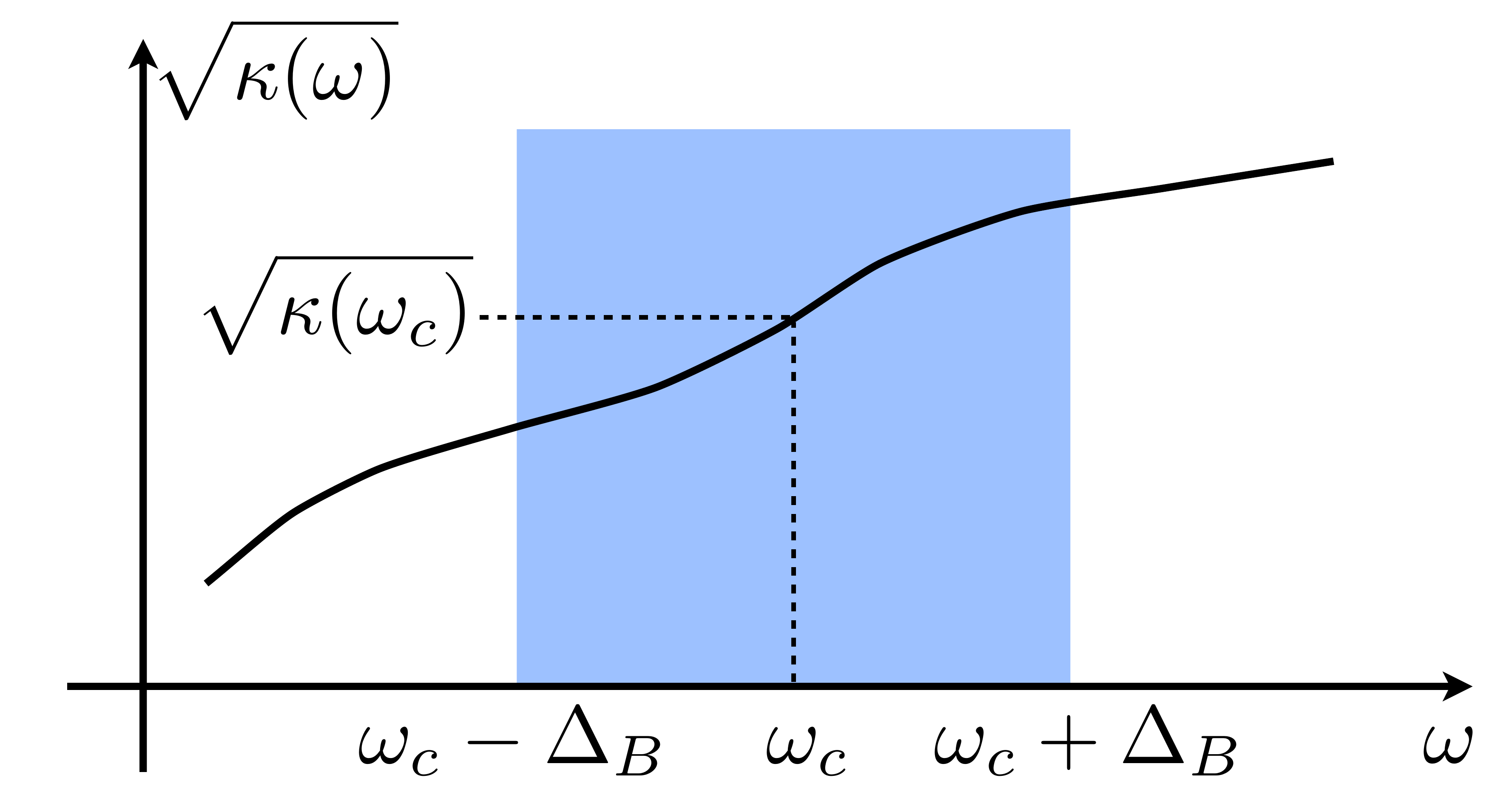}

\caption{\label{fig:sys-bath-coupling}(color online). Schematic illustration of the hierarchy
of frequency scales assumed for the derivation of the quantum Master
equation. Following the standard treatment \cite{yamamoto99}, the
reservoir spectral density $\kappa\left(\omega\right)/2\pi$ is taken
to be a flat function of $\omega$ within the frequency range of interest
$\left[\omega_{c}-\Delta_{B},\omega_{c}+\Delta_{B}\right]$.}
\end{figure}

We consider a generic linear coupling between the resonator mode and
a set of independent harmonic oscillators (representing e.g. the modes
of the free electromagnetic field), as described by the following
textbook system-bath Hamiltonian 
\begin{eqnarray}
H & = & \underset{=H_{0}}{\underbrace{H_{S}+H_{B}}}+H_{I},\\
H_{S} & = & \omega_{c}a^{\dagger}a+g\mathcal{S}\otimes\left(a+a^{\dagger}\right),\\
H_{B} & = & \int_{\omega_{c}-\Delta_{B}}^{\omega_{c}+\Delta_{B}}d\omega\omega b_{\omega}^{\dagger}b_{\omega},\\
H_{I} & = & \int_{\omega_{c}-\Delta_{B}}^{\omega_{c}+\Delta_{B}}d\omega\sqrt{\frac{\kappa\left(\omega\right)}{2\pi}}\left(a^{\dagger}b_{\omega}+ab_{\omega}^{\dagger}\right),
\end{eqnarray}
where $b_{\omega}$ refer to bosonic bath operators obeying standard
commutation relations with $[b_{\omega},b_{\omega'}^{\dagger}]=\delta\left(\omega-\omega'\right)$
etc. and $\Delta_{B}$ denotes the characteristic bandwidth of the
bath \cite{zollerOnline,rable14-lecture,guimond16}. Within a rotating-wave
approximation, we have dropped all energy non-conserving terms, which
is valid if the system's characteristic frequency $\omega_{c}$ is
the largest frequency in the problem \cite{rable14-lecture}. The
bandwidth $\Delta_{B}$ is the frequency range over which the system-bath
coupling is valid; it is closely related to the characteristic memory
or correlation time of the bath $\tau_{c}\sim\Delta_{B}^{-1}$, as
can be readily seen from the relation 
\begin{eqnarray}
\int_{\omega_{c}-\Delta_{B}}^{\omega_{c}+\Delta_{B}}d\omega e^{-i\omega\tau} & = & 2\Delta_{B}e^{-i\omega_{c}\tau}\mathrm{sinc}\left(\Delta_{B}\tau\right)\\
 & = & 2\pi\delta_{\Delta_{B}}\left(\tau\right)e^{-i\omega_{c}\tau},
\end{eqnarray}
as it appears in the standard derivation of the Master equation presented
below (if the spectral noise density $\kappa\left(\omega\right)$
and the thermal occupation number $\bar{n}_{\mathrm{th}}\left(\omega\right)$
are evaluated self-consistently at $\omega=\omega_{c}$). Here, the
function $\delta_{\Delta_{B}}\left(\tau\right)=\pi^{-1}\Delta_{B}\mathrm{sinc}\left(\Delta_{B}\tau\right)$
is a well-known diffraction-like function with a maximal amplitude
$\Delta_{B}/\pi$ at $\tau=0$ and a width of the order of $\tau_{c}\sim2\pi/\Delta_{B}$
\cite{cohen92}. Since the integral equals one, this function
is an approximate delta function which tends to $\delta\left(\tau\right)$
in the so-called white-noise limit $\Delta_{B}\rightarrow\infty$
(that is, $\tau_{c}\rightarrow0$). Intuitively, $\delta_{\Delta_{B}}\left(\tau\right)$
can be seen as a slowly-varying function (on the $\sim\omega_{c}^{-1}$
timescale) that effectively acts as a delta function on timescales
of the system evolution (i.e., much slower than $1/\Delta_{B}$).
Typically, $\Delta_{B}\ll\omega_{c}$ is assumed \cite{zollerOnline,rable14-lecture},
but $\tau_{c}$ is still much shorter than the relevant timescales
of the system dynamics $\tau_{\mathrm{sys}}$ (other than the free
rotation $\omega_{c}$), that is 
\begin{equation}
\omega_{c}\gg\Delta_{B}\gg\tau_{\mathrm{sys}}^{-1}.
\end{equation}
In this case, the bandwidth $\Delta_{B}$ can be much larger than
the spin-resonator coupling strength $g$ (which implies $g\tau_{c}\ll1$,
as required for the standard master equation treatment discussed below),
but still much smaller than the characteristic frequency $\omega_{c}$.
The system-reservoir coupling is usually only valid within a bandwidth
$2\Delta_{B}\ll\omega_{c}$ around $\omega_{c}$ \cite{rable14-lecture}.
Within this frequency range the coupling strength may be approximated
by a constant value as $\kappa\left(\omega\right)\approx\kappa\left(\omega_{c}\right)$,
as schematically depicted in Fig.\ref{fig:sys-bath-coupling}.

\subsection{Microscopic Derivation of the Master Equation \label{sub:Microscopic-Derivation-QME}}

Our analysis is based on the standard Born-Markov framework, where
correlations between the system and the bath are neglected (on relevant
timescales), since the bath is considered to be very large and the
effect of the interaction with the (small) system is negligible. Within
this standard Born-Markov approximation \cite{gardiner00,cohen92},
in the interaction picture the system's dynamics are described by
\begin{equation}
\dot{\tilde{\rho}}=-\int_{0}^{\infty}d\tau\mathrm{Tr}_{B}\left\{ \left[\tilde{H}_{I}\left(t\right),\left[\tilde{H}_{I}\left(t-\tau\right),\tilde{\rho}\left(t\right)\rho_{B}\right]\right]\right\} ,\label{eq:general-QME-microscopic}
\end{equation}
with $\tilde{\rho}=e^{iH_{0}t}\rho\left(t\right)e^{-iH_{0}t}$, $\tilde{H}_{I}\left(t\right)=e^{iH_{0}t}H_{I}e^{-iH_{0}t}$
and $\rho_{B}=Z^{-1}\exp\left[-\beta H_{B}\right]$ refers to a thermal
state of the bath with the standard thermal correlations functions
\cite{gardiner00}
\begin{equation}
\mathrm{Tr}_{B}\left[b_{\omega}^{\dagger}b_{\omega'}\rho_{B}\right]=\bar{n}_{\mathrm{th}}\left(\omega\right)\delta\left(\omega-\omega'\right),\label{eq:thermal-bath-correlation-functions}
\end{equation}
etc. Eq.(\ref{eq:general-QME-microscopic}) can equivalently be expressed
as\begin{widetext} 
\begin{eqnarray}
\dot{\tilde{\rho}} & = & \int_{0}^{\infty}d\tau\mathrm{Tr}_{B}\{\underset{\circled1}{\underbrace{\tilde{H}_{I}\left(t\right)\tilde{\rho}\left(t\right)\rho_{B}\tilde{H}_{I}\left(t-\tau\right)}}-\tilde{H}_{I}\left(t\right)\tilde{H}_{I}\left(t-\tau\right)\tilde{\rho}\left(t\right)\rho_{B}+\mathrm{h.c.}\}
\end{eqnarray}
In the interaction picture, the system-bath coupling reads explicitly
\begin{equation}
\tilde{H}_{I}\left(t\right)=\int_{\omega_{c}-\Delta_{B}}^{\omega_{c}+\Delta_{B}}d\omega\sqrt{\frac{\kappa\left(\omega\right)}{2\pi}}\left\{ e^{-i\omega t}b_{\omega}\left[e^{i\omega_{c}t}\left(a^{\dagger}+\mu\mathcal{S}\right)-\mu\mathcal{S}\right]+e^{i\omega t}b_{\omega}^{\dagger}\left[e^{-i\omega_{c}t}\left(a+\mu\mathcal{S}\right)-\mu\mathcal{S}\right]\right\} ,
\end{equation}
where we have used the fact that the resonator annihilation operators
transform as 
\begin{equation}
\tilde{a}\left(t\right)=e^{iH_{S}t}ae^{-iH_{S}t}=e^{-i\omega_{c}t}\left(a+\mu\mathcal{S}\right)-\mu\mathcal{S},
\end{equation}
while the bath operators transform simply as $\tilde{b}_{\omega}\left(t\right)=e^{iH_{B}t}b_{\omega}e^{-iH_{B}t}=e^{-i\omega t}b_{\omega}$.
Next, let us single out one term explicitly, but all other terms follow
analogously. Using the thermal correlation functions as stated in
Eq.(\ref{eq:thermal-bath-correlation-functions}), we then obtain
\begin{eqnarray}
\mathrm{Tr}_{B}\left\{ \circled1\right\}  & = & \int_{\omega_{c}-\Delta_{B}}^{\omega_{c}+\Delta_{B}}d\omega\frac{\kappa\left(\omega\right)}{2\pi}\bar{n}_{\mathrm{th}}\left(\omega\right)e^{-i\omega\tau}\left[e^{i\omega_{c}t}\left(a^{\dagger}+\mu\mathcal{S}\right)-\mu\mathcal{S}\right]\tilde{\rho}\left(t\right)\left[e^{-i\omega_{c}\left(t-\tau\right)}\left(a+\mu\mathcal{S}\right)-\mu\mathcal{S}\right]\nonumber \\
 &  & +\int_{\omega_{c}-\Delta_{B}}^{\omega_{c}+\Delta_{B}}d\omega\frac{\kappa\left(\omega\right)}{2\pi}\left[\bar{n}_{\mathrm{th}}\left(\omega\right)+1\right]e^{i\omega\tau}\left[e^{-i\omega_{c}t}\left(a+\mu\mathcal{S}\right)-\mu\mathcal{S}\right]\tilde{\rho}\left(t\right)\left[e^{i\omega_{c}\left(t-\tau\right)}\left(a^{\dagger}+\mu\mathcal{S}\right)-\mu\mathcal{S}\right],\label{eq:dissipator-integrand-first-term}
\end{eqnarray}
\end{widetext}and similar expressions for the remaining terms in
Eq.(\ref{eq:general-QME-microscopic}). In the next step, we perform
the integration over the past, using the relation \cite{carmichael02}
\begin{equation}
\int_{0}^{\infty}d\tau e^{\pm i\left(\omega_{c}-\omega\right)\tau}=\pi\delta\left(\omega_{c}-\omega\right)\pm i\mathbb{P}\frac{1}{\omega_{c}-\omega},
\end{equation}
with $\mathbb{P}$ denoting Cauchy's principal value, perform the
integration over frequency, and within a rotating wave approximation
(which is valid for the realistic parameter regime $\mu\kappa\left(\omega_{c}\right)\bar{n}_{\mathrm{th}}\left(\omega_{c}\right)\sqrt{\bar{n}_{\mathrm{th}}\left(\omega_{c}\right)}\ll\omega_{c}$)
drop all fast oscillating terms $\sim\exp\left[\pm i\omega_{c}t\right]$.
After some simple manipulations, we then arrive at the master equation
\begin{eqnarray}
\dot{\tilde{\rho}} & = & \kappa\left(\omega_{c}\right)\left[\bar{n}_{\mathrm{th}}\left(\omega_{c}\right)+1\right]\mathcal{D}\left[a+\mu\mathcal{S}\right]\tilde{\rho}\nonumber \\
 &  & +\kappa\left(\omega_{c}\right)\bar{n}_{\mathrm{th}}\left(\omega_{c}\right)\mathcal{D}\left[a^{\dagger}+\mu\mathcal{S}\right]\tilde{\rho}\nonumber \\
 &  & -i\Delta_{c}\left[\left(a^{\dagger}+\mu\mathcal{S}\right)\left(a+\mu\mathcal{S}\right),\tilde{\rho}\right]\nonumber \\
 &  & +\gamma\mathcal{D}\left[\mathcal{S}\right]\tilde{\rho}-i\Delta_{S}\left[\mathcal{S}^{2},\tilde{\rho}\right].
\end{eqnarray}
Here, we have introduced the decay rate 
\begin{eqnarray}
\gamma & = & \mu^{2}\int_{\omega_{c}-\Delta_{B}}^{\omega_{c}+\Delta_{B}}d\omega\kappa\left(\omega\right)\left[2\bar{n}_{\mathrm{th}}\left(\omega\right)+1\right]\delta\left(\omega-0\right),\label{eq:decay-rate-gamma}
\end{eqnarray}
which derives from the terms in Eq.(\ref{eq:dissipator-integrand-first-term})
rotating at zero frequency, and the Lamb-like energy shifts 
\begin{eqnarray}
\Delta_{c} & = & \mathbb{P}\int_{\omega_{c}-\Delta_{B}}^{\omega_{c}+\Delta_{B}}d\omega\frac{\kappa\left(\omega\right)}{2\pi}\frac{1}{\omega_{c}-\omega},\\
\Delta_{S} & = & \mu^{2}\mathbb{P}\int_{\omega_{c}-\Delta_{B}}^{\omega_{c}+\Delta_{B}}d\omega\frac{\kappa\left(\omega\right)}{2\pi}\frac{1}{\omega}.
\end{eqnarray}
In accordance with the frequency regime $\left(\omega_{c}\gg\Delta_{B}\gg\tau_{\mathrm{sys}}^{-1}\right)$
discussed above, we assume the bandwidth $\Delta_{B}$ to be large,
but finite. In this case, the rate $\gamma$ vanishes $\left(\gamma=0\right)$,
as the integration range does not cover the $\delta$-peak at $\omega=0$.
Physically, the regime where the lower limit of the relevant frequency
range $\omega_{c}-\Delta_{B}$ does not extend all the way down to
zero frequency amounts to the existence of a lower frequency cut-off
$\omega_{\mathrm{cut}}=\omega_{c}-\Delta_{B}$. For example, such
a lower frequency cut-off $\omega_{\mathrm{cut}}$ naturally arises
in the context of a phonon bath where the existence of $\omega_{\mathrm{cut}}\sim\lambda_{\mathrm{cut}}^{-1}$
is due to finite device dimensions (since a phonon wavelength $\lambda$
larger than the device dimensions is not supported by this structure).
Moreover, phonons with a wavelength much larger than the resonator
are not able to resolve the resonator and simply represent a global
shift of the resonator structure as a whole (and therefore do not
linearly couple to the localized resonator mode). On the contrary,
in the limit of infinite bandwidth $\Delta_{B}\rightarrow\infty$,
the decay rate $\gamma$ (as well as the Lamb-like shifts $\Delta_{c},\Delta_{S}$)
will depend on the relevant reservoir spectral density
\begin{equation}
\kappa\left(\omega\right)/2\pi=g^{2}\left(\omega\right)D_{\mathrm{DOS}}\left(\omega\right),
\end{equation}
often abbreviated as $J\left(\omega\right)=\kappa\left(\omega\right)/2\pi$
in the literature \cite{deVega15}. The spectral density $J\left(\omega\right)=\sum_{k}\left|g_{k}\right|^{2}\delta\left(\omega-\omega_{k}\right)$
encodes the features of the environment relevant for the reduced system
description, and depends on both the environmental density of the
modes $D_{\mathrm{DOS}}\left(\omega\right)$ and on how strongly the
system couples to each mode $\sim g\left(\omega\right)$. For concreteness,
let us discuss two particular examples: (i) First, in quantum optical
systems typically $J\left(\omega\right)\sim\omega^{n}$ for a positive
integer $n$ \cite{yamamoto99,carmichael02}; in particular, for coupling
of a harmonic oscillator to the electromagnetic field in three dimensions
in free space the spectral density scales as $J\left(\omega\right)\sim\omega^{3}$
\cite{paavola08}. In this case, even in the absence of a lower frequency
cut-off $\omega_{\mathrm{cut}}$, the rate $\gamma$ vanishes, because
$\kappa\left(\omega\right)\bar{n}_{\mathrm{th}}\left(\omega\right)\sim\omega^{2}\rightarrow0$
in the limit $\omega\rightarrow0$. (ii) Second, a prominent phenomenological
ansatz frequently used in the literature is the so-called Caldeira-Leggett
model, where $J\left(\omega\right)\sim\omega^{\alpha}\Omega_{\mathrm{cut}}^{1-\alpha}e^{-\omega/\Omega_{\mathrm{cut}}}$
for all $\alpha>0$ and some high-frequency cut-off $\Omega_{\mathrm{cut}}$
\cite{deVega15}. Environments with $0<\alpha<1$ are referred to
as sub-ohmic, while those corresponding to $\alpha=1$ and $\alpha>1$
are called ohmic and super-ohmic, respectively \cite{deVega15}. Within
this Caldeira-Leggett model (and for $\Delta_{B}\rightarrow\infty$),
the decay rate $\gamma$ given in Eq.(\ref{eq:decay-rate-gamma})
vanishes for super-ohmic spectral densities with $\alpha>1$, becomes
a constant for $\alpha=1$ and diverges for $\alpha<1$, since $\bar{n}_{\mathrm{th}}\left(\omega\right)\sim k_{B}T/\omega$
for $k_{B}T\gg\omega$. 

Here, we restrict our analysis to the regime where $\gamma$ vanishes,
either because of the existence of a lower frequency cut-off $\omega_{\mathrm{cut}}>0$
or a spectral density with $J\left(\omega\right)\sim\omega^{\alpha}\left(\alpha>1\right)$,
as discussed above. Moreover, following the standard treatment \cite{beaudoin11,cohen92}
we neglect the Lamb shift $\Delta_{S}\sim\mu^{2}$ (typically, it
is assumed that the Cauchy principal part of an integral of the spectral
density is very small compared to the real part expressions \cite{gardiner00,scala07}),
yielding the master equation 
\begin{eqnarray}
\dot{\tilde{\rho}} & = & \kappa\left(\omega_{c}\right)\left[\bar{n}_{\mathrm{th}}\left(\omega_{c}\right)+1\right]\mathcal{D}\left[a+\mu\mathcal{S}\right]\tilde{\rho}\nonumber \\
 &  & +\kappa\left(\omega_{c}\right)\bar{n}_{\mathrm{th}}\left(\omega_{c}\right)\mathcal{D}\left[a^{\dagger}+\mu\mathcal{S}\right]\tilde{\rho}\nonumber \\
 &  & -i\Delta_{c}\left[\left(a^{\dagger}+\mu\mathcal{S}\right)\left(a+\mu\mathcal{S}\right),\tilde{\rho}\right],
\end{eqnarray}
which (due to the interaction-mediated hybridization of spin and resonator
degrees of freedom $\sim g$) displays correlated decay terms of both
resonator and spin degrees of freedom, that are proportional to the
effective rate $\sim\kappa\left(\omega\right)\bar{n}_{\mathrm{th}}\left(\omega\right)$
evaluated at the (large) characteristic system frequency $\omega_{c}$.
Using the relation
\begin{equation}
e^{-iH_{S}t}\left(a+\mu\mathcal{S}\right)e^{iH_{S}t}=e^{i\omega_{c}t}\left(a+\mu\mathcal{S}\right),
\end{equation}
the corresponding master equation in the Schrödinger picture is found
to be 
\begin{eqnarray}
\dot{\rho} & = & \kappa\left(\omega_{c}\right)\left[\bar{n}_{\mathrm{th}}\left(\omega_{c}\right)+1\right]\mathcal{D}\left[a+\mu\mathcal{S}\right]\rho\nonumber \\
 &  & +\kappa\left(\omega_{c}\right)\bar{n}_{\mathrm{th}}\left(\omega_{c}\right)\mathcal{D}\left[a^{\dagger}+\mu\mathcal{S}\right]\rho\nonumber \\
 &  & -i\left[H_{S},\rho\right]-i\Delta_{c}\left[\left(a^{\dagger}+\mu\mathcal{S}\right)\left(a+\mu\mathcal{S}\right),\rho\right].
\end{eqnarray}
In what follows, we restrict our analysis to the experimentally most
relevant regime of weak spin-resonator coupling where $\mu=g/\omega_{c}\ll1$.
Within the corresponding approximation of independent rates of variation
\cite{cohen92}, the interactions with the environment are
treated separately for spin and resonator degrees of freedom; in other
words, they can approximately treated as independent entities and
the terms (rates of variation) due to internal and dissipative dynamics
are added independently. While for ultra-strong coupling the qubit-resonator
system needs to be treated as a whole when studying its interaction
with the environment \cite{beaudoin11}, yielding irreversible dynamics
through jumps between dressed states (rather than bare states), in
the weak coupling regime we recover standard (quantum optical) dissipators,
i.e., 
\begin{eqnarray}
\dot{\rho} & = & -i\left[H_{S},\rho\right]+\kappa\left[\bar{n}_{\mathrm{th}}+1\right]\mathcal{D}\left[a\right]\rho+\kappa\bar{n}_{\mathrm{th}}\mathcal{D}\left[a^{\dagger}\right]\rho.\label{eq:Master-equation-independent-cavity-decay}
\end{eqnarray}
In the last step, we have set $\kappa\equiv\kappa\left(\omega_{c}\right)$,
$\bar{n}_{\mathrm{th}}\equiv\bar{n}_{\mathrm{th}}\left(\omega_{c}\right)$
and dropped the energy shift $\Delta_{c}$ which may be incorporated
into a renormalized cavity frequency $\omega_{c}\rightarrow\omega_{c}+\Delta_{c}$. 

\begin{table}
\begin{tabular}{|l|c|c|c|c|c|c|}
\hline 
$\kappa/\omega_{c}\bar{n}_{\mathrm{th}}\left[10^{-3}\right]$ & 0 & 0.5 & 1 & 1.5 & 2 & 2.5\tabularnewline
\hline 
\hline 
$\xi_{\kappa}\left[\%\right]$ for uncorrelated noise & 0.0 & 0.21 & 0.41 & 0.61 & 0.81 & 1.01\tabularnewline
\hline 
$\xi_{\kappa}\left[\%\right]$ for correlated noise  & 0.0 & 0.10 & 0.20 & 0.30 & 0.40 & 0.50\tabularnewline
\hline 
\end{tabular}

\caption{\label{tab:error-noise-models}Comparison of the rethermalization-induced
error $\xi_{\kappa}\left[\%\right]$ for two different master equations,
namely Eq.(\ref{eq:Master-equation-independent-cavity-decay}) (uncorrelated
noise model) and $\dot{\rho}=-i\left[H_{S},\rho\right]+\kappa\left[\bar{n}_{\mathrm{th}}+1\right]\mathcal{D}\left[a+\mu\mathcal{S}\right]\rho+\kappa\bar{n}_{\mathrm{th}}\mathcal{D}\left[a^{\dagger}+\mu\mathcal{S}\right]\rho$
(correlated noise model). The error found for the uncorrelated noise
model (as used in the main text) is about twice as large as the one
found for the correlated one, and may therefore be seen as a conservative
estimate. Note that Fig.4(c) of the main text is partially based on
the first row (uncorrelated noise model). Other numerical parameters:
$\mu=g/\omega_{c}=1/16$, $\Gamma=0$, $k_{B}T/\omega_{c}=2$ and
$\omega_{q}=0$. }
\end{table}

Note that the approximate replacement of the correlated dissipators
by uncorrelated ones, that is $\mathcal{D}\left[a+\mu\mathcal{S}\right]\rho\rightarrow\mathcal{D}\left[a\right]\rho$
and $\mathcal{D}\left[a^{\dagger}+\mu\mathcal{S}\right]\rho\rightarrow\mathcal{D}\left[a^{\dagger}\right]\rho$,
gives rise to a conservative error estimate for our hot gate. As can
be shown analytically (compare Appendix \ref{sec:Analytical-Expression-for-Rethermalization-Induced-Errors}),
the rethermalization-induced error $\xi_{\kappa}$ induced by independent
decay terms as given in Eq.(\ref{eq:Master-equation-independent-cavity-decay})
is twice as large as the one due to correlated decay terms. This statement
has also been verified numerically; compare Tab.\ref{tab:error-noise-models}.

While Eq.(\ref{eq:Master-equation-independent-cavity-decay}) is not
rigorous (given the approximations made throughout its derivation),
this type of noise model (with independent rather than correlated
decay terms, and complemented by additional dissipators for the qubits)
has been used widely to describe a great variety of relevant spin-resonator
systems (in the regime of weak spin-resonator coupling for values
up to $\mu=g/\omega_{c}\lesssim4\%$ \cite{blais07}), ranging e.g.
from superconducting qubits \cite{blais04,blais07} as well as quantum
dots coupled to transmission line resonators \cite{gullans15,liu14},
to NV-center spins \cite{rabl09} or carbon nanotubes \cite{wang15}
coupled to nanomechanical oscillators. For example, in Refs.\cite{gullans15,liu14}
very good agreement with experimental results has been achieved for
$\mu\sim1\%$. 

We conclude this discussion with a final remark on low-frequency noise:
As shown above, the existence of a low-frequency cut-off does exclude
low-frequency contributions to resonator-mediated dephasing of the
spins (since $\gamma=0$). Still, low-frequency noise (deriving for
example from ambient nuclear spins \cite{hanson08}) may still couple
directly to the qubits. In our model, this type of noise is captured
by the dephasing rate $\Gamma$, which may, however, be mitigated
efficiently by simple spin-echo techniques.

\section{Additional Numerical Results\label{sec:Additional-Numerical-Results}}

Here, we provide further detailed results based on the numerical simulation
of the master equation given in Eq.(7). Just as in the main text,
for all simulations shown below the initial state of the spin-resonator
system has been chosen as $\rho\left(0\right)=\left|\Uparrow\Downarrow\right\rangle \left\langle \Uparrow\Downarrow\right|\otimes\rho_{\mathrm{th}}\left(T\right)$,
with the cavity mode in the thermal state $\rho_{\mathrm{th}}\left(T\right)=Z^{-1}\exp\left[-\beta\omega_{c}a^{\dagger}a\right]$.
Apart from the state fidelity $\mathcal{F}$,we also quantify the
logarithmic negativity $E_{\mathcal{N}}$ (which ranges between $0$
for separable states to at maximum $1$ for two maximally-entangled
qubits) in order to quantify the entanglement between the two qubits.

\begin{figure}
\includegraphics[width=0.9\columnwidth]{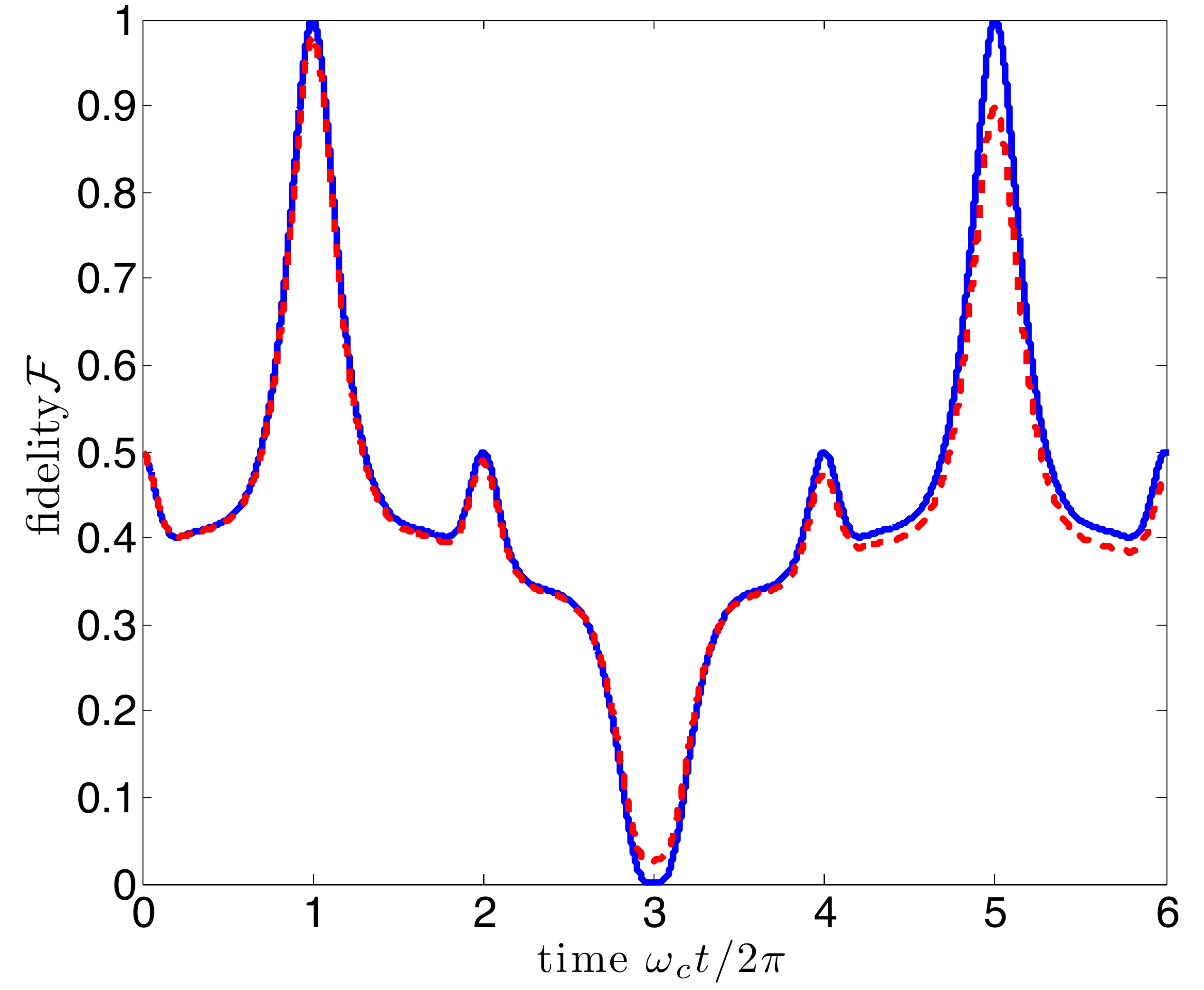}

\caption{\label{fig:fidelity-long-time}(color online). Fidelity $\mathcal{F}$
for the two-qubit state $\rho_{\mathrm{qubits}}$ with the target
state $\left|\Psi_{\mathrm{tar}}\right\rangle =\left(\left|\Uparrow\Downarrow\right\rangle +i\left|\Downarrow\Uparrow\right\rangle \right)/\sqrt{2}$
for $\Gamma/\omega_{c}=0$ (blue solid line) and $\Gamma/\omega_{c}=1\%$
(red dashed line). For sufficiently low noise, at $\omega_{c}t=2\pi$
and $\omega_{c}t=5\times2\pi$ the fidelity with the maximally entangled
state $\left|\Psi_{\mathrm{tar}}\right\rangle $ reaches the maximal
value $\mathcal{F}=1$. Numerical parameters: $\omega_{q}/\omega_{c}=0$,
$k_{B}T/\omega_{c}=2$ $\left(\bar{n}_{\mathrm{th}}\approx1.54\right)$,
$g/\omega_{c}=1/4$, $\kappa/\omega_{c}=Q^{-1}=10^{-5}$. }
\end{figure}

\textit{Periodic recurrence}s.---First, as displayed in Fig.\ref{fig:fidelity-long-time},
we observe periodic recurrences of the maximally-entangling dynamics:
For example, for $g/\omega_{c}=1/4$ (as used in Fig.\ref{fig:fidelity-long-time}),
ideally---apart from $\mathcal{F}=1$ at $(\omega_{c}/2\pi)t=1$---we
find $\mathcal{F}=1$ again at $(\omega_{c}/2\pi)t=5$, since $U_{\mathrm{id}}^{x}\left(m=5,1/4\right)=\exp\left[i\pi\sigma_{1}^{x}\sigma_{2}^{x}\right]U_{\mathrm{id}}^{x}\left(1,1/4\right)=-U_{\mathrm{id}}^{x}\left(1,1/4\right)$.
This statement holds provided that dephasing is negligible on the
relevant timescale; compare the dashed curve in Fig.\ref{fig:fidelity-long-time}
which accounts for dephasing of the qubits. 

\begin{figure}
\includegraphics[width=0.9\columnwidth]{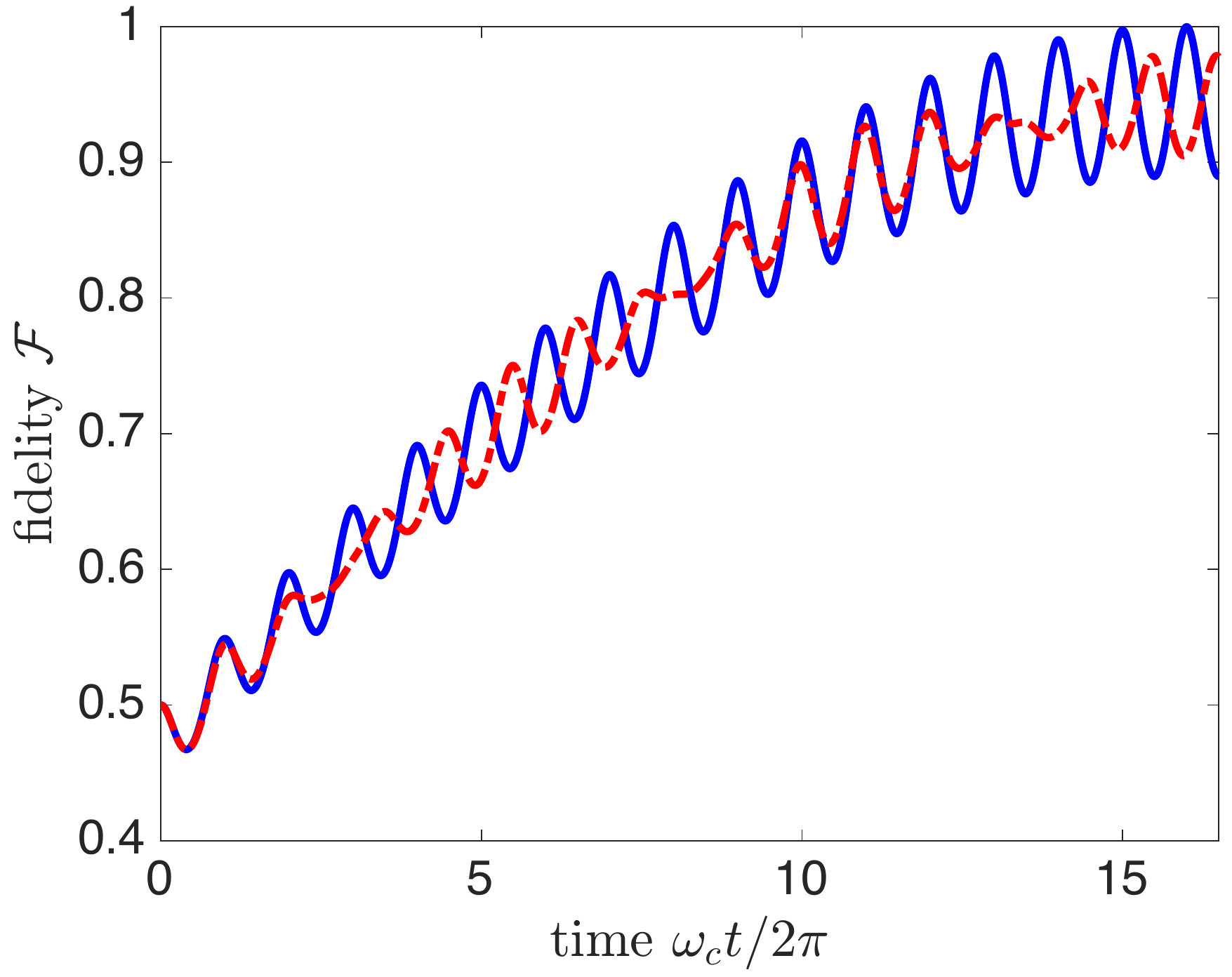}

\caption{\label{fig:fidelity-non-zero-splitting}(color online). Fidelity $\mathcal{F}=\left<\Psi_{\mathrm{tar}}|\rho_{\mathrm{qubits}}|\Psi_{\mathrm{tar}}\right>$
for the two-qubit state $\rho_{\mathrm{qubits}}=\mathrm{Tr}_{\mathrm{cav}}\left[\rho\right]$
with the target state $\left|\Psi_{\mathrm{tar}}\right\rangle =\left(\left|\Uparrow\Downarrow\right\rangle +i\left|\Downarrow\Uparrow\right\rangle \right)/\sqrt{2}$
for both $\omega_{q}/\omega_{c}=0$ (solid blue line) and $\omega_{q}/\omega_{c}=0.1$
(dashed red line); here, $g/\omega_{c}=1/16<0.1$. Other numerical
parameters: $k_{B}T/\omega_{c}=2$ $(\bar{n}_{\mathrm{th}}\approx1.54)$,
$Q=10^{5}$ and $\Gamma/\omega_{c}=0$.}
\end{figure}

\begin{figure}
\includegraphics[width=0.9\columnwidth]{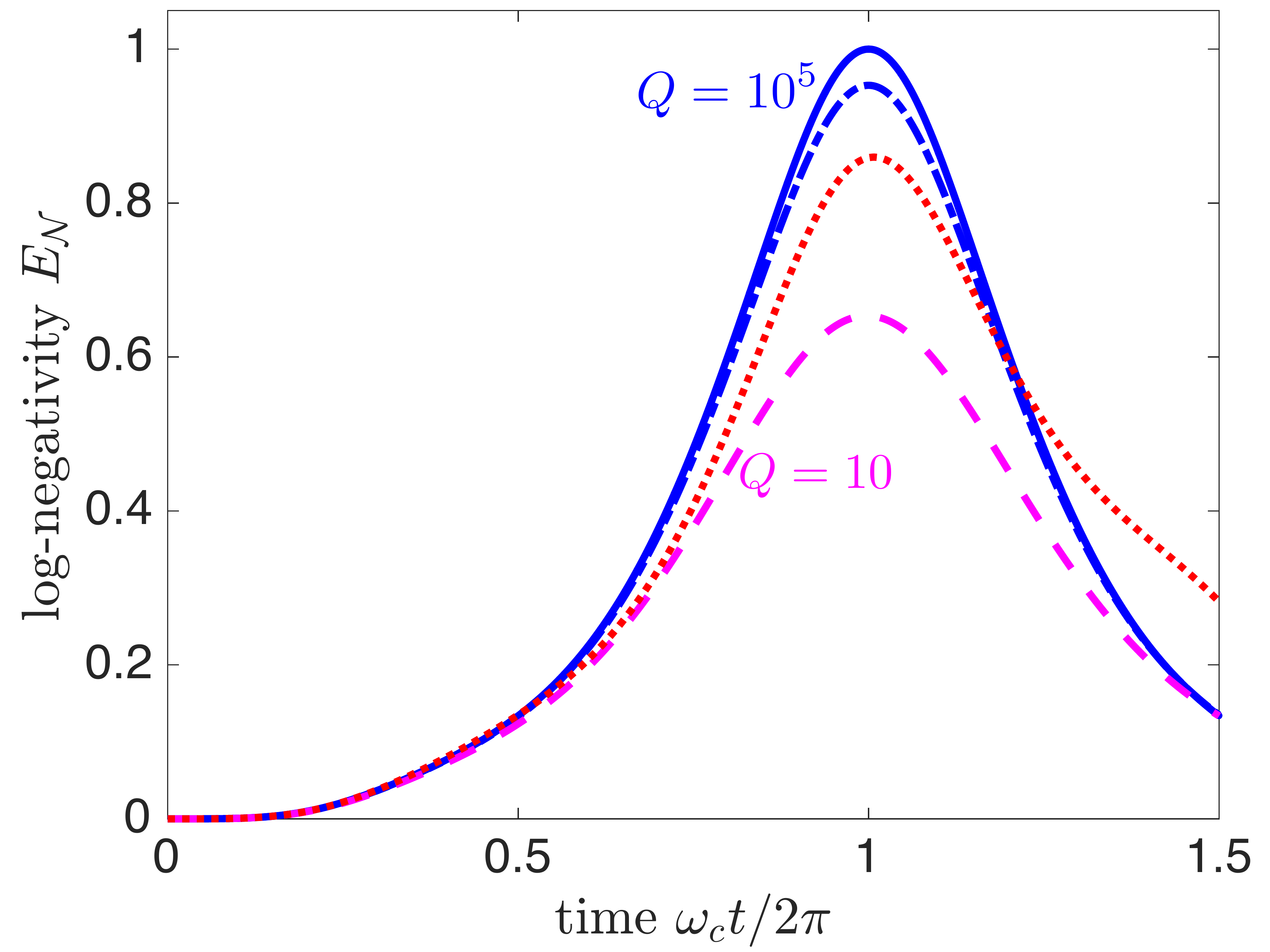}

\caption{\label{fig:log-neg-Q-factor}(color online). Logarithmic negativity
$E_{\mathcal{N}}$ for $k_{B}T/\omega_{c}=1$ and different cavity
quality factors: $Q=10^{5}$ (solid blue), $Q=10^{2}$ (dash-dotted
blue), and $Q=10$ (dashed magenta). A clear reduction of the maximum
entanglement is observed, if the quality factor $Q$ is too low to
satisfy the hot-gate requirement given in Eq.(8). Here, we have $g/\omega_{c}\times g/k_{B}T=1/16=6.25\times10^{-2}$.
The red (dotted) curve refers to $Q=10^{2}$ and $\omega_{q}/\omega_{c}=0.2$.
Other numerical parameters: $g/\omega_{c}=1/4$ and $\Gamma/\omega_{c}=0$.}
\end{figure}

\textit{Non-zero level splitting}.---While our analytical treatment
has assumed $\omega_{q}=0$, in Fig.\ref{fig:fidelity-non-zero-splitting}
we provide exemplary numerical results that explicitly account for
a non-zero qubit level splitting $\omega_{q}>0$, showing that the
proposed protocol can tolerate non-zero level splittings of the qubits
$\omega_{q}/\omega_{c}\lesssim0.1$, without a severe reduction in
the fidelity of the protocol. Again, this numerical finding is corroborated
in Fig.\ref{fig:log-neg-Q-factor}. Here, it is shown explicitly that
a strong entanglement reduction is observed once condition (8) is
violated. Conversely, within the range of parameter values satisfying
Eq.(8), the results are rather insensitive to the particular parameter
values. 

\begin{figure}
\includegraphics[width=0.9\columnwidth]{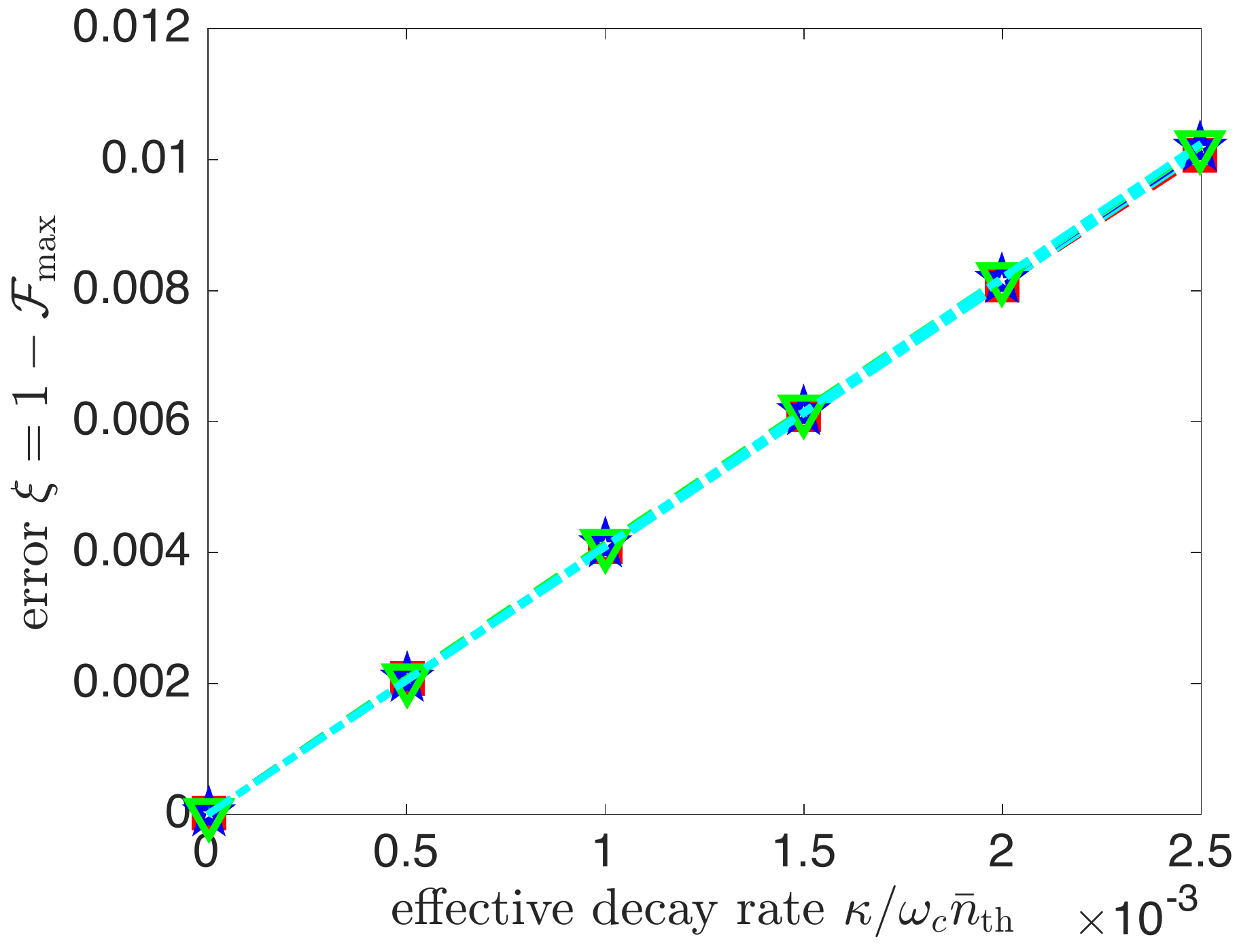}

\caption{\label{fig:error-ODE-small-regime}(color online). Error as a function
of the effective rethermalization rate $\kappa\bar{n}_{\mathrm{th}}$
for $g/\omega_{c}=1/16$ (red squares), $g/\omega_{c}=1/\left(8\sqrt{2}\right)$
(blue stars) and $g/\omega_{c}=1/8$ (green triangles) and $k_{B}T/\omega_{c}=2$
$\left(\bar{n}_{\mathrm{th}}\approx1.54\right)$, within the relevant
small-error regime ($\kappa_{\mathrm{eff}}/g_{\mathrm{eff}}\ll1$).
The dash-dotted lines in cyan refer to linear fits, demonstrating
a linear error scaling in the small error-regime ($\kappa_{\mathrm{eff}}/g_{\mathrm{eff}}\ll1$),
which is independent of $\mu=g/\omega_{c}$. Accordingly, the error
is larger for higher temperatures, but all temperature related effects
are approximately captured by the thermal occupation number $\bar{n}_{\mathrm{th}}$.
Other numerical parameters: $\Gamma=0$ and $\omega_{q}=0$.}
\end{figure}

\textit{Rethermalization-induced errors}.---As illustrated in Fig.\ref{fig:error-ODE-small-regime},
we have numerically checked that (for small infidelities) the rethermalization
induced error $\xi_{\kappa}$ scales linearly with the effective rethermalization
rate $\kappa_{\mathrm{eff}}=\kappa\bar{n}_{\mathrm{th}}$. Notably,
as evidenced in Fig.\ref{fig:error-ODE-small-regime}, the error is
found to be independent of the spin-resonator coupling $g$. As demonstrated
in in Sec. \ref{sec:Analytical-Expression-for-Rethermalization-Induced-Errors},
this numerical result can be corroborated analytically within a perturbative
framework.

\begin{figure}
\includegraphics[width=0.9\columnwidth]{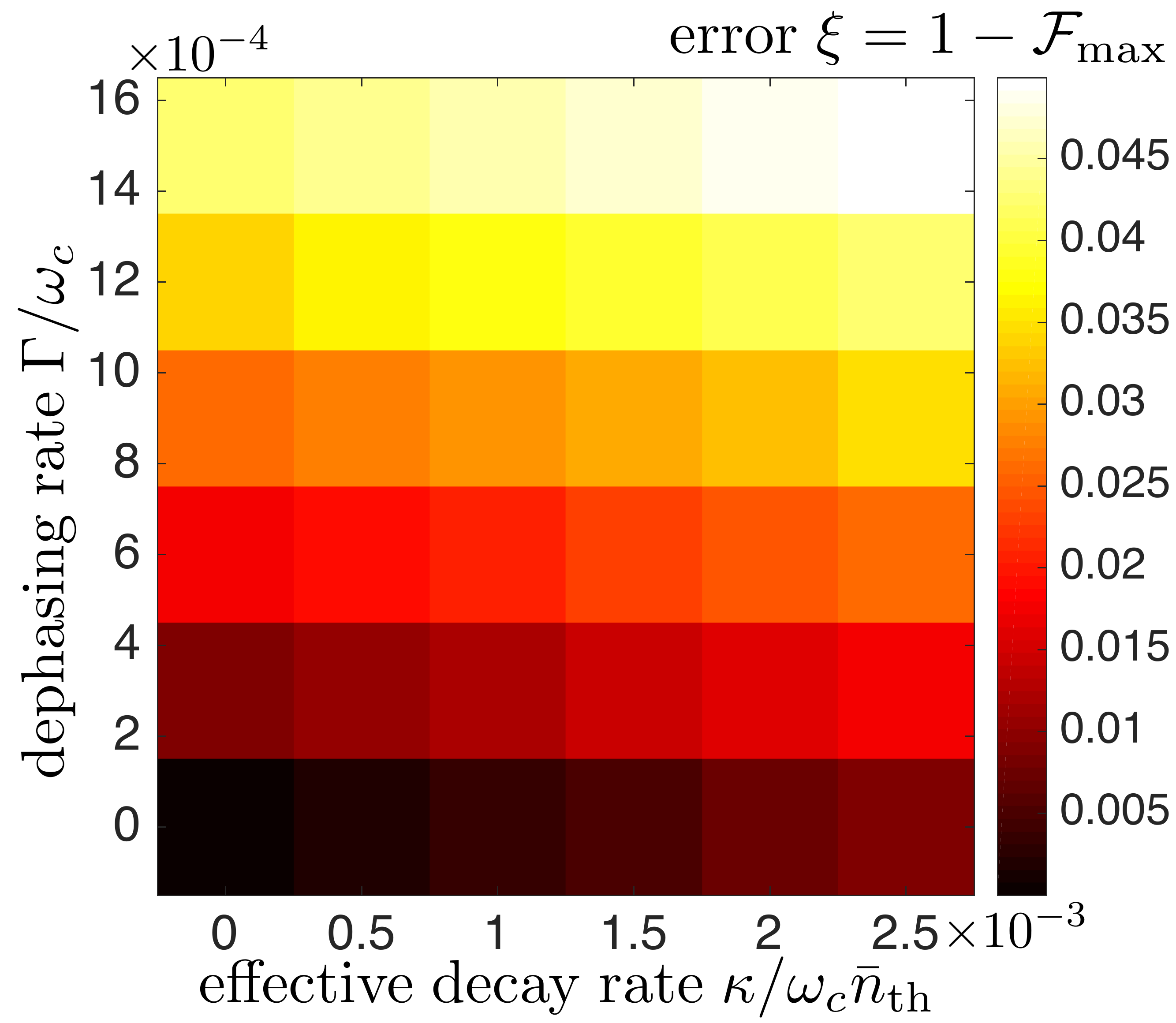}

\caption{\label{fig:total-error-T4}(color online). Total error $\xi$ as a
function of both the effective rethermalization rate $\sim\kappa/\omega_{c}\bar{n}_{\mathrm{th}}\sim\bar{n}_{\mathrm{th}}/Q$
and the spin dephasing rate $\sim\Gamma/\omega_{c}$ for $g/\omega_{c}=1/16$,
$k_{B}T/\omega_{c}=4$ and $\omega_{q}=0$.}
\end{figure}

\begin{figure*}
\includegraphics[width=2\columnwidth]{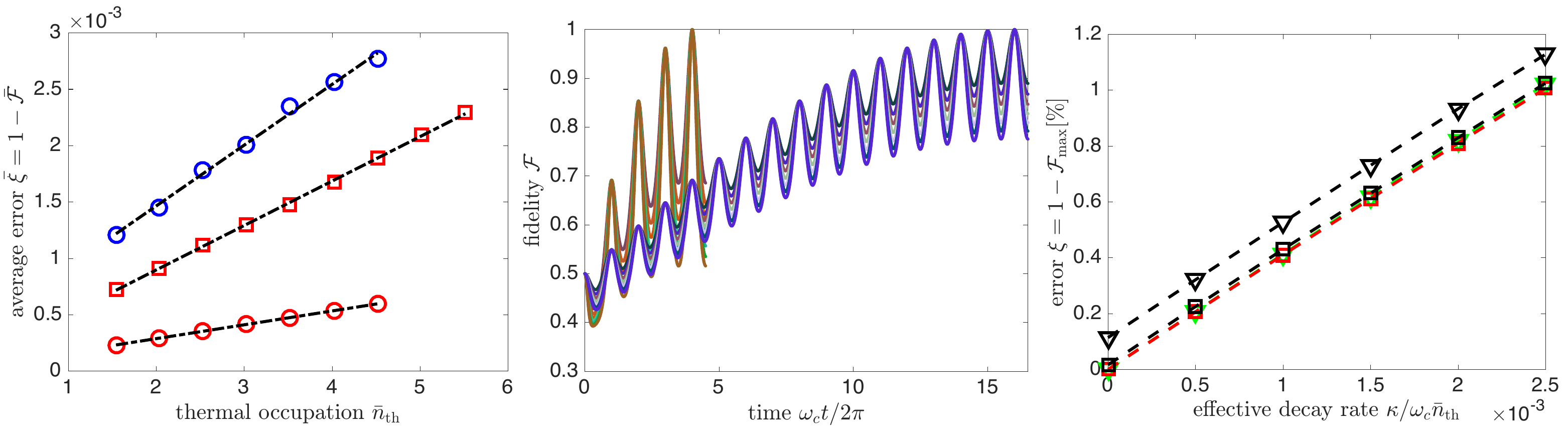}

\caption{\label{fig:average-error}(color online). Timing errors. Left: Total
average error $\bar{\xi}$ as a function of the thermal occupation
number $\bar{n}_{\mathrm{th}}$ for timing windows $\left(\omega_{c}/2\pi\right)\Delta t=5\%$
(circles) and $\left(\omega_{c}/2\pi\right)\Delta t=10\%$ (squares);
here, $g/\omega_{c}=1/16$ (red symbols) and $g/\omega_{c}=1/8$ (blue
symbols, upper curve). All curves can be fit very well to linear error
models (see black dashed lines). Center: Set of underlying (temperature-dependent)
simulations for both $g/\omega_{c}=1/16$ (terminating at $\omega_{c}t/2\pi=16.5$)
and $g/\omega_{c}=1/8$ (terminating at $\omega_{c}t/2\pi=4.5$).
Note that larger amplitudes are observed for larger values of $\mu=g/\omega_{c}$.
Other numerical parameters: $Q=10^{5}$, $\Gamma=0$ and $\omega_{q}=0$.
Right: Same analysis as done in Fig.\ref{fig:error-ODE-small-regime}
for $g/\omega_{c}=1/8$ (triangles) and $g/\omega_{c}=1/16$ (squares).
The black curves account for a finite timing accuracy $\left(\omega_{c}/2\pi\right)\Delta t=5\%$,
showing that the detrimental effects of time jitter are less pronounced
for smaller values of $\mu=g/\omega_{c}$.}
\end{figure*}

\textit{Full error analysis}.---Similar to Fig.4(c) in the main text,
in Fig.\ref{fig:total-error-T4} we provide numerical results that
fully account for higher-order, correlated errors (beyond the linear
error approximation). Here, we have chosen a temperature $k_{B}T/\omega_{c}=4$,
a factor two larger than the one used in Fig.4(c) in the main text.
Still, if the rethermalization induced error is scaled in terms of
the effective decay rate $\kappa_{\mathrm{eff}}=\kappa\bar{n}_{\mathrm{th}}$,
we obtain (approximately) the same total error $\xi$, independently
of the temperature $k_{B}T$, showing that the effective decay rate
$\kappa_{\mathrm{eff}}=\kappa\bar{n}_{\mathrm{th}}$ captures well
any temperature-related effects. This is evidenced numerically in
Fig.\ref{fig:total-error-T4} which approximately coincides with the
results displayed in Fig.4(c) in the main text and is line with our
simple error estimate for rethermalization induced errors; compare
Eq.(10) in the main text. 

\textit{Timing errors}.---Finally, we consider errors (infidelities)
due to limited timing accuracies. To do so, we take the average fidelity
of our protocol $\bar{\mathcal{F}}$ within a certain timing window
$\Delta t$ centered around the stroboscopic time $t_{\mathrm{max}}$
for which maximum fidelity (minimal infidelity) is achieved; for example,
in quantum dot systems timing accuracies $\Delta t$ of a few picoseconds
have been demonstrated experimentally \cite{bocquillon13-sm}. For
$g/2\pi=10\mathrm{MHz}$ and $\omega_{c}/2\pi=160\mathrm{MHz}$ (that
is, $\mu=g/\omega_{c}=1/16$) as used in the main text, the pulse
time lies in the microsecond regime $\left(t_{\mathrm{max}}=\pi/8g_{\mathrm{eff}}\approx0.2\mu\mathrm{s}\right)$,
for which $\Delta t\approx1\mathrm{ps}$ is feasible; for this relatively
long pulse, the relative time jitter is well below the percent level,
i.e., $\left(\omega_{c}/2\pi\right)\Delta t\approx10^{-4}$. Based
on our numerical simulations, we make the following observations:
(i) As demonstrated in Fig.\ref{fig:average-error}, we find an average
error scaling linearly with $\sim\bar{n}_{\mathrm{th}}$, that is
$\bar{\xi}=1-\bar{\mathcal{F}}\sim\bar{n}_{\mathrm{th}}$. (ii) More
precisely, the error expressions given in the main text can be generalized
to 
\begin{equation}
\bar{\xi}=\bar{\alpha}_{\text{\ensuremath{\kappa}}}\frac{\kappa}{\omega_{c}}\bar{n}_{\mathrm{th}}+\bar{\alpha}_{\Gamma}\frac{\Gamma}{\omega_{c}}+\bar{\beta}_{\kappa}+\bar{\beta}_{\Gamma}.
\end{equation}
Here, the unit-less quantities $\bar{\alpha}_{\gamma},\bar{\beta}_{\gamma}$
for $\gamma=\kappa,\Gamma$ depend on the timing window $\Delta t$.
For example, for $g/\omega_{c}=1/16$ and $\left(\omega_{c}/2\pi\right)\Delta t=5\%$,
we then extract $\bar{\alpha}_{\text{\ensuremath{\kappa}}}\approx4.03$,
$\bar{\beta}_{\kappa}\approx2.2\times10^{-4}$, $\bar{\alpha}_{\Gamma}\approx24.22$
and $\bar{\beta}_{\Gamma}\approx5.1\times10^{-4}$. (iii) As shown
in Fig.\ref{fig:average-error}, for the experimentally most relevant
regime where $\left(\omega_{c}/2\pi\right)\Delta t\ll1$ (such that
the timing window covers a small range of the oscillations only),
this error is found to decrease for a smaller spin-resonator coupling
strength $g/\omega_{c}$, because larger values of $g/\omega_{c}$
imply larger oscillation amplitudes within the relevant range over
which we have to average; compare the center and right plots in Fig.\ref{fig:average-error}.
Therefore, for the experimentally most relevant regime where $\left(\omega_{c}/2\pi\right)\Delta t\ll1$
and $g/\omega_{c}\lesssim1/16$, the effects of time jitter should
be negligible.

\section{Analytical Expression for Rethermalization-Induced Errors \label{sec:Analytical-Expression-for-Rethermalization-Induced-Errors}}

In this Appendix we derive an analytical expression for rethermalization-induced
errors. In particular we show that this expression is independent
of the spin-resonator coupling strength $g$. 

Our analysis starts out from the master equation 
\begin{equation}
\dot{\rho}=-i\left[H,\rho\right]+\sum_{j=1,2}\mathcal{D}\left[L_{j}\right]\rho,
\end{equation}
where the Hamiltonian $H=\omega_{c}a^{\dagger}a+g\mathcal{S}\otimes\left(a+a^{\dagger}\right)$
refers to the ideal (noise-free) dynamics and the jump-operators $L_{1}=\sqrt{\kappa_{1}}a$,
$L_{2}=\sqrt{\kappa_{2}}a^{\dagger}$ with $\kappa_{1}=\kappa\left(\bar{n}_{\mathrm{th}}+1\right)$
and $\kappa_{2}=\kappa\bar{n}_{\mathrm{th}}$ describe rethermalization
of the resonator mode with a rate $\kappa=\omega_{c}/Q$ that is enhanced
by the thermal occupation number $\bar{n}_{\mathrm{th}}$. It is convenient
to move to an interaction picture, defined by $\tilde{\rho}\left(t\right)=\exp\left[iHt\right]\rho\left(t\right)\exp\left[-iHt\right]$.
In this interaction picture, the system's dynamics is described by
\begin{equation}
\dot{\tilde{\rho}}=\sum_{j=1,2}\mathcal{D}\left[\tilde{L}_{j}\right]\tilde{\rho},\label{eq:Master-equation-kappa-noise-interaction-pic}
\end{equation}
with time-dependent jump operators $\tilde{L}_{j}=\exp\left[iHt\right]L_{j}\exp\left[-iHt\right]$.
Using the exact relation $\exp\left[-iHt\right]=U\exp\left[-i\omega_{c}ta^{\dagger}a\right]U^{\dagger}U_{\mathrm{sp}}\left(t\right)$,
with the polaron transformation $U=\exp\left[\mu\mathcal{S}\left(a-a^{\dagger}\right)\right]$
and the pure spin (entangling) gate $U_{\mathrm{sp}}\left(t\right)=\exp\left[i\mu^{2}\omega_{c}t\mathcal{S}^{2}\right]$,
the time-dependent jump operators $\tilde{L}_{j}$ take on a simple
form 
\begin{eqnarray}
\tilde{L}_{1}\left(\tau\right) & = & \sqrt{\kappa_{1}}\left[e^{-i\omega_{c}\tau}a+\left(e^{-i\omega_{c}\tau}-1\right)\mu\mathcal{S}\right],\nonumber \\
\tilde{L}_{2}\left(\tau\right) & = & \sqrt{\kappa_{2}}\left[e^{i\omega_{c}\tau}a^{\dagger}+\left(e^{i\omega_{c}\tau}-1\right)\mu\mathcal{S}\right].\label{eq:time-dependent-jump-operators}
\end{eqnarray}
The formal solution to Eq.(\ref{eq:Master-equation-kappa-noise-interaction-pic})
reads 
\begin{equation}
\tilde{\rho}\left(t\right)=\tilde{\rho}\left(0\right)+\sum_{j}\int_{0}^{t}d\tau\mathcal{D}\left[\tilde{L}_{j}\left(\tau\right)\right]\tilde{\rho}\left(\tau\right),\label{eq:Master-equation-kappa-noise-interaction-pic-formal-solution}
\end{equation}
where in the interaction picture the zeroth-order solution $\tilde{\rho}_{0}\left(t\right)=\tilde{\rho}\left(0\right)=\rho\left(0\right)$
stays inert, and accounts for the ideal (noise-free) dynamics only
in the lab frame, $\rho_{0}\left(t\right)=\exp\left[-iHt\right]\tilde{\rho}_{0}\left(t\right)\exp\left[iHt\right]=\exp\left[-iHt\right]\rho\left(0\right)\exp\left[iHt\right]$.
To obtain the first-order correction $\tilde{\rho}_{1}\left(t\right)$
within a perturbative framework, we re-insert the zeroth-order solution
into the dissipator of Eq.(\ref{eq:Master-equation-kappa-noise-interaction-pic-formal-solution}),
i.e. effectively we take $\tilde{\rho}\left(\tau\right)\rightarrow\rho\left(0\right)$,
which yields $\tilde{\rho}\left(t\right)\approx\rho\left(0\right)+\tilde{\rho}_{1}\left(t\right)$,
with
\begin{equation}
\tilde{\rho}_{1}\left(t\right)=\sum_{j}\int_{0}^{t}d\tau\mathcal{D}\left[\tilde{L}_{j}\left(\tau\right)\right]\rho\left(0\right).\label{eq:rho-first-order-correction-formal}
\end{equation}
Inserting the expressions given in Eq.(\ref{eq:time-dependent-jump-operators})
into Eq.(\ref{eq:rho-first-order-correction-formal}) and performing
the integration, with $\int_{0}^{t}d\tau\left|1-e^{\pm i\omega_{c}\tau}\right|^{2}=2(t-\frac{\sin\left(\omega_{c}t\right)}{\omega_{c}})$
and $\int_{0}^{t}d\tau\left(1-e^{\pm i\omega_{c}\tau}\right)=t\pm i\frac{e^{\pm i\omega_{c}t}-1}{\omega_{c}}$
, one arrives at \begin{widetext} 
\begin{eqnarray}
\tilde{\rho}_{1}\left(t\right) & = & \kappa_{1}t\mathcal{D}\left[a\right]\rho\left(0\right)+\kappa_{2}t\mathcal{D}\left[a^{\dagger}\right]\rho\left(0\right)+2\left(\kappa_{1}+\kappa_{2}\right)\mu^{2}\left(t-\frac{\sin\left(\omega_{c}t\right)}{\omega_{c}}\right)\mathcal{D}\left[\mathcal{S}\right]\rho\left(0\right)\nonumber \\
 &  & +\left[\kappa_{1}\mu\left(t-i\frac{e^{-i\omega_{c}t}-1}{\omega_{c}}\right)\left\{ a\rho\left(0\right)\mathcal{S}-\frac{1}{2}\left\{ a\mathcal{S},\rho\left(0\right)\right\} \right\} +\mathrm{h.c.}\right]\nonumber \\
 &  & +\left[\kappa_{2}\mu\left(t+i\frac{e^{i\omega_{c}t}-1}{\omega_{c}}\right)\left\{ a^{\dagger}\rho\left(0\right)\mathcal{S}-\frac{1}{2}\left\{ a^{\dagger}\mathcal{S},\rho\left(0\right)\right\} \right\} +\mathrm{h.c.}\right].\label{eq:rho-first-order-correction-explicit}
\end{eqnarray}
which, for stroboscopic times $t_{m}=2\pi m/\omega_{c}$ (with $m$
integer), simplifies to 
\begin{eqnarray*}
\tilde{\rho}_{1}\left(t_{m}\right) & = & \kappa_{1}t_{m}\mathcal{D}\left[a\right]\rho\left(0\right)+\kappa_{2}t_{m}\mathcal{D}\left[a^{\dagger}\right]\rho\left(0\right)+2\left(\kappa_{1}+\kappa_{2}\right)\mu^{2}t_{m}\mathcal{D}\left[\mathcal{S}\right]\rho\left(0\right)\\
 &  & +\left[\kappa_{1}\mu t_{m}\left\{ a\rho\left(0\right)\mathcal{S}-\frac{1}{2}\left\{ a\mathcal{S},\rho\left(0\right)\right\} \right\} +\mathrm{h.c.}\right]+\left[\kappa_{2}\mu t_{m}\left\{ a^{\dagger}\rho\left(0\right)\mathcal{S}-\frac{1}{2}\left\{ a^{\dagger}\mathcal{S},\rho\left(0\right)\right\} \right\} +\mathrm{h.c.}\right].
\end{eqnarray*}
\end{widetext}Next, we perform a transformation back to the lab frame,
with $\rho\left(t\right)=\exp\left[-iHt\right]\tilde{\rho}\left(t\right)\exp\left[iHt\right]$.
As discussed in the main text, for stroboscopic times the ideal evolution
simplifies to $\exp\left[-iHt_{m}\right]=\exp\left[i\mu^{2}2\pi m\mathcal{S}^{2}\right]=\exp\left(-i\phi_{\mathrm{gp}}\right)\exp\left[i4\pi m\mu^{2}\sigma_{1}^{x}\sigma_{2}^{x}\right]$.
The ideal (noise-free) evolution is given by $\rho_{\mathrm{id}}\left(t_{m}\right)=\exp\left[-iHt_{m}\right]\rho\left(0\right)\exp\left[iHt_{m}\right]=\varrho_{\mathrm{id}}\left(t_{m}\right)\otimes\rho_{\mathrm{th}}$,
where $\varrho_{\mathrm{id}}\left(t_{m}\right)=\exp\left[i4\pi m\mu^{2}\sigma_{1}^{x}\sigma_{2}^{x}\right]\varrho\left(0\right)\exp\left[-i4\pi m\mu^{2}\sigma_{1}^{x}\sigma_{2}^{x}\right]$
is the ideal qubit's state at time $t_{m}$, starting from the initial
state $\rho\left(0\right)=\varrho\left(0\right)\otimes\rho_{\mathrm{th}}$.
Then, the system's density matrix at time $t_{m}$ is approximately
given by
\begin{eqnarray*}
\rho\left(t_{m}\right) & = & \rho_{\mathrm{id}}\left(t_{m}\right)+\kappa_{1}t_{m}\mathcal{D}\left[a\right]\rho_{\mathrm{id}}\left(t_{m}\right)\\
 &  & +\kappa_{2}t_{m}\mathcal{D}\left[a^{\dagger}\right]\rho_{\mathrm{id}}\left(t_{m}\right)\\
 &  & +2\left(\kappa_{1}+\kappa_{2}\right)\mu^{2}t_{m}\mathcal{D}\left[\mathcal{S}\right]\rho_{\mathrm{id}}\left(t_{m}\right)\\
 &  & +\left[\kappa_{1}\mu t_{m}\left\{ a\rho_{\mathrm{id}}\left(t_{m}\right)\mathcal{S}-\frac{1}{2}\left\{ a\mathcal{S},\rho_{\mathrm{id}}\left(t_{m}\right)\right\} \right\} \right.\\
 &  & \left.+\kappa_{2}\mu t_{m}\left\{ a^{\dagger}\rho\left(0\right)\mathcal{S}-\frac{1}{2}\left\{ a^{\dagger}\mathcal{S},\rho\left(0\right)\right\} \right\} +\mathrm{h.c.}\right].
\end{eqnarray*}
Note that, in the limit $\kappa_{i}\rightarrow0$, one retrieves the
ideal result $\rho\left(t_{m}\right)=\rho_{\mathrm{id}}\left(t_{m}\right)$.
Next, we trace out the resonator mode. Assuming the state of the resonator
mode to be diagonal in the occupation number basis (in particular,
this holds for a thermal state $\rho_{\mathrm{th}}$), none of the
cross-terms contribute to the partial trace, and for stroboscopic
times $t_{m}$ the state of the qubits is given by 
\begin{equation}
\varrho\left(t_{m}\right)=\varrho_{\mathrm{id}}\left(t_{m}\right)+2\kappa\left(2\bar{n}_{\mathrm{th}}+1\right)t_{m}\mu^{2}\mathcal{D}\left[\mathcal{S}\right]\varrho_{\mathrm{id}}\left(t_{m}\right).\label{eq:qubit-density-matrix-kappa-noise-strobo-times}
\end{equation}
As expected naïvely, the error term scales with $\sim\kappa\bar{n}_{\mathrm{th}}t_{m}$,
but it is further reduced by the factor $\mu^{2}=\left(g/\omega_{c}\right)^{2}$.
Eq.(\ref{eq:qubit-density-matrix-kappa-noise-strobo-times}) holds
for stroboscopic times $t_{m}=2\pi m/\omega_{c}$, with $m$ integer.
If $m\mu^{2}=1/16$, the ideal evolution $\exp\left[-iHt_{m}\right]=\exp\left(-i\phi_{\mathrm{gp}}\right)\exp\left[i\frac{\pi}{4}\sigma_{1}^{x}\sigma_{2}^{x}\right]$
equals a maximally-entangling gate, which (for an initial pure state
like $\left|\Psi\right\rangle _{0}=\left|\Downarrow\Downarrow\right\rangle $)
yields the desired ideal qubit target state $\left|\Psi_{\mathrm{tar}}\right\rangle =\exp\left[i\frac{\pi}{4}\sigma_{1}^{x}\sigma_{2}^{x}\right]\left|\Psi\right\rangle _{0}$.
Then, in the presence of noise, at the nominally ideal time $t_{\mathrm{max}}=\pi/8\mu^{2}\omega_{c}=\pi/8g_{\mathrm{eff}}$
the qubit's density matrix reads 
\begin{eqnarray}
\varrho\left(t_{\mathrm{max}}\right) & = & \left|\Psi_{\mathrm{tar}}\right\rangle \left\langle \Psi_{\mathrm{tar}}\right|\\
 &  & +\frac{\pi}{4}\frac{\kappa}{\omega_{c}}\left(2\bar{n}_{\mathrm{th}}+1\right)\mathcal{D}\left[\mathcal{S}\right]\left|\Psi_{\mathrm{tar}}\right\rangle \left\langle \Psi_{\mathrm{tar}}\right|.\nonumber 
\end{eqnarray}
Therefore, to first order rethermalization-induced noise leads to
dephasing dynamics in the eigenbasis of $\mathcal{S}$ with a single
such phase flip. Since neither the desired target state $\left|\Psi_{\mathrm{tar}}\right\rangle $
nor the initial state $\left|\Psi\right\rangle _{0}$ is an eigenstate
of $\mathcal{S}$, the system looses fidelity with a probability $\frac{\pi}{4}\frac{\kappa}{\omega_{c}}\left(2\bar{n}_{\mathrm{th}}+1\right)$;
notably, this expression is independent of the spin-resonator coupling
strength $g$. 

\begin{figure}
\includegraphics[width=1\columnwidth]{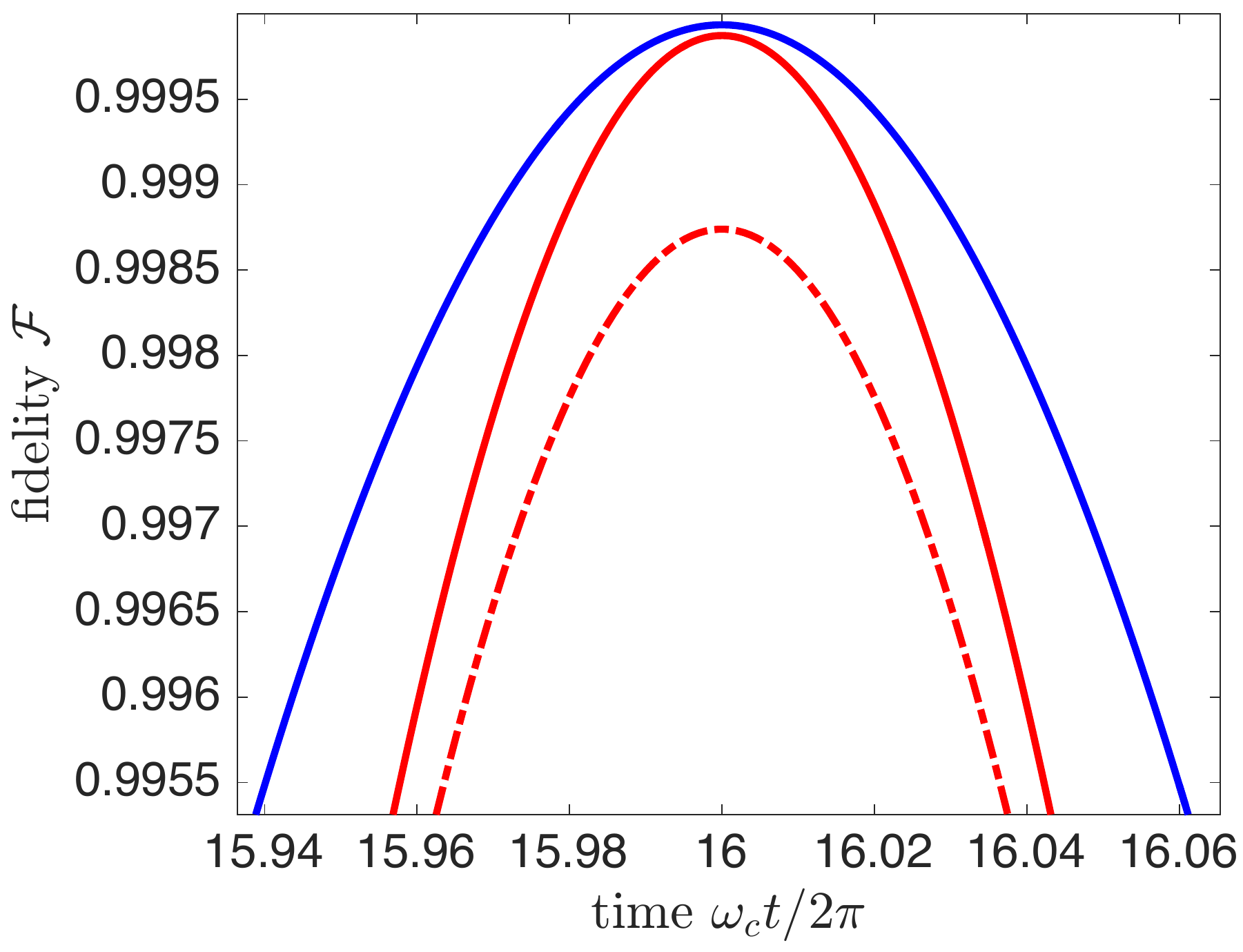}

\caption{\label{fig:fidelity-zoom-in}(color online). Fidelity $\mathcal{F}$
close to the ideal time $t_{\mathrm{max}}$ for $g/\omega_{c}=1/16$.
The different curves refer to $Q=10^{5}$, $k_{B}T/\omega_{c}=2$,
i.e. $\bar{n}_{\mathrm{th}}\approx1.54$, (blue solid, top curve),
$Q=10^{5}$, $k_{B}T/\omega_{c}=4$, i.e. $\bar{n}_{\mathrm{th}}\approx3.52$,
(red solid) and $Q=10^{4}$, $k_{B}T/\omega_{c}=4$ (red dash-dotted).
The error $\xi=1-\mathcal{F}$ can be estimated well with the formula
$\xi_{k}\approx4\bar{n}_{\mathrm{th}}/Q$, giving (for example) $\mathcal{F}\approx1-4\times3.52/10^{4}\approx0.9986$.
Other numerical parameters: $\Gamma=0$ and $\omega_{q}=0$.}
\end{figure}

For the fidelity with the maximally entangled target state, we then
obtain 
\begin{equation}
\mathcal{F}=\left<\Psi_{\mathrm{tar}}|\varrho\left(t_{\mathrm{max}}\right)|\Psi_{\mathrm{tar}}\right>=1-\frac{\pi}{2}\frac{\kappa}{\omega_{c}}\left(2\bar{n}_{\mathrm{th}}+1\right),
\end{equation}
with a thermalization-induced error term given by 
\begin{equation}
\xi_{\kappa}=\pi\left(\kappa/\omega_{c}\right)\bar{n}_{\mathrm{th}}+\frac{\pi}{2}Q^{-1}.\label{eq:thermalization-induced-error}
\end{equation}
This analytical result is in good agreement with our numerical findings
(from which we have deduced $\xi_{\kappa}\approx\alpha_{\kappa}\left(\kappa/\omega_{c}\right)\bar{n}_{\mathrm{th}}$,
with $\alpha_{\kappa}\approx4$), showing (i) a linear scaling with
the effective rethermalization rate $\sim\kappa\bar{n}_{\mathrm{th}}$,
(ii) with a pre-factor $\alpha_{\kappa}=\pi$ (close to $\sim4$)
that is \textit{independent} of the spin-resonator coupling strength
$g$ and (iii) a constant offset $\sim Q^{-1}$ which is negligible
for realistic quality factors $Q\approx10^{5}-10^{6}$. The latter
is due to photon/phonon emission with a rate $\sim\kappa=\omega_{c}/Q$
at $T\rightarrow0$. As illustrated further in Fig.\ref{fig:fidelity-zoom-in}
with a close-up of the fidelity $\mathcal{F}\left(t\right)$ around
the optimal point $t_{\mathrm{max}}$, the error $\xi_{\kappa}$ can
be estimated well with this simple formula, where all temperature
related effects are captured by the simple linear expression in the
thermal occupation number $\bar{n}_{\mathrm{th}}$. 

\textit{Correlated noise model}.---An analog analysis to the one presented
above can be performed for a master equation with correlated (rather
than uncorrelated) noise. In this case, as shown in Appendix \ref{sub:Microscopic-Derivation-QME},
the jump-operators are given by $L_{1}=\sqrt{\kappa_{1}}\left(a+\mu\mathcal{S}\right)$,
$L_{2}=\sqrt{\kappa_{2}}\left(a^{\dagger}+\mu\mathcal{S}\right)$,
which take on a simple form in the interaction picture, namely 
\begin{eqnarray}
\tilde{L}_{1}\left(\tau\right) & = & \sqrt{\kappa_{1}}e^{-i\omega_{c}\tau}\left(a+\mu\mathcal{S}\right),\\
\tilde{L}_{2}\left(\tau\right) & = & \sqrt{\kappa_{2}}e^{i\omega_{c}\tau}\left(a^{\dagger}+\mu\mathcal{S}\right),
\end{eqnarray}
as compared to Eq.(\ref{eq:time-dependent-jump-operators}) within
the uncorrelated noise model discussed above. Then, following the
same steps as above, the integration $\int_{0}^{t}d\tau\left|1-e^{\pm i\omega_{c}\tau}\right|^{2}=2(t-\frac{\sin\left(\omega_{c}t\right)}{\omega_{c}})$
is simply replaced by $\int_{0}^{t}d\tau=t$; accordingly, in this
scenario, the pre-factor of the spin dephasing term $\mathcal{D}\left[\mathcal{S}\right]\rho\left(0\right)$
simplifies to $\sim\left(\kappa_{1}+\kappa_{2}\right)\mu^{2}t$, which
for stroboscopic times $t_{m}=2\pi m/\omega_{c}$ is exactly a factor
of two smaller than the corresponding rate in Eq.(\ref{eq:rho-first-order-correction-explicit})
for the uncorrelated noise model. In summary, along the lines of our
previous analysis, for a correlated noise model Eqs.(\ref{eq:qubit-density-matrix-kappa-noise-strobo-times})
and (\ref{eq:thermalization-induced-error}) should be replaced by
\begin{equation}
\varrho\left(t_{m}\right)=\varrho_{\mathrm{id}}\left(t_{m}\right)+\kappa\left(2\bar{n}_{\mathrm{th}}+1\right)t_{m}\mu^{2}\mathcal{D}\left[\mathcal{S}\right]\varrho_{\mathrm{id}}\left(t_{m}\right),
\end{equation}
 and 
\begin{equation}
\xi_{\kappa}=\frac{\pi}{2}\left(\kappa/\omega_{c}\right)\bar{n}_{\mathrm{th}}+\frac{\pi}{4}Q^{-1},
\end{equation}
respectively, showing that for uncorrelated spin-resonator noise the
rethermalization-induced error is approximately twice as large as
for correlated spin-resonator noise; also compare the numerical results
presented in Tab. \ref{tab:error-noise-models}.

\section{Analytical Model for Dephasing-Induced Errors \label{sec:Analytical-Model-for-dephasing-errors}}

In this Appendix we provide an analytical model for dephasing-induced
errors. Neglecting rethermalization-induced errors for the moment,
here we consider the following master equation 
\begin{equation}
\dot{\rho}=\underset{\mathcal{L}_{0}\rho}{\underbrace{-i\left[H_{\mathrm{id}},\rho\right]}}+\underset{\mathcal{L}_{1}\rho}{\underbrace{\gamma_{\phi}\left[\mathcal{D}\left[\sigma_{1}^{z}\right]\rho+\mathcal{D}\left[\sigma_{2}^{z}\right]\rho\right]}},\label{eq:Master-equation-pure-dephasing-analytical-1}
\end{equation}
where $H_{\mathrm{id}}=\omega_{c}a^{\dagger}a+g\left(\sigma_{1}^{z}+\sigma_{2}^{z}\right)\otimes\left(a+a^{\dagger}\right)$
describes the ideal (error-free), coherent evolution for longitudinal
coupling between the qubits and the resonator mode, and $\gamma_{\phi}$
is the pure dephasing rate. Since the superoperators $\mathcal{L}_{0}$
and $\mathcal{L}_{1}$ as defined in Eq.(\ref{eq:Master-equation-pure-dephasing-analytical-1})
commute, that is $\left[\mathcal{L}_{0},\mathcal{L}_{1}\right]=0$
(since $\left[H_{\mathrm{id}},\mathcal{D}\left[\sigma_{i}^{z}\right]X\right]=\mathcal{D}\left[\sigma_{i}^{z}\right]\left[H_{\mathrm{id}},X\right]$
for any operator $X$), the full evolution simplifies to 
\begin{equation}
\rho\left(t\right)=e^{\mathcal{L}_{1}t}e^{\mathcal{L}_{0}t}\rho\left(0\right)=e^{\mathcal{L}_{1}t}\rho_{\mathrm{id}}\left(t\right),
\end{equation}
where we have defined the ideal target state at time $t$ as $\rho_{\mathrm{id}}\left(t\right)=\exp\left[\mathcal{L}_{0}t\right]\rho\left(0\right)$,
which, starting from the initial state $\rho\left(0\right)$, exclusively
accounts for the ideal (error-free), coherent evolution. For small
infidelities $\left(\gamma_{\phi}t\ll1\right)$, the deviation from
the ideal dynamics $\Delta\rho=\rho-\rho_{\mathrm{id}}$ is approximately
given by 
\begin{equation}
\Delta\rho\left(t\right)\approx\gamma_{\phi}t\sum_{i}\mathcal{D}\left[\sigma_{i}^{z}\right]\rho_{\mathrm{id}}\left(t\right),\label{eq:linear-dephasing-error-analytical}
\end{equation}
showing that (in the regime of interest where $\gamma_{\phi}t\ll1$)
the dominant dephasing induced errors are linearly proportional to
$\sim\gamma_{\phi}t_{g}\sim\gamma_{\phi}/g_{\mathrm{eff}}=\gamma_{\phi}/\mu^{2}\omega_{c}$,
as expected; here, $t_{g}\sim g_{\mathrm{eff}}$ is the relevant gate
time which has to be short compared to $\gamma_{\phi}^{-1}$. 

In what follows, for completeness we derive the same result within
a quantum jump approach. Eq.(\ref{eq:Master-equation-pure-dephasing-analytical-1})
can be rewritten as 
\begin{eqnarray}
\dot{\rho} & = & -iH\rho+i\rho H^{\dagger}+\mathcal{J}\rho,\label{eq:Master-equation-pure-dephasing-analytical-2}
\end{eqnarray}
where $H=H_{\mathrm{id}}-i\gamma_{\phi}$ and $\mathcal{J}\rho=\gamma_{\phi}\sum_{i}\sigma_{i}^{z}\rho\sigma_{i}^{z}$.
The formal solution to Eq.(\ref{eq:Master-equation-pure-dephasing-analytical-2})
reads 
\begin{equation}
\rho\left(t\right)=e^{-iHt}\rho\left(0\right)e^{iH^{\dagger}t}+\int_{0}^{t}d\tau e^{-iH\left(t-\tau\right)}\mathcal{J}\rho\left(\tau\right)e^{iH^{\dagger}\left(t-\tau\right)}.\label{eq:Master-equation-pure-dephasing-formal-solution}
\end{equation}
Defining the ideal target state at time $t$ as 
\begin{equation}
\rho_{\mathrm{id}}\left(t\right)=e^{-iH_{\mathrm{id}}\left(t-\tau\right)}\rho\left(\tau\right)e^{iH_{\mathrm{id}}\left(t-\tau\right)},
\end{equation}
the exact solution given in Eq.(\ref{eq:Master-equation-pure-dephasing-formal-solution})
can be iterated, giving an illustrative expansion in terms of the
jumps $\mathcal{J}$. It reads 
\begin{eqnarray*}
\rho\left(t\right) & = & \mathcal{U}\left(t\right)\rho\left(0\right)+\int_{0}^{t}d\tau_{1}\mathcal{U}\left(t-\tau_{1}\right)\mathcal{J}\mathcal{U}\left(\tau_{1}\right)\rho\left(0\right)\\
 &  & +\int_{0}^{t}d\tau_{2}\int_{0}^{\tau_{2}}d\tau_{1}\mathcal{U}\left(t-\tau_{2}\right)\mathcal{J}\mathcal{U}\left(\tau_{2}-\tau_{1}\right)\times\\
 &  & \mathcal{J}\mathcal{U}\left(\tau_{1}\right)\rho\left(0\right)+\dots
\end{eqnarray*}
Here, the $n$-th order term comprises $n$ jumps $\mathcal{J}$ with
free evolution $\mathcal{U}\left(t\right)\rho=e^{-iHt}\rho e^{iH^{\dagger}t}$
between the jumps. Up to second order in $\mathcal{J}$ we then find
\begin{eqnarray}
\rho\left(t\right) & = & \mathcal{U}\left(t\right)\rho\left(0\right)+e^{-2\gamma_{\phi}t}\gamma_{\phi}t\sum_{i}\sigma_{i}^{z}\rho_{\mathrm{id}}\left(t\right)\sigma_{i}^{z}\\
 &  & +\frac{1}{2}e^{-2\gamma_{\phi}t}\gamma_{\phi}^{2}t^{2}\sum_{i,j}\sigma_{i}^{z}\sigma_{j}^{z}\rho_{\mathrm{id}}\left(t\right)\sigma_{j}^{z}\sigma_{i}^{z}+\dots\nonumber 
\end{eqnarray}
For the regime of interest where $\gamma_{\phi}t\ll1$, we then obtain
again the result given in Eq.(\ref{eq:linear-dephasing-error-analytical}),
where the dominant error term scales linearly with $\sim\gamma_{\phi}t$.

\section{Relaxation-Induced Errors \label{sec:Relaxation-Induced-Errors}}

In this Appendix we address in detail errors induced by relaxation
processes, typically characterized by the timescale $T_{1}$. First,
we discuss typical relaxation timescales for different physical platforms,
with particular emphasis on their dependence on both temperature $T$
and qubit-level splitting $\omega_{q}$. We conclude that inter-level
scattering processes typically play a minor role as compared to pure
dephasing induced errors, even in our regime of interest with elevated
temperatures of a few Kelvin and small qubit level splittings. Second,
for completeness, we numerically verify the expected linear error
scaling $\sim T_{1}^{-1}$ and---using the fundamental relation $T_{2}^{-1}=1/2T_{1}+1/T_{\phi}$
\cite{kornich14-sm}, with $T_{2}^{-1}(T_{\phi}^{-1})$ referring
to the decoherence (pure-dephasing) rate---give an upper bound on
decoherence-induced errors.

\subsection{Experimental Relaxation Timescales}

Let us first discuss spin qubits in quantum dots where decoherence
predominantly results from spin-orbit interaction and hyperfine interaction
with nuclear spins \cite{hanson07-sm,kloeffel13}. Thereafter we discuss
yet another candidate system for the implementation of the proposed
hot gate, consisting of nitrogen-vacancy centers coupled to the vibrational
mode of a diamond mechanical nano-resonator via strain \cite{bennett13,barfuss15,teissier14}. 

\textit{(i) Single-electron spin qubits}.---For single-electron spins
in GaAs quantum dots the inter Zeeman level spin scattering is typically
dominated by spin-orbit interaction in combination with the emission
of single piezoelectric phonons, while other relaxation processes
are usually negligible \cite{hanson07-sm,kloeffel13}. At low temperatures,
the corresponding phonon-mediated spin relaxation rate $\gamma_{1}$
shows a well-known, pronounced dependence on magnetic field $B$,
namely 
\begin{equation}
\gamma_{1}=T_{1}^{-1}=A\left(g_{s}\mu_{B}B\right)^{5}/\omega_{0}^{4},
\end{equation}
where $A$ is a material-specific constant reflecting the effectiveness
of the spin-phonon coupling strength, $\omega_{q}=g_{s}\mu_{B}B$
is the Zeeman splitting (with the $g$-factor $g_{s}$ and Bohr magneton
$\mu_{B}$) and $\omega_{0}$ refers to the quantum dot single-particle
level spacing; compare Refs.\cite{kloeffel13,hanson07-sm} and references
therein. As usual, for elevated temperatures $k_{B}T\geq\omega_{q}$
this relaxation rate is enhanced by a (bosonic) thermal occupation
factor $\bar{n}_{\mathrm{th}}\left(\omega_{q}\right)\approx k_{B}T/\omega_{q}$
(describing stimulated emission of phonons), yielding a linear scaling
with temperature, that is an effective relaxation rate $\gamma_{1}\sim\omega_{q}^{4}\times k_{B}T$
for temperatures much larger than the Zeeman splitting ($k_{B}T\gg\omega_{q}$)
\cite{kroutvar04}. Both, the strong dependence on the magnetic field
$B$ and the linear dependence on temperature $\sim T$ have been
confirmed experimentally \cite{kroutvar04,hanson07-sm}, showing extremely
long relaxation times of $T_{1}>1\mathrm{s}$ at $B=1\mathrm{T}$
and $T=120\mbox{\ensuremath{\mathrm{mK}}}$ \cite{amasha08}, and
$T_{1}>20\mathrm{ms}$ at $B=4\mathrm{T}$ and $T=1\mathrm{K}$ \cite{kroutvar04}.
For very small magnetic fields $B$, this expression for $T_{1}$
diverges $\left(\gamma_{1}\rightarrow0\right)$, because it accounts
for single-phonon processes only (with single phonons in resonance
with the Zeeman energy $\omega_{q}$, as required by energy conservation)
and Kramer's theorem does not allow for spin-orbit-induced spin relaxation
in the absence of a magnetic field \cite{kloeffel13,hanson07-sm}.
When accounting for two-phonon processes, however, $T_{1}$ does converge
to a finite value \cite{kloeffel13}. As shown theoretically in Refs.\cite{khaetskii01-sm,woods02-sm},
the corresponding two-phonon spin flip rate becomes the dominating
(phonon-mediated) scattering mechanism for sufficiently small magnetic
fields $\lesssim0.4\mathrm{T}$, with a corresponding two-phonon mediated
scattering rate of $\sim1\mathrm{kHz}$ ($T_{1}\sim1\mathrm{ms}$)
for $T\approx4\mathrm{K}$ in GaAs, reaching very long relaxation
times of $T_{1}\sim1\mathrm{s}$ for $T\approx1\mathrm{K}$ and sufficiently
small magnetic fields of $B\lesssim0.1\mathrm{T}$. Similarly, experiments
on the relaxation rate from the two-electron triplet to singlet states
as a function of the singlet-triplet energy splitting $\Delta E_{\mathrm{ST}}$
(referred to as $\omega_{q}$ in our analysis) show relaxation times
well below $1\mathrm{ms}$ as $\Delta E_{\mathrm{ST}}$ approaches
zero \cite{meunier07}, due to a vanishing phonon density of states;
compare Fig.21 in Ref.\cite{hanson07-sm}. Finally, near zero magnetic
field ($\omega_{q}=0$), in GaAs energy relaxation is known to be
dominated by direct hyperfine-mediated electron-nuclear flip-flops
\cite{hanson07-sm}. For a (relatively small) magnetic field $B\gg B_{n}\approx3\mathrm{mT}$
(with $B_{n}$ denoting the effective nuclear magnetic field caused
by ambient nuclear spins), however, this mechanism is suppressed efficiently
by the mismatch between nuclear and electron Zeeman energies \cite{amasha08},
effectively leaving the hyperfine interaction as the well-known, dominating
pure-dephasing mechanism for the electron spin qubit \cite{hanson07-sm}.
Therefore, as soon as the qubit level splitting $\omega_{q}=g_{s}\mu_{B}B$
exceeds the typical hyperfine energy-scale in GaAs $g_{\mathrm{hf}}/2\pi\approx25\mathrm{MHz}$,
one reaches a regime, where $T_{1}$ processes can be neglected compared
to pure-dephasing $\sim T_{2}^{\star}$ (even at temperatures of a
few Kelvin), while easily satisfying the inequality $g_{\mathrm{hf}}\ll\omega_{q}\ll\omega_{c}$
for typical resonator frequencies $\omega_{c}/2\pi\sim\mathrm{GHz}$,
as required for the implementation of the proposed hot gate. The prospects
for a faithful implementation of the proposed hot gate are potentially
even more promising when switching to materials such as Si and Ge
where both hyperfine interactions with the ambient nuclei (since these
materials can be grown nuclear-spin free) and piezoelectric electron-phonon
coupling (due to bulk inversion symmetry) are absent \cite{kornich14-sm,kloeffel13};
note that the latter typically dominates spin relaxation in GaAs-based
systems \cite{woods02-sm,khaetskii01-sm,hanson07-sm}. In fact, silicon-based
experiments have demonstrated $T_{1}\sim3\mathrm{s}$ at $B=1.85\mathrm{T}$
and $T=0.15\mathrm{K}$ \cite{simmons11-sm}, suggesting (according
to the usual thermal enhancement) $T_{1}\sim0.3\mathrm{s}$ for $T\approx1\mathrm{K}$,
which is still much longer than the spin-dephasing timescale $T_{2}^{\star}\sim100\mu\mathrm{s}$
quoted in the main text and agrees with the common wisdom that spin
lifetimes are orders of magnitude longer than the ones reported for
GaAs \cite{kornich14-sm,prance12-sm}; compare our subsequent discussion
on singlet-triplet qubits. 

\textit{(ii) Singlet-triplet spin qubits}.---For singlet-triplet qubits
in silicon relaxation times of $T_{1}\sim10\mathrm{ms}$ have been
demonstrated at zero magnetic field for cryostat temperatures $T\sim15\mathrm{mK}$
\cite{prance12-sm}, which exceeds the $B=0$ lifetimes measured in
comparable GaAs setups by about two orders of magnitude. As discussed
in detail in Appendix \ref{sub:Double-Quantum-Dot}, in this system
the qubit splitting $\omega_{q}$ is set by the well-controlled exchange
splitting $J$, which can be tuned to very small values. For example,
in Ref.\cite{prance12-sm} $\omega_{q}/2\pi\approx16\mathrm{MHz}$,
which is much smaller than any relevant resonator frequency $\omega_{c}$.
As argued in Ref.\cite{prance12-sm}, the measured lifetimes of $T_{1}\sim10\mathrm{ms}$
(at $B=0$) are limited by the (small) hyperfine interaction in natural
(i.e., not purified) silicon with $g_{\mathrm{hf}}\sim3\mathrm{neV}$.
Since the effective relaxation rate at elevated temperatures is determined
by integrated auto-correlation functions of the bath operators (yielding
for example the thermal enhancement factor $\bar{n}_{\mathrm{th}}\left(\omega_{q}\right)\approx k_{B}T/\omega_{q}$
when coupling to a bosonic bath, as discussed above), very long lifetimes
of $T_{1}\sim10\mathrm{ms}$ (at $B=0$) can still be expected, even
at higher temperatures $T\sim\mathrm{K}$, because the autocorrelation
functions of the relevant nuclear spin bath operators do not show
a bosonic thermal enhancement factor; conversely, due to their extremely
small magnetic moment, nuclear spins can be treated as an infinite
temperature bath, even at ultra-low temperatures $\sim100\mathrm{mK}$
and strong magnetic fields \cite{christ08-sm}. Therefore, singlet-triplet
qubits in silicon should be well suited for the implementation of
the proposed hot gate, with tunable qubits splittings much smaller
than relevant resonator frequencies ($\omega_{q}\ll\omega_{c}$) and
relaxation times $T_{1}$ much longer than $T_{2}^{\star}$, even
at elevated temperatures of a few Kelvin. 

\textit{(iii) NV-centers}.---Since for nitrogen-vacancy (NV) centers
in diamond the spin $T_{1}$ time can be several seconds or longer
\cite{ams=0000FCss11,harrison06,jarmola12}, even at temperatures
of a few Kelvin, it is common practice to neglect the spin decay;
compare for example Ref.\cite{bennett13}, which may serve as a potential
platform for a proof-of-principle implementation of the proposed hot
gate. The electronic ground state of the negatively charged NV center
is a spin $S=1$ triplet with spin states $\left|m_{s}=0,\pm1\right\rangle $,
where the levels $\left|\pm1\right\rangle $ are split off from $\left|0\right\rangle $
by the zero-field splitting $D/2\pi=2.88\mathrm{GHz}$. In the absence
of an external magnetic field the states $\left|\pm1\right\rangle $
are degenerate. As discussed in detail in Refs.\cite{bennett13,barfuss15,teissier14},
such a electronic spin can be coupled to the motion of a mechanical
resonator through lattice strain, with perpendicular strain mixing
the $\left|\pm1\right\rangle $ states, which is otherwise a dipole-forbidden
transition $\left(\Delta m_{s}=2\right)$ \cite{bennett13,barfuss15,teissier14}.
If the system is prepared in the $\left|\pm1\right\rangle $ subspace,
the state $\left|0\right\rangle $ remains unpopulated and the effect
of parallel strain plays no role \cite{bennett13}, yielding an effective
qubit with qubit splitting $\omega_{q}=2\gamma_{\mathrm{NV}}B$ (with
$\gamma_{\mathrm{NV}}/2\pi=2.8\mathrm{MHz/G}$), that is coupled to
the mechanical resonator mode of frequency $\omega_{c}\gg\omega_{q}$.
Then, in the absence of an external magnetic field ($\omega_{q}=0$),
the effective Hamiltonian $H_{\mathrm{eff}}$ for this spin-resonator
system takes on the desired form, that is $H_{\mathrm{eff}}=\omega_{c}a^{\dagger}a-g_{\bot}\sigma^{x}\otimes\left(a+a^{\dagger}\right)$,
where $\sigma^{x}=\left|+1\right\rangle \left\langle -1\right|+\mathrm{h.c.}$
and $g_{\bot}$ is the transverse single-phonon strain-coupling strength
\cite{barfuss15}. At first sight, in this setup the spin-resonator
coupling $g_{\bot}$ is static and not easily tunable; hence, while
it does not provide an universal two-qubit primitive, it can nevertheless
be used to generate entanglement at elevated temperatures. The spin-resonator
coupling may, however, effectively be switched on and off by making
use of the hyperfine coupling to adjacent single nuclear spins where
quantum information can be stored with qubit memory lifetimes exceeding
one second \cite{maurer12}.

\subsection{Error Scaling}

\begin{figure}
\includegraphics[width=0.9\columnwidth]{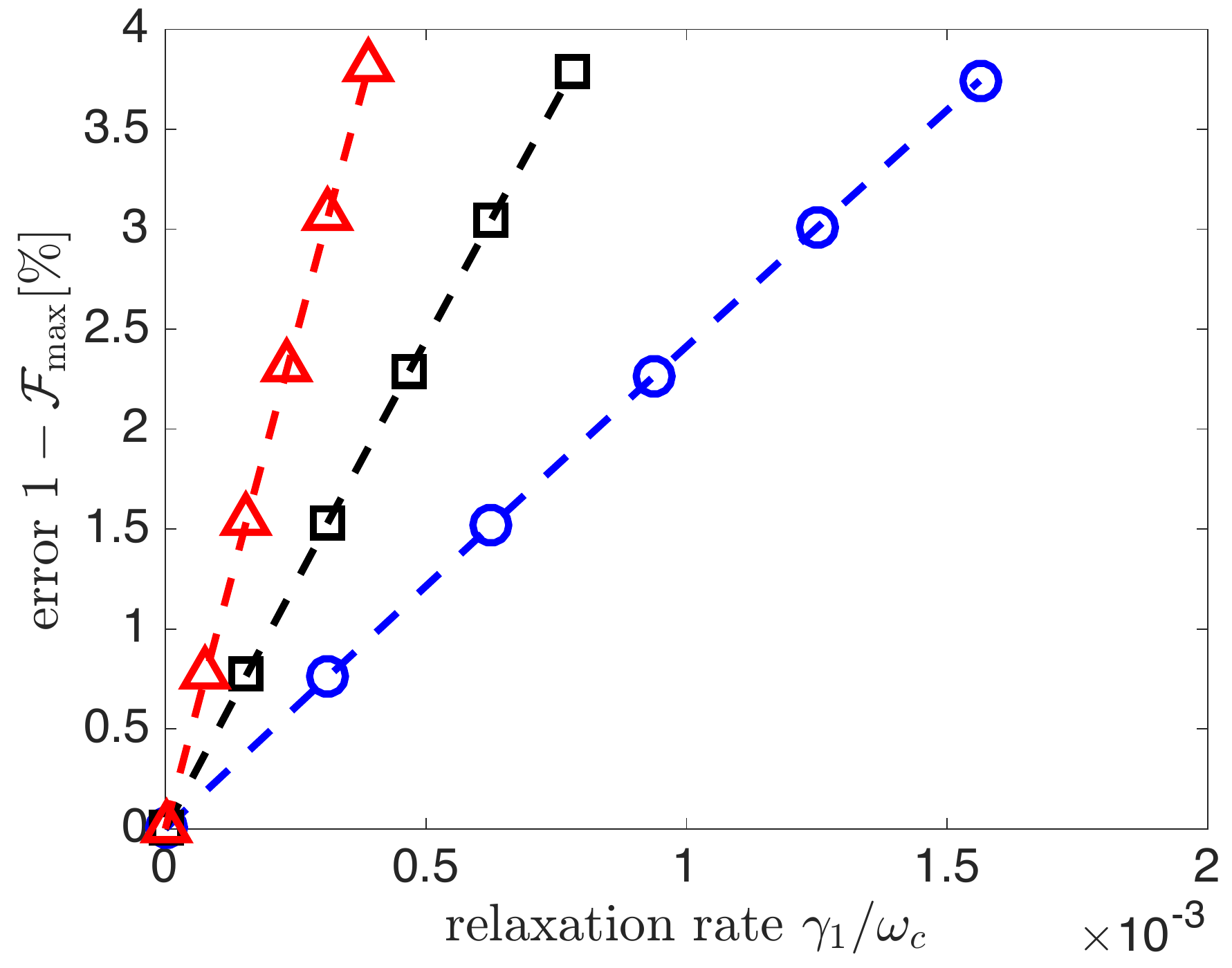}

\caption{\label{fig:relaxation-error}(color online). Relaxation-induced error
$\xi_{\gamma}$ for $g/\omega_{c}=1/8$ (blue circles), $g/\omega_{c}=1/\left(8\sqrt{2}\right)$
(black squares) and $g/\omega_{c}=1/16$ (red diamonds). Other numerical
parameters: $k_{B}T/\omega_{c}=0.01$, $\kappa=0$, $\Gamma=0$ and
$\omega_{q}=0$.}
\end{figure}

To quantitatively capture the effect of relaxation-induced errors,
we have analyzed the master equation
\begin{equation}
\dot{\rho}=-i\left[H,\rho\right]+\gamma_{1}\sum_{i}\mathcal{D}\left[\sigma_{i}^{-}\right]\rho,\label{eq:QME-relaxation-errors}
\end{equation}
where the first term refers to the ideal, coherent dynamics and the
second term describes single-spin relaxation with a rate $\gamma_{1}=T_{1}^{-1}$;
incoherent excitation processes could be included as well, with additional
terms of the same form with the appropriate replacement $\sigma_{i}^{-}\rightarrow\sigma_{i}^{+}$,
but are omitted here for clarity. Along the lines of our analysis
for dephasing-induced errors, the relaxation-induced error is expected
to scale linearly with the relaxation rate as $\xi_{\gamma}\sim\gamma_{1}/g_{\mathrm{eff}}$,
that is 
\begin{equation}
\xi_{\gamma}\approx\alpha_{\gamma}\frac{\gamma_{1}}{\omega_{c}},
\end{equation}
with the pre-factor $\alpha_{\gamma}=c_{\gamma}/\mu^{2}$, where $\mu=g/\omega_{c}$.
As shown in Fig.\ref{fig:relaxation-error}, based on numerical simulations
of Eq.(\ref{eq:QME-relaxation-errors}), this linear error scaling
has been verified numerically, yielding the numerical pre-factor $c_{\gamma}\approx0.38$,
that is $\alpha_{\gamma}\approx0.38/\mu^{2}$. This numerical pre-factor
coincides very well with the value obtained for the dephasing-induced
error $\sim\Gamma$ (when properly accounting for the factor of four
in our definition $\Gamma=2/T_{2}^{\star}$; compare the corresponding
master equation Eq.(7) in the main text); recall $\xi_{\Gamma}\approx\alpha_{\Gamma}\Gamma/\omega_{c}=4\alpha_{\Gamma}\gamma_{\phi}/\omega_{c}$,
with $4\alpha_{\Gamma}\approx0.4/\mu^{2}$ and $\gamma_{\phi}\equiv\Gamma/4$
(to match with our definition of $\gamma_{1}$). Accordingly, in the
typical scenario where $T_{2}^{\star}\ll T_{1}$ (as discussed in
the previous subsection), indeed relaxation-induced errors (as well
as similar incoherent excitation processes) can be safely neglected.
In the opposite regime, where pure-dephasing processes are negligible
(such that the decoherence timescale reaches its fundamental upper
limit $T_{2}\leq2T_{1}$, i.e. the qubit coherence is limited by spin
flips), the total error $\xi_{\mathrm{dec}}$ induced by qubit decoherence
is simply given by $\xi_{\mathrm{dec}}\approx\xi_{\gamma}\approx\alpha_{\gamma}\gamma_{1}/\omega_{c}$.
Finally, in the worst-case regime where the pure-dephasing rate and
the relaxation rate are comparable $\left(\gamma_{\phi}\approx\gamma_{1}\right)$,
the total error due to qubit decoherence amounts to $\xi_{\mathrm{dec}}=\xi_{\gamma}+\xi_{\Gamma}\approx2\alpha_{\gamma}\gamma_{1}/\omega_{c}\approx2\alpha_{\Gamma}\Gamma/\omega_{c}$,
i.e. just a factor of two larger than the decoherence-induced error
considered in the main text.

\section{Average Gate Fidelity \label{sec:Average-Gate-Fidelity}}

\begin{figure}
\includegraphics[width=0.9\columnwidth]{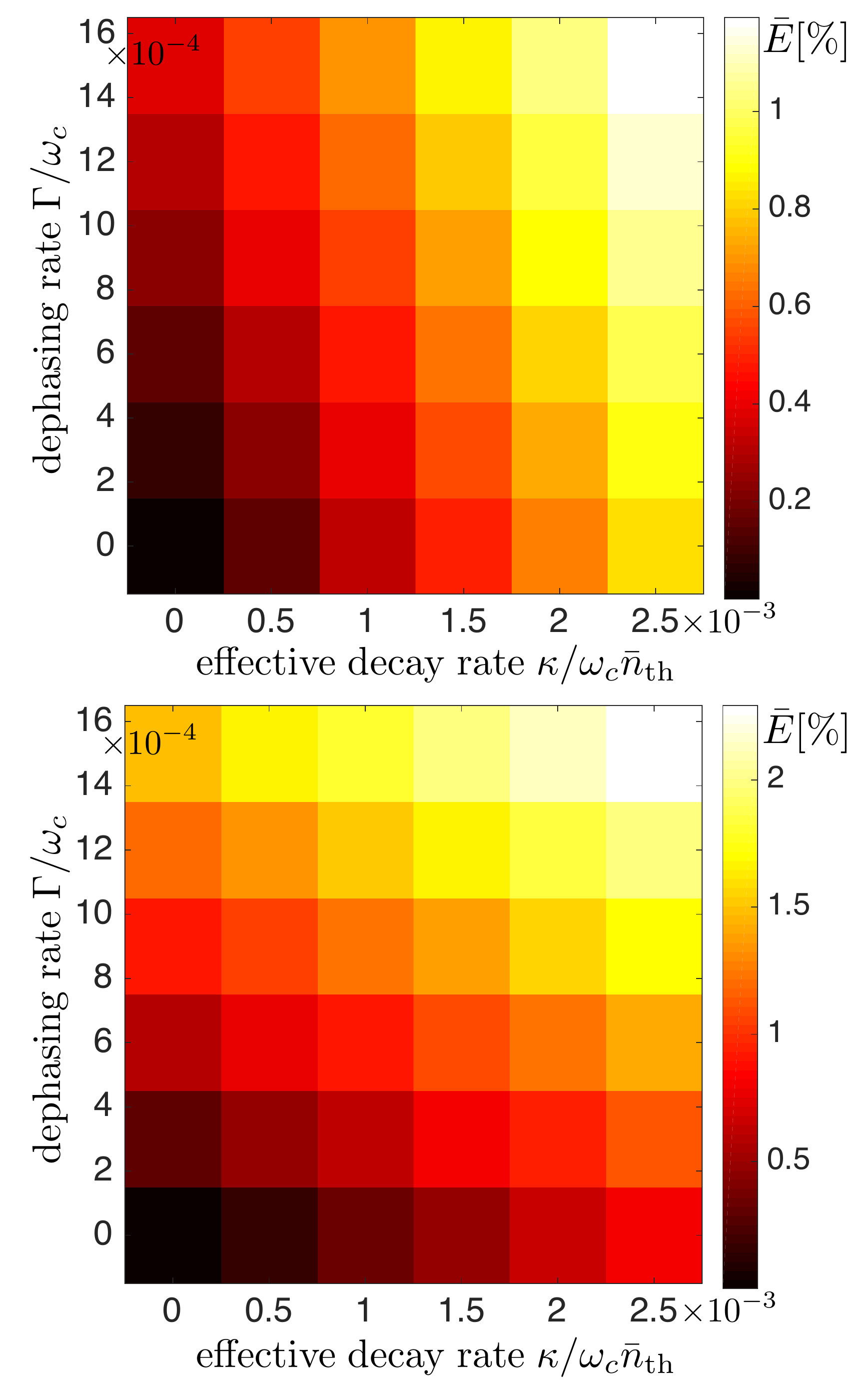}

\caption{\label{fig:total-average-gate-error-T2}(color online). Total average
gate error $\bar{E}$ (in percent) as a function of both the effective
rethermalization rate $\sim\kappa/\omega_{c}\bar{n}_{\mathrm{th}}\sim\bar{n}_{\mathrm{th}}/Q$
and the spin dephasing rate $\sim\Gamma/\omega_{c}$ for $g/\omega_{c}=1/4$
(top) and $g/\omega_{c}=1/8$ (bottom). Other numerical parameters:
$k_{B}T/\omega_{c}=2$ and $\omega_{q}=0$.}
\end{figure}

The average gate fidelity $\bar{F}$ is a useful measure in order
to quantify how well the completely-positive, trace-preserving quantum
operation $\mathcal{M}$ (in the presence of noise) approximates a
given unitary gate $U_{\mathrm{id}}$, which represents the ideal
(noise-free) evolution. Formally, it is defined as 
\begin{equation}
\bar{F}=\int d\psi\left<\psi\right|U_{\mathrm{id}}^{\dagger}\mathcal{M}\left(\left|\psi\right\rangle \left\langle \psi\right|\right)U_{\mathrm{id}}\left|\psi\right>,
\end{equation}
where the integral runs over the uniform (Haar) measure $d\psi$ on
state space, with $\int d\psi=1$ \cite{nielsen02-sm}. As shown in
Ref.\cite{nielsen02-sm}, $\bar{F}$ may be re-expressed as 
\begin{equation}
\bar{F}=\frac{dF_{\mathrm{ent}}+1}{d+1},\label{eq:average-gate-fid-entanglement-fid}
\end{equation}
where $d$ is the dimension of the Hilbert space ($d=4$ for two qubits)
and the entanglement fidelity $F_{\mathrm{ent}}$ is the fidelity
of the state obtained when $\mathcal{M}$ acts on one half of a maximally
entangled state with the state obtained from the action of the ideal
evolution; it is given by 
\begin{equation}
F_{\mathrm{ent}}=\frac{1}{d^{3}}\sum_{P\in G}\mathrm{tr}\left[P^{\dagger}U_{\mathrm{id}}^{\dagger}\mathcal{M}\left(P\right)U_{\mathrm{id}}\right].\label{eq:entanglement-fid}
\end{equation}
Here, $G$ is a set of $d\times d$ unitary operators, forming a basis
for a qudit, i.e., $\mathrm{tr}\left[P_{j}^{\dagger}P_{k}\right]=\delta_{jk}d$,
$j,k=1,\dots,d^{2}$. For two qubits we may take the set of Pauli
matrices modulo phase, comprising in total 16 operators $G=\left\{ \mathds1,\sigma_{i}^{\alpha},\sigma_{1}^{\alpha}\sigma_{2}^{\beta}\right\} $,
with $i=1,2$, $\alpha=x,y,z$. Experimentally, $\bar{F}$ may be
determined using standard state tomography \cite{nielsen02-sm}. 

\textit{Errors}.---The average gate error (infidelity) is defined
as $\bar{E}=1-\bar{F}$. As follows directly from Eq.(\ref{eq:average-gate-fid-entanglement-fid}),
it is related to the entanglement infidelity $E_{\mathrm{ent}}=1-F_{\mathrm{ent}}$
via $\bar{E}=d/\left(d+1\right)\times E_{\mathrm{ent}}$; thus, for
two qubits $\bar{E}=\left(4/5\right)E_{\mathrm{ent}}$. 

\textit{Numerical results}.---Numerical results for the average gate
error $\bar{E}$ are presented in Fig.\ref{fig:total-average-gate-error-T2}.
Here, the map $\mathcal{M}\left(P\right)$ is given implicitly as
$\mathcal{M}\left(P\right)=\mathrm{tr}_{a}\left[e^{\mathcal{L}t_{\mathrm{max}}}P\otimes\rho_{\mathrm{th}}\right]$,
where the superoperator $\mathcal{L}\bullet=-i\left[H,\bullet\right]+\mathcal{L}_{\mathrm{noise}}\bullet$
is the Liouvillian associated with the master equation given in Eq.(7)
in the main text, which includes undesired processes due to rethermalization
of the cavity mode and dephasing of the spins. Broadly speaking, our
numerical results for the (average) gate error $\bar{E}$ are comparable
to the ones obtained for the state infidelity $\xi=1-\mathcal{F}$,
as discussed in the main text. First, comparison of our results for
$g/\omega_{c}=1/4$ and $g/\omega_{c}=1/8$ shows that rethermalization-induced
errors are approximately independent of the spin-resonator coupling
$g$; for example, for $\Gamma=0$ and $\kappa/\omega_{c}\bar{n}_{\mathrm{th}}=2.5\times10^{-3}$
we find $\bar{E}_{\kappa}\approx0.82\%$ for both $g/\omega_{c}=1/4$
and $g/\omega_{c}=1/8$, respectively. Second, as expected, the dephasing
induced error scales as $\bar{E}_{\Gamma}\sim1/g^{2}\sim1/\mu^{2}$;
for example, as shown in Fig.\ref{fig:total-average-gate-error-T2},
for $\kappa=0$ and $\Gamma/\omega_{c}=1.5\times10^{-3}$, we find
$\bar{E}_{\Gamma}\approx0.376\%$ and $\bar{E}_{\Gamma}\approx1.49\%\approx4\times0.376\%$
for $g/\omega_{c}=1/4$ and $g/\omega_{c}=1/8$, respectively.


\begin{thebibliography}{10}

\bibitem{hanson08}R. Hanson and D. D. Awschalom, Nature \textbf{453},
1043 (2008). 

\bibitem{nielsen10}M. A. Nielsen and I. L. Chuang, Quantum Computation
and Quantum Information (Cambridge University Press, 2010). 

\bibitem{schreiber14}L. R. Schreiber and H. Bluhm, Nature Nanotech.
\textbf{9}, 966 (2014). 

\bibitem{knill05}E. Knill, Nature \textbf{434}, 39 (2005). 

\bibitem{nickerson13}N. H. Nickerson, Y. Li, and S. C. Benjamin,
Nat. Commun. \textbf{4}, 1756 (2013).

\bibitem{majer07}J. Majer \textit{et al.}, Nature \textbf{449}, 443
(2007).

\bibitem{silanp07}M. A. Sillanpää, J. I. Park, and R. W. Simmonds,
Nature \textbf{449}, 438 (2007). 

\bibitem{schmidt-kaler03}F. Schmidt-Kaler \textit{et al.}, Nature
\textbf{422}, 408 (2003).

\bibitem{soerensen99}A. Sorensen and K. Molmer, Phys. Rev. Lett.
\textbf{82}, 1971 (1999).

\bibitem{soerensen00}A. Sorensen and K. Molmer, Phys. Rev. A \textbf{62},
022311 (2000).

\bibitem{moelmer99}K. Molmer and A. Sorensen, Phys. Rev. Lett.
\textbf{82}, 1835 (1999).

\bibitem{milburn99}G. J. Milburn, arXiv:quant-ph/9908037 (unpublished). 

\bibitem{milburn00}G. J. Milburn, S. Schneider, and D. F. V. James,
Fortschr. Phys. \textbf{48}, 801 (2000).

\bibitem{poyatos98}J. F. Poyatos, J. I. Cirac, and P. Zoller, Phys.
Rev. Lett. \textbf{81}, 1322 (1998).

\bibitem{cirac00}J. I. Cirac and P. Zoller, Nature \textbf{404},
579 (2000).

\bibitem{garcia-ripoll03}J. J. Garcia-Ripoll, P. Zoller, and J. I.
Cirac, Phys. Rev. Lett. \textbf{91}, 157901 (2003).

\bibitem{garcia-ripoll05}J. J. Garcia-Ripoll, P. Zoller, and J. I.
Cirac, Phys. Rev. A \textbf{71}, 062309 (2005).

\bibitem{porras04}D. Porras and J. I. Cirac, Phys. Rev. Lett. \textbf{92},
207901 (2004).

\bibitem{leibfried03}D. Leibfried \textit{et al.}, Nature \textbf{422},
412 (2003).

\bibitem{kirchmair09}G. Kirchmair, J. Benhelm, F. Zähringer, R. Gerritsma,
C. F. Roos and R. Blatt, New J. Phys. \textbf{11}, 023002 (2009).

\bibitem{kerman13}A. J. Kerman, 
New Journal of Physics \textbf{15}, 123011 (2013). 

\bibitem{royer16}B. Royer, A. L. Grimsmo, N. Didier, and A. Blais,
arXiv:1603.04424 (unpublished). 

\bibitem{reilly15}D. J. Reilly, NPJ Quantum Information \textbf{1},
15011 (2015).

\bibitem{treutlein14}P. Treutlein, C. Genes, K. Hammerer, M. Poggio,
and P. Rabl, \textquotedbl{}Hybrid Mechanical Systems\textquotedbl{},
in: \textquotedbl{}Cavity Optomechanics\textquotedbl{}, ed. by M.
Aspelmeyer, T. J. Kippenberg, and F. Marquardt (Springer, Berlin 2014)
pp. 327-351. 

\bibitem{blais04}A. Blais, R.-S. Huang, A. Wallraff, S. M. Girvin,
and R. J. Schoelkopf, Phys. Rev. A \textbf{69}, 062320 (2004).

\bibitem{childress04}L. Childress, A. S. Sorensen, and M. D. Lukin,
Phys. Rev. A \textbf{69}, 042302 (2004).

\bibitem{taylor06}J. M. Taylor and M. D. Lukin, arXiv:cond-mat/0605144
(unpublished). 

\bibitem{jin12}P.-Q. Jin, M. Marthaler, A. Shnirman, and G. Schon,
Phys. Rev. Lett. \textbf{108}, 190506 (2012).

\bibitem{hu12}X. Hu, Y.-x. Liu, and F. Nori, Phys. Rev. B \textbf{86},
035314 (2012).

\bibitem{trif08}M. Trif, V. N. Golovach, and D. Loss, Phys. Rev.
B \textbf{77}, 045434 (2008).

\bibitem{gullans15}M. J. Gullans, Y.-Y. Liu, J. Stehlik, J. R. Petta,
and J. M. Taylor, Phys. Rev. Lett. \textbf{114}, 196802 (2015).

\bibitem{jin11}P.-Q. Jin, M. Marthaler, J. H. Cole, A. Shnirman,
and G. Schon, Phys. Rev. B \textbf{84}, 035322 (2011).

\bibitem{kulkarni14}M. Kulkarni, O. Cotlet, and H. E. Tureci, Phys.
Rev. B \textbf{90}, 125402 (2014).

\bibitem{srinivasa16}V. Srinivasa, J. M. Taylor, and C. Tahan, Phys. Rev. B \textbf{94}, 205421 (2016). 

\bibitem{frey12}T. Frey, P. J. Leek, M. Beck, A. Blais, T. Ihn, K.
Ensslin, and A. Wallraff, Phys. Rev. Lett. \textbf{108}, 046807 (2012).

\bibitem{petersson12}K. D. Petersson, L. W. McFaul, M. D. Schroer,
M. Jung, J. M. Taylor, A. A. Houck, and J. R. Petta, Nature \textbf{490},
380 (2012).

\bibitem{liu14}Y.-Y. Liu, K. D. Petersson, J. Stehlik, J. M. Taylor,
and J. R. Petta, Phys. Rev. Lett. \textbf{113}, 036801 (2014).

\bibitem{toida13}H. Toida, T. Nakajima, and S. Komiyama, Phys. Rev.
Lett. \textbf{110}, 066802 (2013). Also see: A. Wallraff, A. Stockklauser,
T. Ihn, J. R. Petta, and A. Blais, Phys. Rev. Lett. \textbf{111},
249701 (2013).

\bibitem{delbecq11}M. R. Delbecq, V. Schmitt, F. D. Parmentier, N.
Roch, J. J. Viennot, G. Feve, B. Huard, C. Mora, A. Cottet, and T.
Kontos, Phys. Rev. Lett. \textbf{107}, 256804 (2011).

\bibitem{viennot14}J. J. Viennot, M. R. Delbecq, M. C. Dartiailh,
A. Cottet, and T. Kontos, Phys. Rev. B \textbf{89}, 165404 (2014).

\bibitem{viennot15}J. J. Viennot, M. C. Dartiailh, A. Cottet, and
T. Kontos, Science \textbf{349}, 408 (2015).

\bibitem{zou14}L. J. Zou, D. Marcos, S. Diehl, S. Putz,
J. Schmiedmayer, J. Majer, and P. Rabl, Phys. Rev. Lett. \textbf{113},
023603 (2014). 

\bibitem{beaudoin16}F. Beaudoin, D. Lachance-Quirion, W. A. Coish,
M. Pioro-Ladriere, Nanotechnology \textbf{27}, 464003 (2016).

\bibitem{mi17}X. Mi, J. V. Cady, D. M. Zajac, P. W. Deelman, and J. R. Petta, 
Science \textbf{355}, 126 (2017).

\bibitem{stockklauser17}A. Stockklauser, P. Scarlino, J. V. Koski, S. Gasparinetti, C. K. Andersen, C. Reichl, W. Wegscheider, T. Ihn, K. Ensslin, and A. Wallraff,
Phys. Rev. X \textbf{7}, 011030 (2017).

\bibitem{schuetz15}M. J. A. Schuetz, E. M. Kessler, G. Giedke, L.
M. K. Vandersypen, M. D. Lukin, and J. I. Cirac, Phys. Rev. X \textbf{5},
031031 (2015).

\bibitem{chen15}J. C. Chen, Y. Sato, R. Kosaka, M. Hashisaka, K.
Muraki, and T. Fujisawa, Sci. Rep. \textbf{5}, 15176 (2015).

\bibitem{golter16}D. A. Golter, T. Oo, M. Amezcua, K. A. Stewart,
and H. Wang, Phys. Rev. Lett. \textbf{116}, 143602 (2016).

\bibitem{rabl09}P. Rabl, P. Cappellaro, M. V. Gurudev Dutt, L. Jiang,
J. R. Maze, and M. D. Lukin, Phys. Rev. B \textbf{79}, 041302(R) (2009).

\bibitem{rabl10}P. Rabl, S. J. Kolkowitz, F. H. L. Koppens, J. G.
E. Harris, P. Zoller, and M. D. Lukin, Nat. Phys. \textbf{6}, 602
(2010).

\bibitem{kepesidis13}K. V. Kepesidis, S. D. Bennett, S. Portolan,
M. D. Lukin, and P. Rabl, Phys. Rev. B \textbf{88}, 064105 (2013). 

\bibitem{bennett13}S. D. Bennett, N. Y. Yao, J. Otterbach, P. Zoller,
P. Rabl, and M. D. Lukin, Phys. Rev. Lett. \textbf{110}, 156402 (2013).

\bibitem{palyi12}A. Palyi, P. R. Struck, M. Rudner, K. Flensberg,
and G. Burkard, Phys. Rev. Lett. \textbf{108}, 206811 (2012).


\bibitem{cohen92}C. Cohen-Tannoudji, J. Dupont-Roc, and G. Grynberg, 
\textit{Atom-Photon Interactions: Basic Processes and Applications} (Wiley, New York, 1992).

\bibitem{schoelkopf08}R. J. Schoelkopf and S. M. Girvin, Nature \textbf{451},
664 (2008).

\bibitem{samkharadze15}N. Samkharadze, A. Bruno, P. Scarlino, G.
Zheng, D. P. DiVincenzo, L. DiCarlo, and L. M. K. Vandersypen, Phys.
Rev. Applied \textbf{5}, 044004 (2016). 

\bibitem{barends08}R. Barends, J. J. A. Baselmans, S. J. C. Yates,
J. R. Gao, J. N. Hovenier, and T. M. Klapwijk, Phys. Rev. Lett. \textbf{100},
257002 (2008). 

\bibitem{veldhorst14}M. Veldhorst \textit{et al.}, Nature Nano. \textbf{9},
981 (2014). 

\bibitem{chow12}Chow \textit{et al.}, Phys. Rev. Lett. \textbf{109}, 060501 (2012).















\bibitem{hanson07-sm}R. Hanson, L. P. Kouwenhoven, J. R. Petta, S.
Tarucha, and L. M. K. Vandersypen, Rev. Mod. Phys. \textbf{79}, 1217
(2007).

\bibitem{levy02}J. Levy, Phys. Rev. Lett. \textbf{89}, 147902 (2002).


\bibitem{cahill69}K. E. Cahill and R. J. Glauber, Phys. Rev. \textbf{177},
1857 (1969). 



\bibitem{gustafsson12-sm}M. V. Gustafsson, P. V. Santos, G. Johansson,
and P. Delsing, Nat. Phys.\textbf{ 8}, 338 (2012).

\bibitem{manenti16-sm}R. Manenti, M. J. Peterer, A. Nersisyan, E.
B. Magnusson, A. Patterson, and P. J. Leek, Phys. Rev. B \textbf{93},
041411(R) (2016).


\bibitem{malinowski16-sm}F. K. Malinowski \textit{et al.}, Nat. Nanotechnol. \textbf{12}, 16 (2017). 


\bibitem{zollerOnline}P. Zoller, \textit{Quantum Optics: Continous
Measurement, Stochastic Schrödinger Equations, Master Equations etc.},
lecture notes, available online (http://www.coqus.at/fileadmin/quantum/coqus\linebreak{}
/documents/Klemens\_Hammerer/QO\_Zoller\_Lecture2-Continous\_Measurement.pdf). 

\bibitem{rable14-lecture}P. Rabl, \textit{Advanced Quantum Optics},
lecture notes TU Vienna (2014).

\bibitem{guimond16}P.-O. Guimond, H. Pichler, A. Rauschenbeutel,
and P. Zoller, Phys. Rev. A \textbf{94}, 033829 (2016).


\bibitem{gardiner00}C. W. Gardiner and P. Zoller, \textit{Quantum
Noise} (Springer, Berlin, 2000).

\bibitem{yamamoto99}Y. Yamamoto and A. Imamoglu, \textit{Mesoscopic
Quantum Optics} (Wiley, New York, 1999).

\bibitem{carmichael02}H. J. Carmichael, \textit{Statistical Methods
in Quantum Optics 1} (Springer, Berlin, 2002).

\bibitem{deVega15}I. de Vega and D. Alonso, Rev. Mod. Phys. \textbf{89}, 15001 (2017).

\bibitem{paavola08}J. Paavola, \textit{Role of the environmental
spectrum in the decoherence and dissipation of a quantum Brownian
particle}, Master\textquoteright s thesis, University of Turku, 2008.

\bibitem{beaudoin11}F. Beaudoin, J. M. Gambetta, and A. Blais, Phys.
Rev. A \textbf{84}, 043832 (2011).

\bibitem{scala07}M. Scala, B. Militello, A. Messina, S. Maniscalco,
J. Piilo, and K. A. Suominen, J. Phys. A: Math. Theor. \textbf{40},
14527 (2007). 


\bibitem{blais07}A. Blais, J. Gambetta, A. Wallraff, D. I. Schuster,
S. M. Girvin, M. H. Devoret, and R. J. Schoelkopf, Phys. Rev. A \textbf{75},
032329 (2007).




\bibitem{wang15}H. Wang and G. Burkard, Phys. Rev. B \textbf{92},
195432 (2015).


\bibitem{bocquillon13-sm}E. Bocquillon \textit{et al.}, Science \textbf{339},
1054 (2013).

\bibitem{kornich14-sm}V. Kornich, C. Kloeffel, and D. Loss, Phys.
Rev. B \textbf{89}, 085410 (2014).

\bibitem{kloeffel13}C. Kloeffel and D. Loss, Annu. Rev. Condens.
Matter Phys. \textbf{4}, 51 (2013).

\bibitem{kroutvar04}M. Kroutvar, Y. Ducommun, D. Heiss, M. Bichler,
D. Schuh, G. Abstreiter, and J. J. Finley, Nature \textbf{432}, 81
(2004). 

\bibitem{amasha08}S. Amasha, K. MacLean, I. P. Radu, D. M. Zumbuhl,
M. A. Kastner, M. P. Hanson, and A. C. Gossard, Phys. Rev. Lett. \textbf{100},
046803 (2008).

\bibitem{woods02-sm}L. M. Woods, T. L. Reinecke, and Y. Lyanda-Geller,
Phys. Rev. B \textbf{66}, 161318 (2002). 

\bibitem{khaetskii01-sm}A. V. Khaetskii and Y. V. Nazarov, Phys.
Rev. B \textbf{64}, 125316 (2001). 

\bibitem{meunier07}T. Meunier \textit{et al.}, Phys. Rev. Lett. \textbf{98},
126601 (2007).

\bibitem{simmons11-sm}C. B. Simmons \textit{et al.}, Phys. Rev. Lett.
\textbf{106}, 156804 (2011).

\bibitem{prance12-sm}J. R. Prance \textit{et al.}, Phys. Rev. Lett.
\textbf{108}, 046808 (2012).

\bibitem{christ08-sm}H. Christ, J. I. Cirac, and G. Giedke, Phys.
Rev. B \textbf{78}, 125314 (2008). 

\bibitem{ams=0000FCss11}R. Amsüss \textit{et al.}, Phys. Rev. Lett.
\textbf{107}, 060502 (2011).

\bibitem{harrison06}J. Harrison, M. Sellars, and N. Manson, Diam.
Relat. Mater. \textbf{15}, 586 (2006). 

\bibitem{jarmola12}A. Jarmola, V. M. Acosta, K. Jensen, S. Chemerisov,
and D. Budker, Phys. Rev. Lett. \textbf{108}, 197601 (2012).


\bibitem{barfuss15}A. Barfuss, J. Teissier, E .Neu, A. Nunnenkamp,
and P. Maletinsky, Nature Phys. \textbf{11}, 820 (2015).

\bibitem{teissier14}J. Teissier, A. Barfuss, P. Appel, E. Neu, and
P. Maletinsky, Phys. Rev. Lett. \textbf{113}, 020503 (2014).

\bibitem{maurer12}P. C. Maurer et al., Science \textbf{336}, 1283
(2012).

\bibitem{nielsen02-sm}M. A. Nielsen, Phys. Lett. A \textbf{303},
249 (2002).

\end{thebibliography}
\end{document}